\documentclass{emulateapj}

\newcommand\gsim{\,\lower3pt\hbox{$\sim$}\llap{\raise2pt\hbox{$>$}}\,}
\newcommand\lsim{\,\lower3pt\hbox{$\sim$}\llap{\raise2pt\hbox{$<$}}\,}

\usepackage{graphicx}
\usepackage{epsfig}
\usepackage{subfigure}
\usepackage{natbib}

\shortauthors{LUGAZ ET AL.}

\shorttitle{CME-CME INTERACTION: INFLUENCE OF ORIENTATION}

\begin{document}

%% ------------------------------------------------------------------------ %%
%
%  TITLE
%
%% ------------------------------------------------------------------------ %%

\title{The Interaction of Two Coronal Mass Ejections: Influence of Relative Orientation}

\author{N.\ Lugaz\altaffilmark{1}, C.~J.~Farrugia\altaffilmark{1}, W. B. Manchester, IV\altaffilmark{2}, N. Schwadron\altaffilmark{1}}
\altaffiltext{1}{Space Science Center and Department of Physics, University of New Hampshire, Durham, NH, USA}
\altaffiltext{2}{Center for Space Environment Modeling, University of Michigan, Ann Arbor, MI, USA}

%% ------------------------------------------------------------------------ %%
%
%  ABSTRACT
%
%% ------------------------------------------------------------------------ %%

\begin{abstract}
We report on a numerical investigation of two coronal mass ejections (CMEs) which interact as they propagate in the inner heliosphere. We focus on the effect of the orientation of the CMEs relative to each other by performing four different simulations with the axis of the second CME rotated by 90$^\circ$ from one simulation to the next. Each magneto-hydrodynamic (MHD) simulation is performed in three dimensions (3-D) with the Space Weather Modeling Framework (SWMF) in an idealized setting reminiscent of solar minimum conditions. We extract synthetic satellite measurements during and after the interaction and compare the different cases. We also analyze the kinematics of the two CMEs, including the evolution of  their widths and aspect ratios. We find that the first CME contracts radially as a result of the interaction in all cases, but the amount of subsequent radial expansion depends on the relative orientation of the two CMEs. Reconnection between the two ejecta and between the ejecta and the interplanetary magnetic field (IMF) determines the type of structure resulting from the interaction. When a CME with a high inclination with respect to the ecliptic overtakes one with a low inclination, it is possible to create a compound event with a smooth rotation in the magnetic field vector over more than 180$^\circ$. Due to reconnection, the second CME only appears as an extended ``tail'', and the event may be mistaken for a glancing encounter with an isolated CME. This configuration differs significantly from the one usually studied of a multiple-magnetic cloud event, which we found to be associated with the interaction of  two CMEs with the same orientation.
\end{abstract}

\section{INTRODUCTION} \label{intro}

With an average of more than three coronal mass ejections (CMEs) per day near solar maximum, and a typical CME transit time from Sun to 1 AU between one and four days, CME-CME interaction should occur regularly in the inner heliosphere. Some of the earliest reports of likely CME-CME interaction were associated with the series of events in early August 1972 \citep[]{Intriligator:1976, Ivanov:1982} measured {\it in situ} by Pioneer 9 at 0.8~AU and near 1~AU by Soviet, European and American satellites. \citet{Ivanov:1982}, in particular, discusses the increase in geo-effective potential due to the interaction of shock waves with other shock waves and previous ejections. Multi-spacecraft measurements, especially with Helios in the inner heliopshere, revealed additional instances of CME-CME interaction \citep[]{Burlaga:1987, Farrugia:2004}. With the improvement of coronagraph observations, and specifically the large field-of-view of LASCO/C3 covering distances up to 32~$R_\odot$, the 1990s witnessed the first direct observations of CME-CME interactions \citep[]{Gopalswamy:2001} and further confirmations that they are associated with complex ejecta or compound streams at 1~AU \citep[]{Burlaga:2002,Burlaga:2003, Wang:2002}. It has also been proposed that some seemingly isolated CMEs measured {\it in situ} may in fact result from the interaction of multiple CMEs on their way to Earth \citep[]{Dasso:2009, Lugaz:2012b}. 
The past six years have seen a similar increase in detection of CME-CME interactions thanks to the wide field-of-view of the Heliospheric Imagers (HIs) \citep[]{Eyles:2009} onboard the {\it Solar-Terrestrial Relations Observatory} \citep[STEREO, see:][]{Kaiser:2008}. Recent HI observations of CME-CME interaction include the 2008 November CMEs \citep[]{CShen:2012}, the 2010 May CMEs \citep[]{Lugaz:2012b} and the 2010 August CMEs \citep[]{Harrison:2012, Liu:2012, Temmer:2012}. Analysis of these events often combine remote-sensing observations with {\it in situ} measurements and sometimes numerical simulations \citep[]{Lugaz:2009b, Webb:2009,Webb:2013}.

The interaction of successive CMEs provides one of the best situations to study the physics of complex space plasma phenomena, such as the propagation of a shock wave inside a low-$\beta$ magnetically dominated structure, the merging of shock waves, and the reconnection between the large-scale structures of magnetic clouds. Additionally, interacting CMEs are often associated at Earth with extended periods of strong southward $B_z$ \citep[e.g.][]{Wang:2003a,Farrugia:2006} and with intense and long-lived geomagnetic storms \citep[]{Burlaga:1987, Farrugia:2006, Farrugia:2006b, Xie:2006}. 

Because direct heliospheric observations were rare until 10 years ago, and because of the complexity of the interaction process, numerical simulations have been one of the methods of choice to investigate CME-CME interaction. The first simulations were done in idealized settings, often without a solar wind \citep[]{Vandas:1997,Schmidt:2004} and more recent simulations have been performed using 2.5-D and 3-D magneto-hydrodynamical (MHD) codes with solar wind models \citep[e.g.]{Wu:2002, Odstrcil:2003, Lugaz:2005b, Xiong:2006,Xiong:2009, FShen:2011} or even for real events \citep[]{Lugaz:2007, WuCC:2007, FShen:2013}. Most of these simulations were performed with relative coarse grids with minimum cell size in the heliosphere (past 20~$R_\odot$) close to 1~$R_\odot$. Using adaptive mesh refinement (AMR), we previously performed two high-resolution 3-D simulations, one with a minimum resolution along the Sun-Earth line of 0.06~$R_\odot$ from the Sun to 0.6~AU \citep[]{Lugaz:2005b} and another with a minimum resolution along the Sun-Earth line of at least 0.12~$R_\odot$ from the Sun to 1~AU \citep[]{Lugaz:2007}.
  
There has only been one series of parametric studies performed in 2.5-D with a relative coarse uniform resolution of 1.5~$R_\odot$ to investigate the effects of the different speed and direction of a shock  overtaking a CME in 50 simulations \citep[]{Xiong:2006b} and the effects of the different speed and direction of a CME overtaking another one in 34 additional simulations \citep[]{Xiong:2009}. Because of the continuous improvement in computing capabilities, it is now possible to perform a series of global 3-D numerical simulations with AMR and minimum cell size of the order of 0.1~$R_\odot$  or less. Higher resolution is extremely important, in particular to resolve the jump in the plasma quantities across fast shocks and also to better resolve the reconnection between the magnetic ejecta. It should be noted that a spatial resolution of 1~$R_\odot$ at 1~AU corresponds to a 20-minute temporal resolution in synthetic satellite measurements for a typical speed of 580~km~s$^{-1}$. Setting the cell size 10 times smaller in each dimension results in a temporal resolution of 2 minutes. This should be compared with the 3-second temporal resolution of {\it Wind} measurements, for example. By performing simulations in idealized settings and not focusing on reproducing observations and measurements from an actual event, we are able to focus on the physical causes of the CME evolution during and after their interaction.

In this article, we present the first four of a series of 3-D MHD simulations of the interaction of two CMEs. We focus on the effect of different orientations of the overtaking CME, since this can only be studied accurately with 3-D simulations.  In section \ref{model}, we give an overview of the numerical set-up and models used, followed by a description of the evolution of the first CME until the launch of the second one. In section \ref{CME2a}, we give an overview of one case, which will serve as the ``base'' case in the rest of our investigations. We analyze and compare the three other simulations in \ref{CME2bd} and discuss synthetic satellite measurements at 34 and 50~$R_\odot$ in section \ref{satellite}. We discuss our results and conclude in section \ref{conclusion}.

%%%%%%%%%%%%%%%%%%%%%%%%%%%%%%%%%%%%%%%%%%%%%%%%%%%%%%%%%%%%%%%%%%%%%%%%%%%%
\section{Numerical Models and Set-up} \label{model}
%%%%%%%%%%%%%%%%%%%%%%%%%%%%%%%%%%%%%%%%%%%%%%%%%%%%%%%%%%%%%%%%%%%%%%%%%%%%%
\subsection{Simulation Set-up}

We use the Space Weather Modeling Framework \citep[SWMF,][]{Toth:2007, Toth:2012} with the Solar Corona (SC) and Inner Heliosophere (IH) modules to perform the simulations. The SC and IH components of the SWMF rely on the 3-D MHD code, BATS-R-US \citep[]{Gombosi:2001}. In addition to other space plasma physics questions, it has been used previously to study CME initiation and propagation, including the interaction of CMEs \citep[]{Roussev:2003a,Manchester:2004b, Manchester:2012, Lugaz:2005b,Lugaz:2007, Cohen:2008a}. The simulation domains are cubes of 48 $\times$ 48 $\times$ 48 $R_\odot$ and 440 $\times$ 440 $\times$ 440 $R_\odot$ centered at the Sun, for SC and IH, respectively. The coupling between the two domains occur on a sphere at 19~$R_\odot$. We use a Cartesian grid with initially about 166,000 blocks of $4 \times 4 \times 4$ cells each in SC and about 223,000 blocks of  $4 \times 4 \times 4$ cells each in IH, corresponding to a initial number of cells of 24.9 millions, ranging in size from 0.023~$R_\odot$ in a shell of 1.2~$R_\odot$ around the solar surface to 3.44~$R_\odot$ in IH away from the Sun-Earth line. We initially resolve the Sun-Earth line in the direction of the eruptions as shown in the left panel of Figure~1. During the course of the simulation, we perform AMR in SC until the launch of the second CME for a maximum number of blocks of 240,000 corresponding to 15.3 million cells. We perform AMR in IH starting 2 hours after the launch of the second CME and resolve the Sun-Earth line until 0.25~AU for a maximum number of blocks of 282,000 corresponding to 18 million cells. The final grid structures in SC and IH are shown in the middle and right panels of Figure~1. Note that, in all the panels of this Figure, blocks are plotted; each block is divided into $4^3$ cells.

 %%%%%%%%%%%%%%%%%%%%%%%
\begin{figure*}[ht*]
\centering
{\includegraphics*[width=5.4cm]{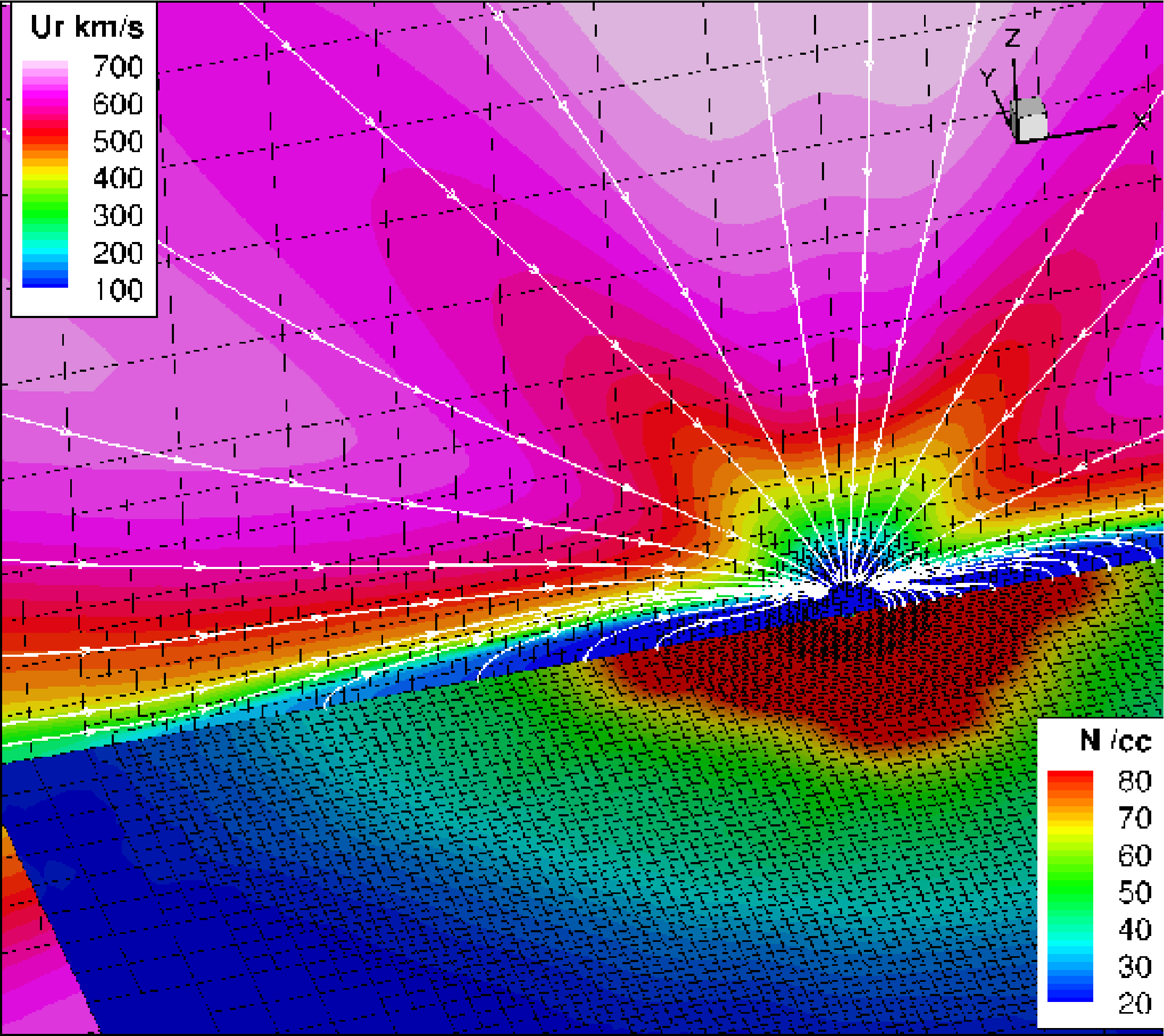}}
{\includegraphics*[width=5.4cm]{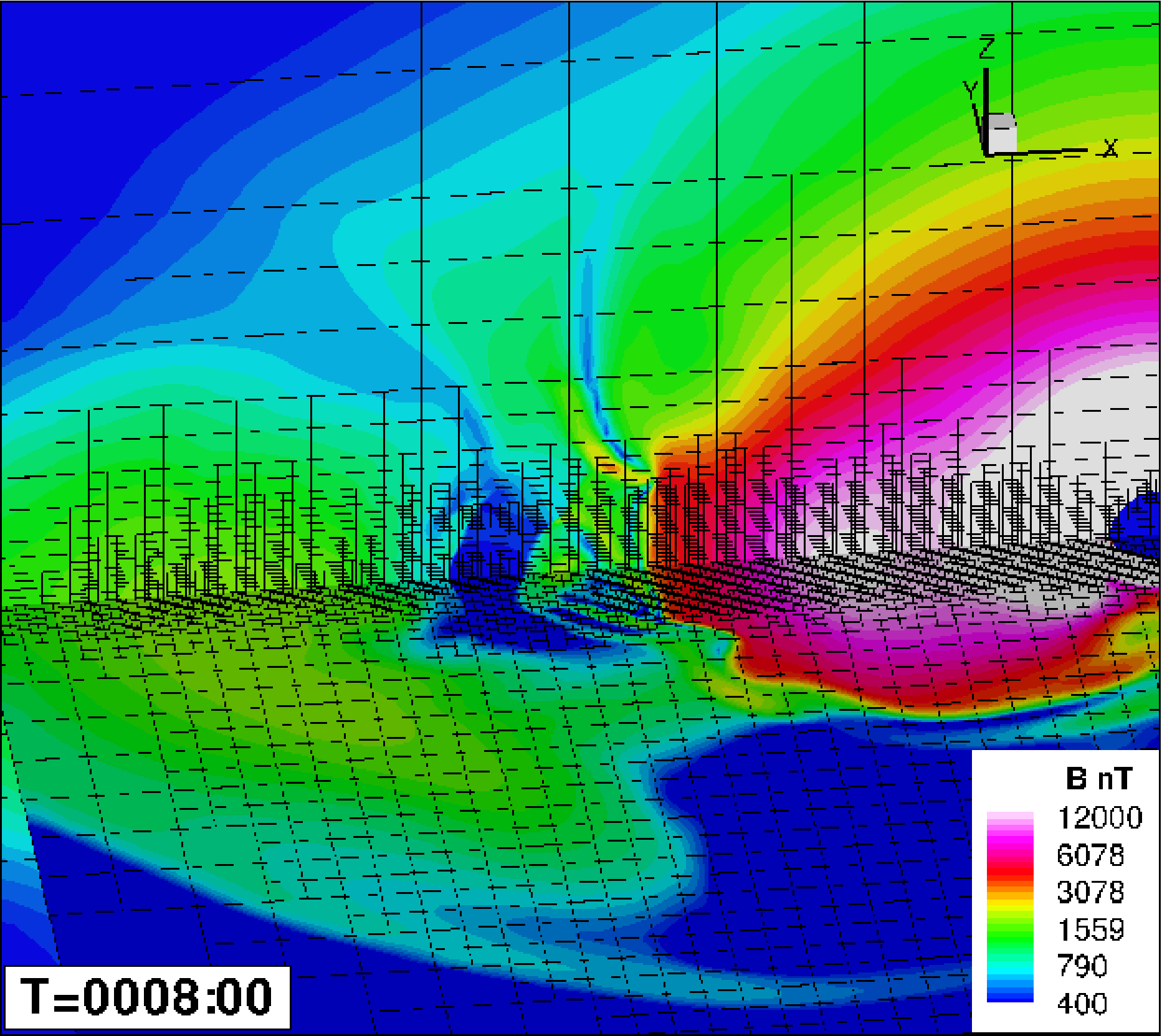}}
{\includegraphics*[width=5.4cm]{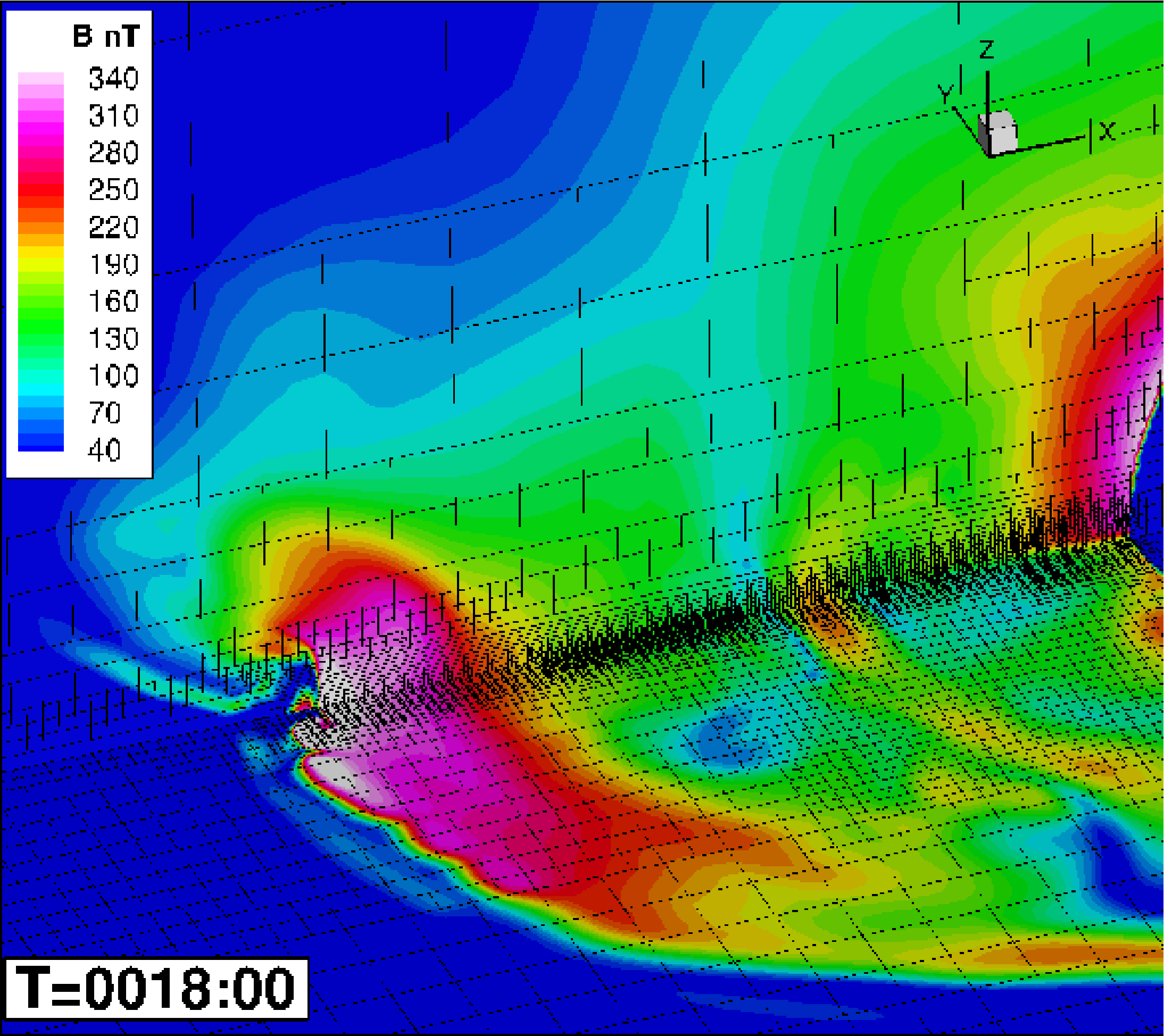}}
\caption{Left: Initial solar wind and grid structures before the launch of CME1. The radial velocity is shown on $x-z$ cut passing through the central meridian with projections of the magnetic field lines as white lines. The density scaled by 1/$r^2$ is shown in an ecliptic $x-y$ cut. The block structure is shown in black. Middle: Grid structure at the maximum of the block number in SC (at time t = 8 hours) with the magnetic field strength color-coded in a meridian and ecliptic cuts. Right: Same as center but for IH at time t = 18 hours.}
\end{figure*}
%%%%%%%%%%%%%%%%%%%%%%%%

\subsection{Solar Wind and CME Models}

We use the solar wind model of \citet{Holst:2010} where Alfv{\'e}n waves drive the solar wind \citep[see also][]{Evans:2012}. Contrary to other works with this model \citep[]{Manchester:2012, Jin:2013}, the simulations presented here were performed with a single fluid (same electron and proton temperature) and no heat conduction. While some of the shock physics is not expected to be as well reproduced with this version of the model, it has the notable advantage of allowing for a faster timestep. A more complete comparison of the effect of the solar wind model on the CME properties can be found in \citet{Pomoell:2012}. 
To set-up the solar magnetic field, we use a simple non-tilted dipole with an octopole component resulting in a maximum magnetic field strength of 5.5~G at the solar poles with a polarity corresponding to that of even solar cycles (such as the current solar cycle 24). The only free parameters of the model are the dissipation length which we set at a value of 380~km and a normalization parameter to control the mass flux at 1~AU. Both parameters were chosen in order to reproduce  values of the bimodal solar wind density and velocity typical of solar minimum conditions. Due to our choice of solar magnetic field, the solar wind is axi-symmetric. The initial density and radial velocity of the solar wind in the corona are shown in the left panel of Figure~1.

To initiate the CMEs, we use the flux rope model of \citet{Gibson:1998} (GL), which we used previously to study isolated and interacting CMEs \citep[]{Manchester:2004a,Lugaz:2005b}. A numerical comparison of this model with the flux rope model of \citet{Titov:1999} (TD) modified in \citet{Lugaz:2007} was performed with the SWMF by \citet{Loesch:2011}. The GL flux rope has a larger initial spatial extent but is less twisted than the modified TD flux rope. The flux rope is inserted in a state of force imbalance due to the increase magnetic pressure in the flux rope. %which is not balanced by the coronal magnetic field.
Therefore, it erupts as soon as it is inserted. Previous studies have shown that the evolution after the first 15 to 20 minutes is realistic, while the early rise phase is obviously not captured by this implementation of the numerical model \citep[e.g., see][]{Manchester:2004a}. We do not attempt to understand how successive CMEs may occur in rapid succession \citep[for example as sympathetic eruptions, as recently studied by][]{Torok:2011,Lynch:2013}.

In this series of simulations, both CMEs are inserted, seven hours after each other, at the same location on the solar surface: at the solar equator towards the $-x$ direction. Due to the solar rotation, the two CMEs propagate about 4$^\circ$ from each other. The parameters of the GL flux rope for CME1 are chosen as follows: $r_0 = 1~R_\odot, r_1 = 1.5~R_\odot, a = 0.25~R_\odot, a_1 = 0.45$, where $r_0$ and $r_1$ are the radius of the flux rope and heliocentric distance of the center of the flux rope, respectively. $a$ and $a_1$ are parameters controlling the CME initial internal energy (through stretching in the radial direction) and magnetic field strength. Full details on the GL flux rope implementation in BATS-R-US can be found in \citet{Manchester:2004a} and we have used the same convention to name the parameters. With these parameters, CME1 is initiated with a center at 1.74~$R_\odot$ and is initially about 0.96~$R_\odot$ wide in the radial direction. The maximum magnetic field inside the flux rope is 3.75~G. CME1 is initiated with an orientation of 180$^\circ$ corresponding to an axial magnetic field in the $+y$ direction and a north-to-south rotation of the poloidal field. Following the notation of \citet{Bothmer:1998}, this is a NES, right-handed CME, with low inclination with respect to the ecliptic.
The parameters of the GL flux rope for CME2 are $r_0 = 1~R_\odot, r_1 = 1.5~R_\odot, a = 0.25~R_\odot, a_1 = 0.6$ and we only varied the orientation of the flux rope with values of 180$^\circ$, 0$^\circ$, 90$^\circ$ and $-90^\circ$, corresponding to cases A, B, C and D described below, respectively. In all four cases, CME2 is also right-handed and the maximum magnetic field strength inside CME2 is initially 5.8~G.

\subsection{Evolution of CME1 Until the Launch of CME2}
The evolution of CME1 is presented in Figure~2. A snapshot of the scaled density of CME1 in the $x-z$ meridional plane, 30 minutes after its initiation with 3-D magnetic field lines color-coded with the radial velocity is shown in the left panel of Figure~2. This shows the core of the GL flux rope with the eastward axial field and a high density region surrounded by a low-density cavity with stronger poloidal field. A snapshot of the radial velocity of CME1 in the $x-z$ meridian plane, three hours after its initiation with 3-D magnetic field lines is shown in the middle panel of Figure~2. It shows that CME1 drives a shock and the core of axial field is still confined into a relatively compact area. The right panel of Figure~2 shows the time-height evolution of CME1 in black. The vertical lines mark the start of the interaction with CME2, discussed in the next section. We track the CME front, center and back as well as the position of the shock wave. Here, we define the center of the magnetic ejecta as the location of the maxima of the axial magnetic field strength, $B_y$, and the boundaries of the ejecta as the locations where the ratio of poloidal to axial magnetic fields, $\sqrt{B_x^2 + B_z^2}/|B_y|$ for CME1, is maximum.  For CME1, this is typically obtained where $B_y$ vanishes.

We also calculate the CME angular width as the ratio of its maximum latitudinal extent to the heliocentric position of its center, as well as the aspect ratio of CME1, defined as the ratio of the latitudinal extent to the radial width of the CME. For a discussion of the expected evolution of the angular width and aspect ratio of a CME, see for example \citet{Savani:2011a}. Note that here we only focus on the magnetic ejecta, not the density structure which is what is usually tracked in coronagraphic images to determine the CME angular width and aspect ratio \citep[see][for a discussion of the appearance of magnetic ejecta in Thomson scattered light]{THoward:2012b}.

%%%%%%%%%%%%%%%%%%%%%%%
\begin{figure*}[ht*]
\begin{minipage}[b]{0.59\linewidth}\centering
{\includegraphics*[width=4.8cm]{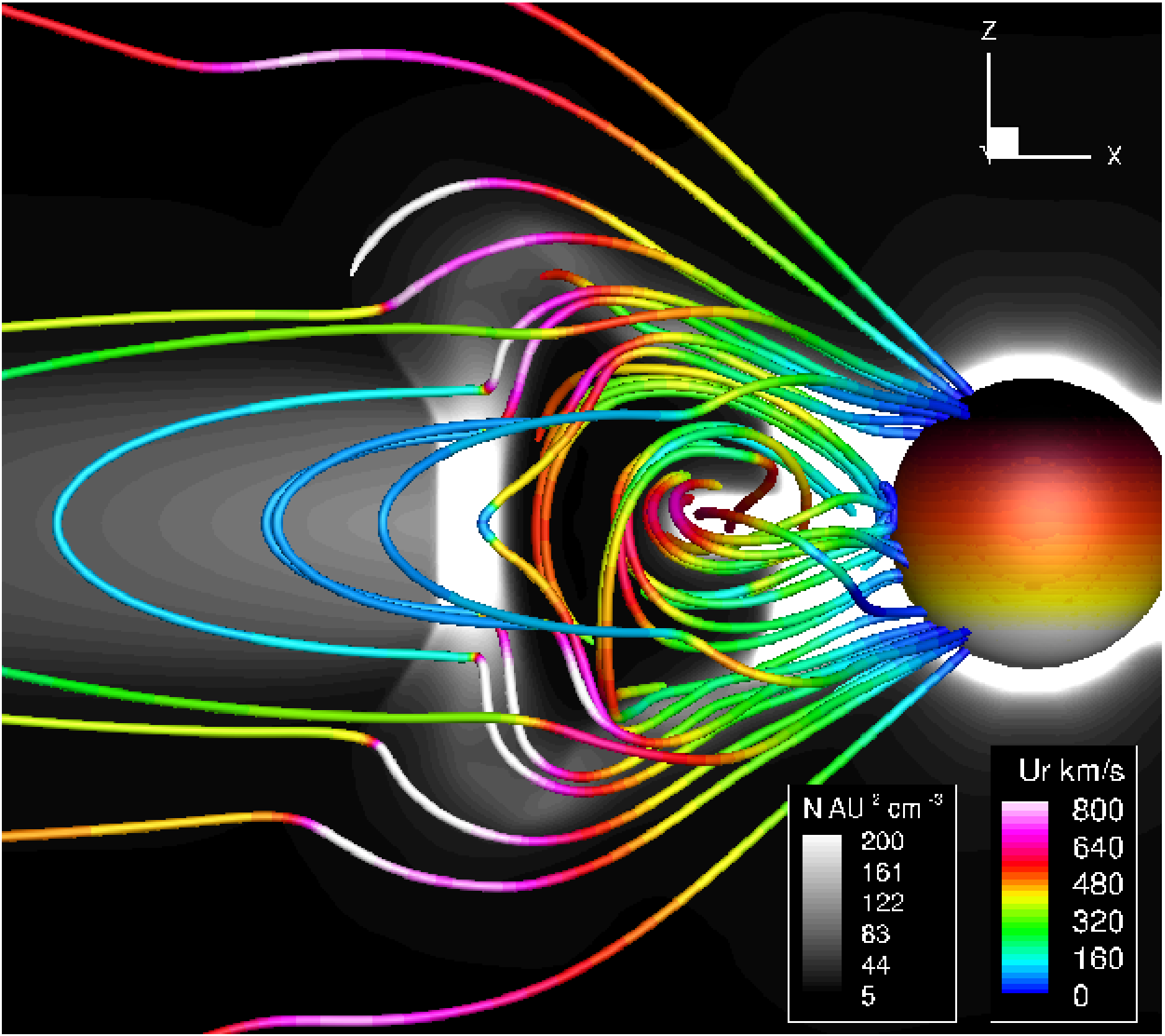}}
{\includegraphics*[width=4.8cm]{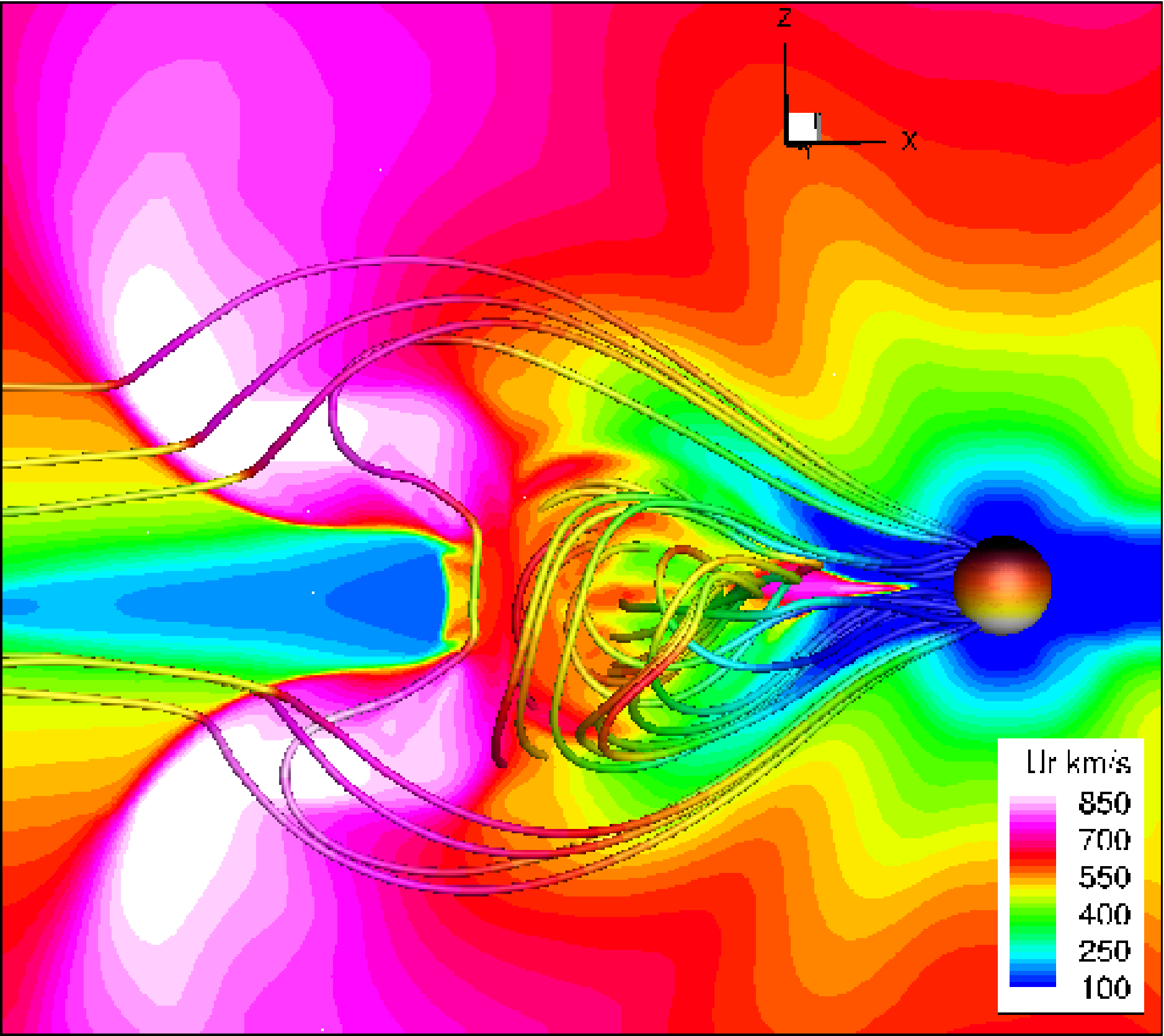}}
\end{minipage}
\begin{minipage}[b]{0.41\linewidth}\centering
{\includegraphics*[width=6.6cm]{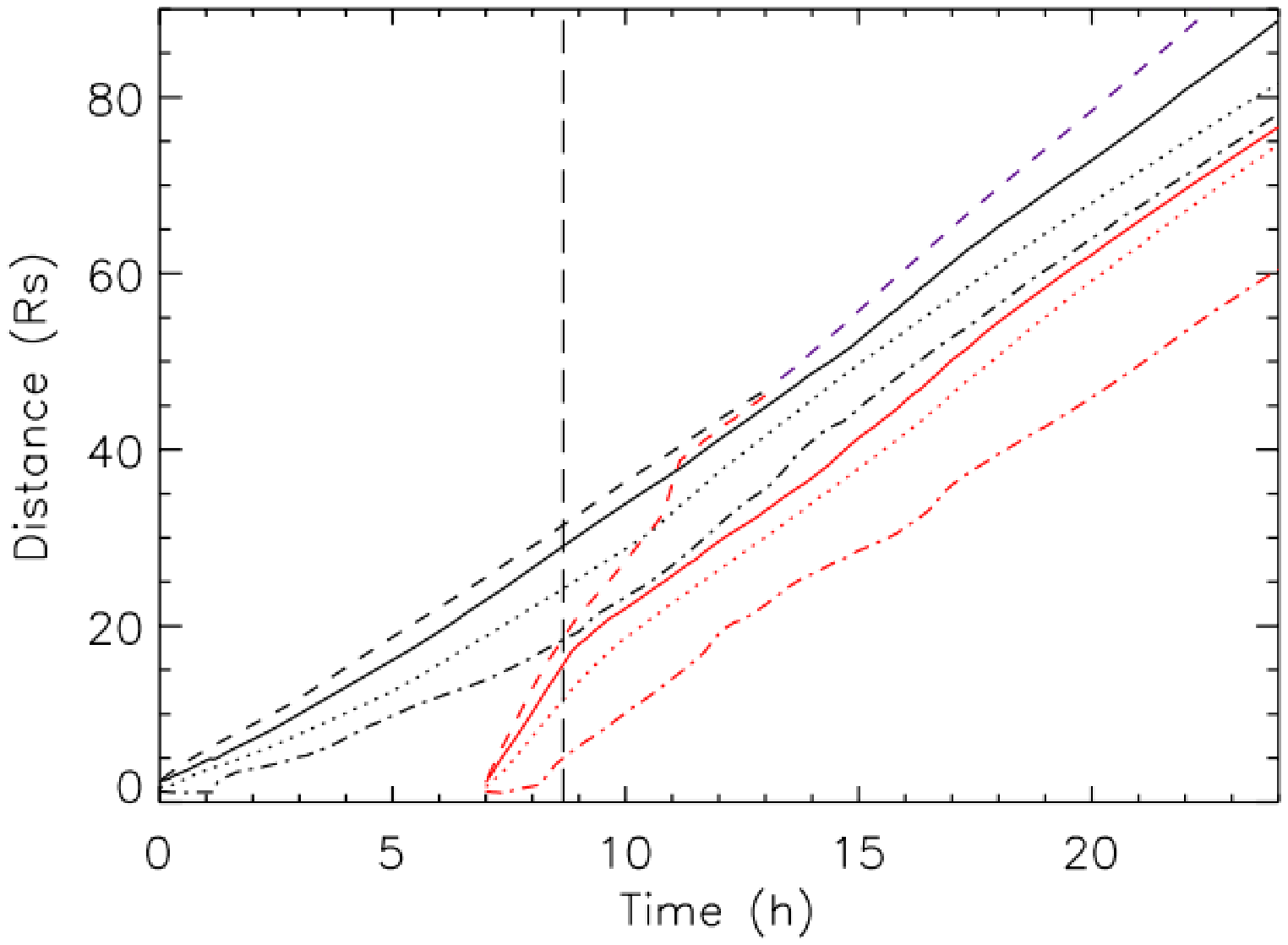}}
\end{minipage}
\caption{Evolution of the first CME showing 2 meridional cuts at 30 minutes ({\it left}) and 3 hours ({\it middle}) with 3-D magnetic field lines color-coded with the radial velocity. In the left panel, the density scaled by 1/$r^2$ is shown in black and white. An animated version of these panels is available online. The right panel shows a time-height plot of the two CMEs for Case A with the shocks, CME front, center and back plotted with dashed, solid, dotted and dash-dotted lines, respectively. CME2a is plotted in red and the vertical dashed line points to the beginning of the interaction at 8.67 hours. For all cases, the time-height plot of CME1 is the same up to 8 hours.}
\end{figure*}
%%%%%%%%%%%%%%%%%%%%%%%%

We estimate the errors in determining direct position measurements in the ecliptic plane to be about $\pm 0.1~R_\odot$, which means that the radial width has uncertainties of $\pm 0.2~R_\odot$. The error in measuring the CME latitudinal extent is larger since the CMEs are fully inside the most resolved area only up to a distance of 7-10~$R_\odot$ (see grid structure in middle panel of Figure~1). The uncertainties in the aspect ratio increases from $\pm 0.05$ early on to about $\pm 0.4$ at the end of the simulation. Similarly, the error in the CME angular width increases from $\pm 0.2^\circ$ to about $\pm 3^\circ$ at the end of the simulation. We estimate the errors in the propagation speed to be about $\pm 25$~km~s$^{-1}$.

For the first 70 minutes after the initiation, the aspect ratio decreases from an initial value of 1.5 to a minimum of 0.7. This happens because the CME is initiated with a large angular width of 45$^\circ$. As the CME propagates, its angular width remains at first nearly constant, decreasing to only 39$^\circ$, while its radial width increases. This first phase does not reproduce the observed behavior of CMEs in the low corona \citep[for example the over-expansion discussed in][]{Spiro:2010b}. This is a direct consequence of the chosen initiation mechanism with an out-of-equilibrium flux rope; however, we believe it does not affect the CME behavior at later times. %In the meantime, the CME over-expands in the radial direction resulting in a decreasing aspect ratio. 

After 70 minutes, the CME radial expansion slows down and the CME continues to expand in the latitudinal direction; the aspect ratio increases and stabilizes at a value of about 1.2 from about four to nine hours after the CME initiation. Therefore, based on the magnetic field, the CME is only slightly elliptical, whereas it appears to have a much more pronounced  pancake shape based on the density structure. A similar result was recently reported in a simulation by \citet{Savani:2013} where the magnetic flux remains more concentrated than azimuthal cuts make it appear as the CME propagates from the Sun to 1~AU. 

If we focus on the first four hours of the CME propagation, before any interaction, the radial width, $W$, of CME1 can be fitted by a power-law of the form $W = 0.27~r^{0.77}$, with both quantities expressed in AU and where $r$ is the radial distance of the center of the magnetic ejecta. This is similar, both for the width at 1~AU and the radial dependence, to previous theoretical works \citep[]{Demoulin:2009}, statistical works based on {\it in situ} measurements \citep[]{Bothmer:1998, Liu:2005, Gulisano:2010} and recent works based on the analysis of STEREO images \citep[]{Savani:2009, Nieves:2012, Lugaz:2012b}. It illustrates the fact that the radial expansion in our simulation, before the interaction, is realistic, validating both the CME and solar wind models.

Seven hours after its initiation, at the time of the launch of CME2, the shock driven by CME1 is at 25.6~$R_\odot$, the boundaries of the magnetic ejecta are at 22.9 and 13.9~$R_\odot$ with the center at 18.8~$R_\odot$. The propagation speed of the center of CME1 is about 600~km~s$^{-1}$ and the the upper and lower boundaries of the magnetic ejecta propagate with speeds of about 650~km~s$^{-1}$ and 480~km~s$^{-1}$, respectively. The latitudinal extent of CME1 is about 10.8~$R_\odot$ corresponding to an angular width of 30$^\circ$ (or a half-cone angle of about 15$^\circ$ in latitude); its aspect ratio is 1.2.

%%%%%%%%%%%%%%%%%%%%%%%
\begin{figure*}[ht*]
\centering
{\includegraphics*[width=6.5cm]{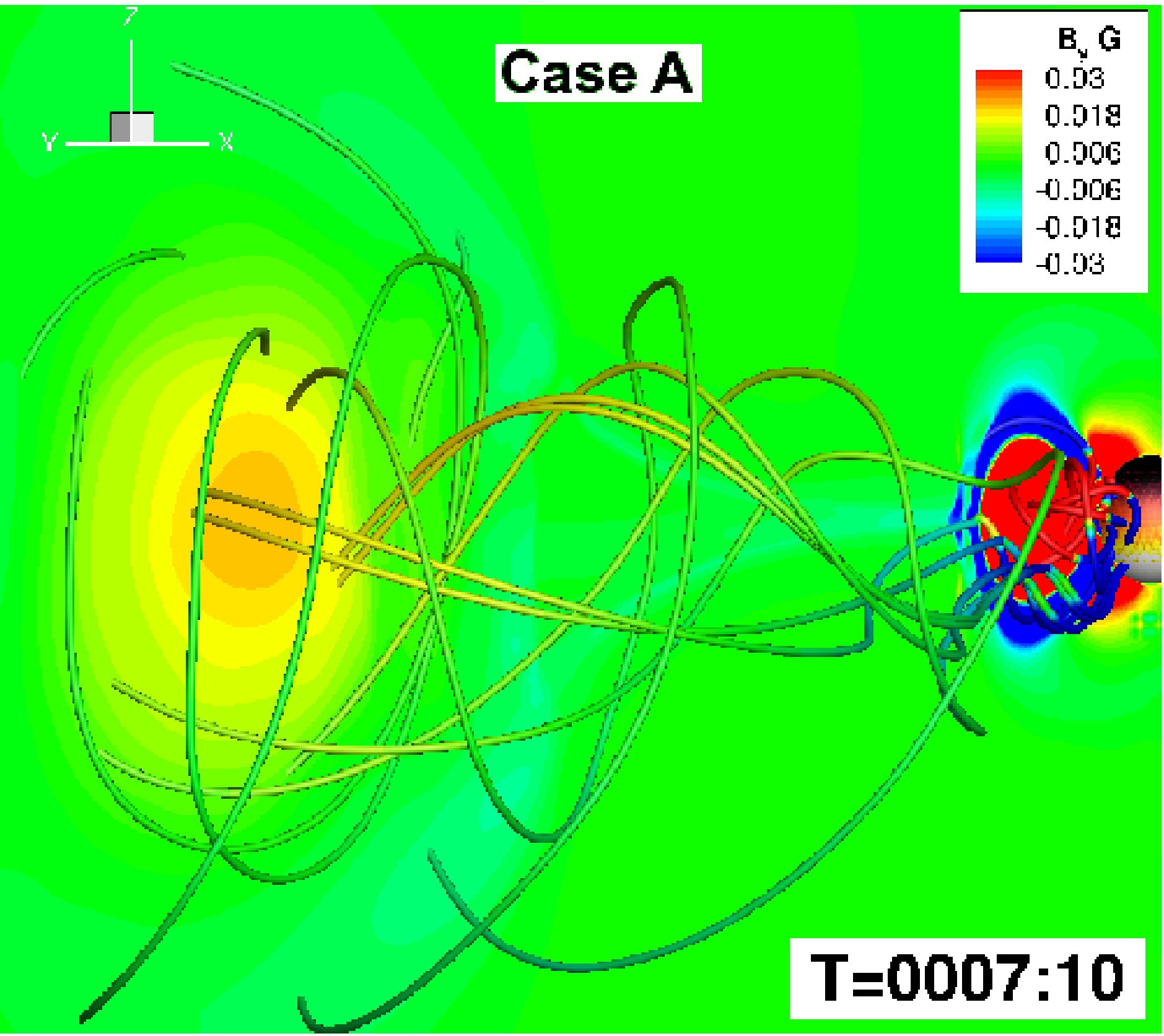}}
{\includegraphics*[width=6.5cm]{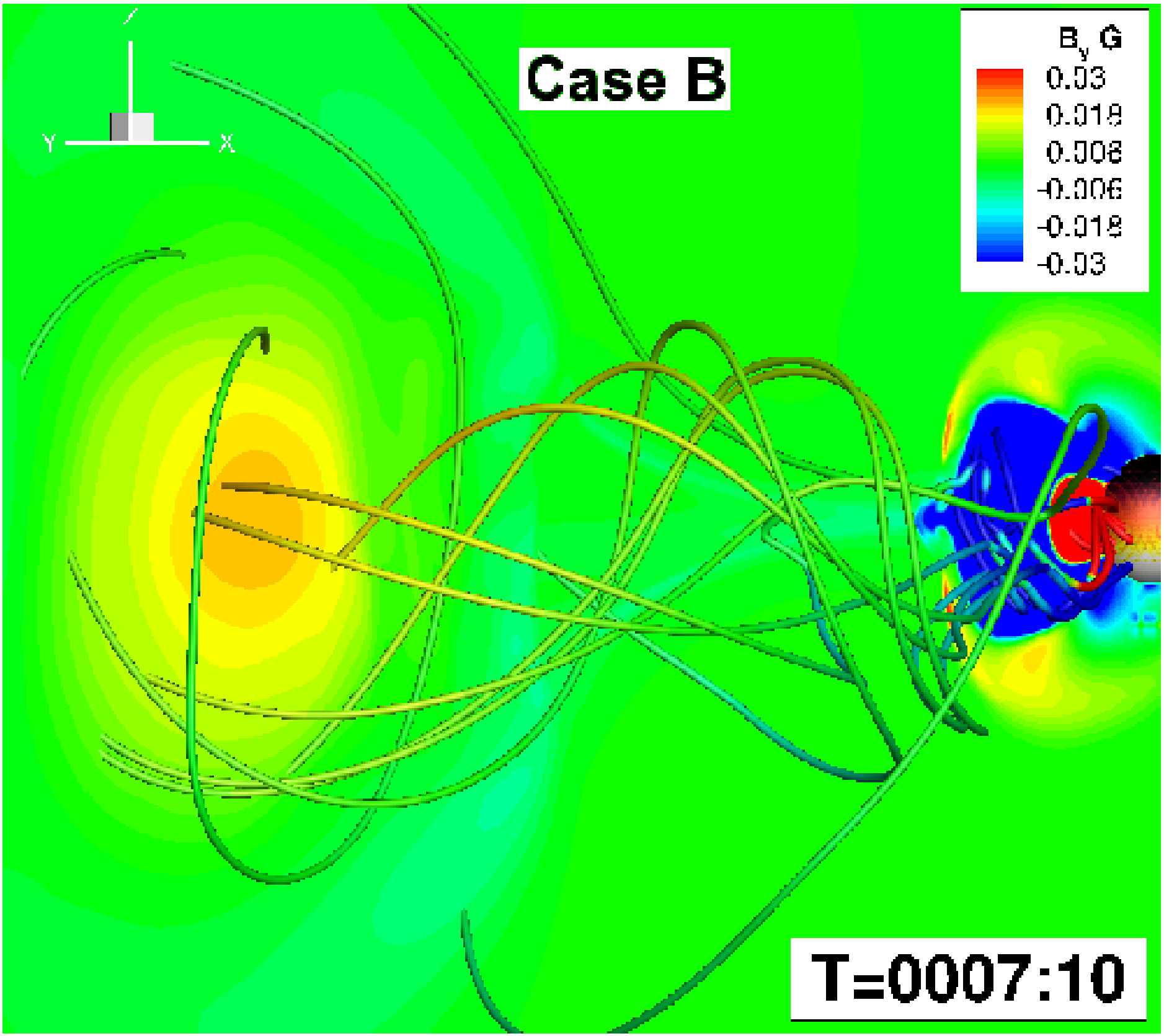}}\\
{\includegraphics*[width=6.5cm]{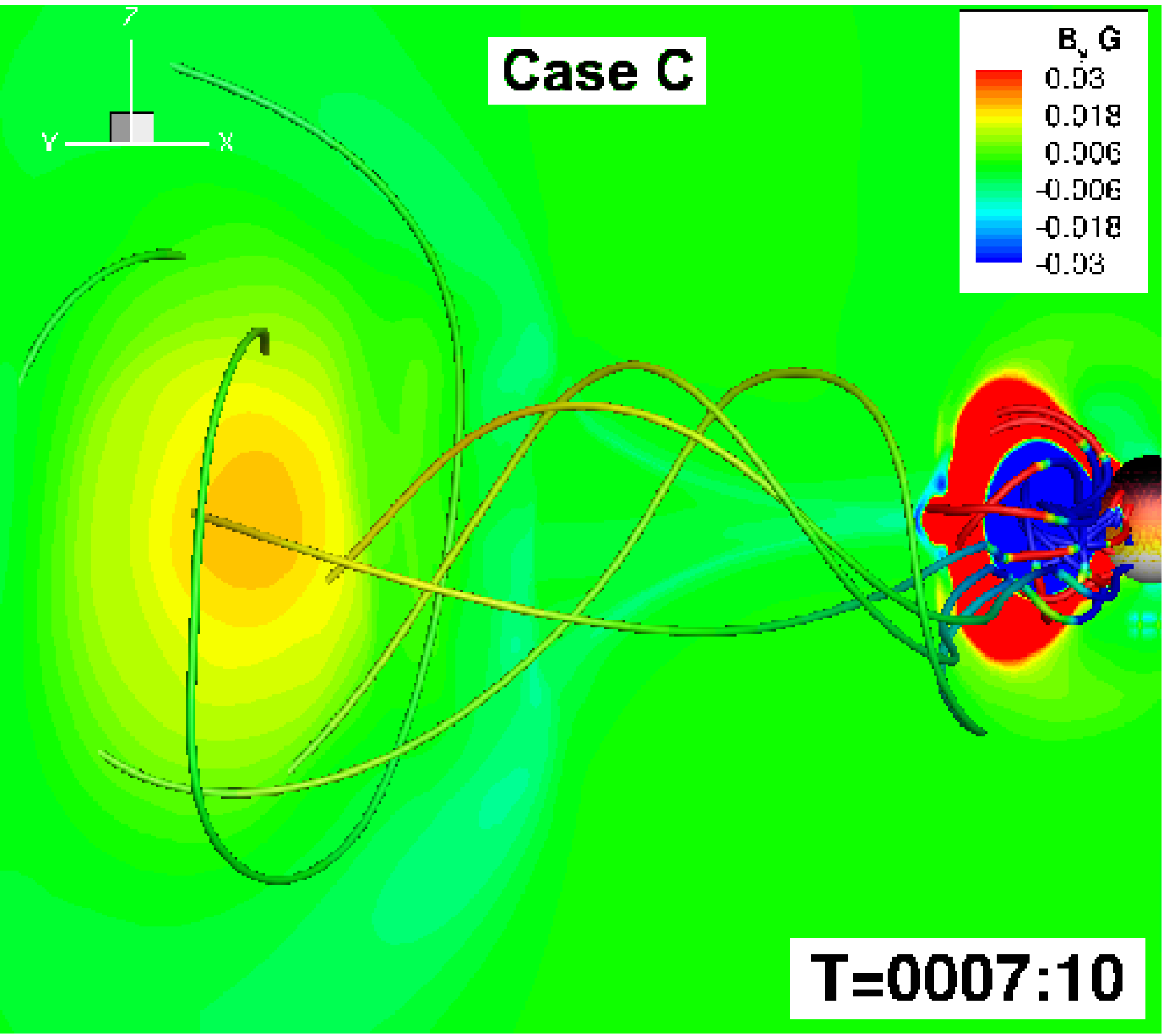}}
{\includegraphics*[width=6.5cm]{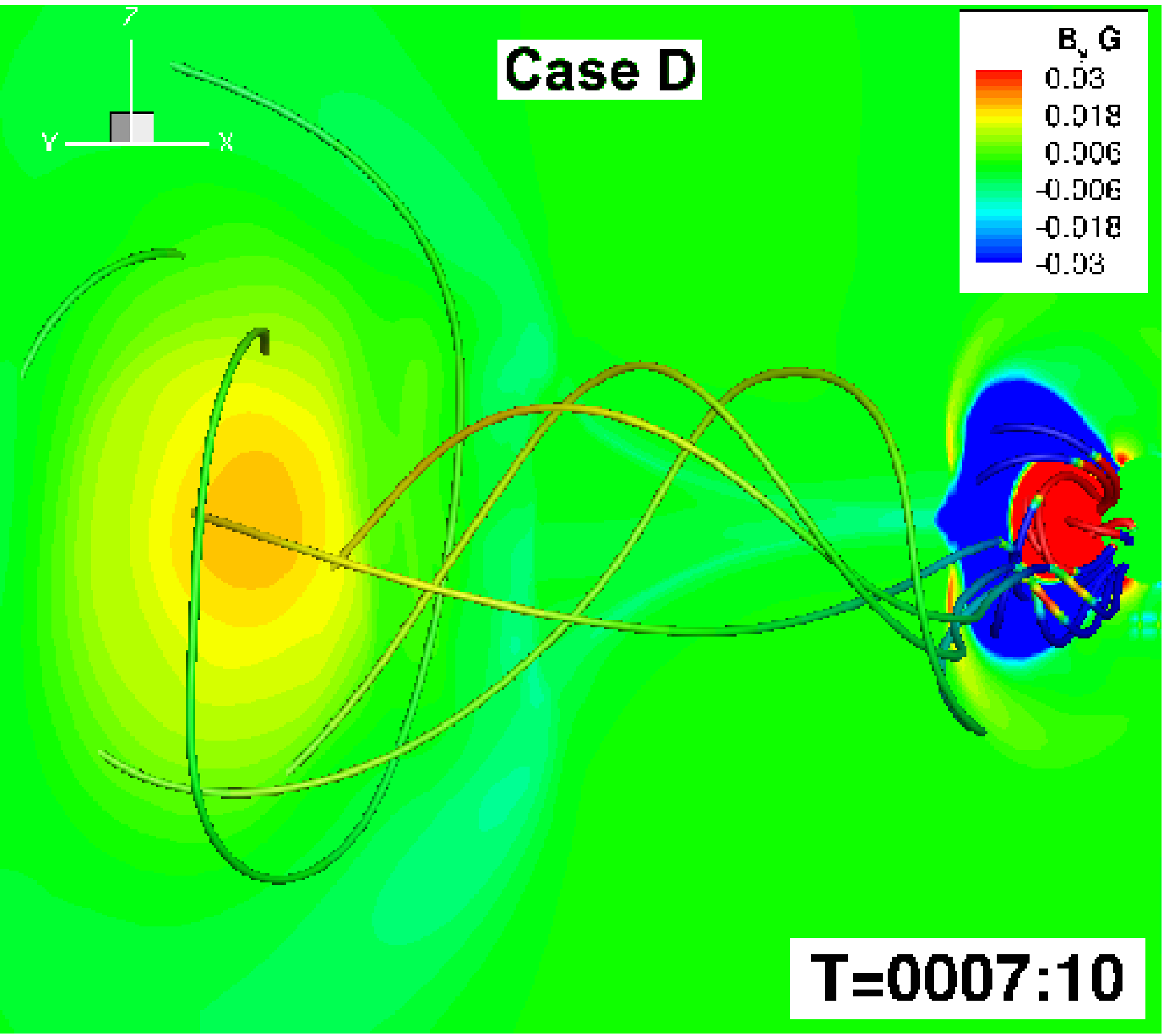}}
\caption{The four cases 10 minutes after the launch of the second CME. The panels show 3-D magnetic field lines and a 2-D meridional cut containing the Sun-Earth line, both color-coded with the east-west component of the magnetic field, $B_y$. Cases A, B, C and D are on the top left, top right, bottom left and bottom right, respectively.}
\end{figure*}

\section{Case A: Same Orientation} \label{CME2a}

In the next two sections, we discuss the interaction of CME1 with CME2 in four cases corresponding to different orientations of CME2. Figure~3 and 4 show the four cases 10 minutes after the launch of the second CME and at time t = 8 hours (in SC), respectively. Figure~3 shows the east-west component of the magnetic field, $B_y$, (axial field for CME1 and for CME2 for cases A and B) for the different cases. Figure~4 shows the north-south component of the magnetic field, $B_z$, (poloidal field for CME1 and for CME2 for cases A and B) just before the beginning of the interaction.
We first discuss in detail the case where the second CME (hereafter CME2a) has the same orientation as the first CME (hereafter CME1a). This is the typical case studied previously in 3-D numerical simulations by \citet{Lugaz:2005b} and \citet{FShen:2011}. It can also be studied in 2.5-D as was done by \citet{Xiong:2007}. 

The center of CME2a initially propagates with a speed of about 1200~km~s$^{-1}$ with a strong expansion, as the front of the magnetic ejecta propagates with a speed of about 1600~km~s$^{-1}$ and the back of the magnetic ejecta does not leave the solar surface until 30 minutes after the launch of CME2a (at time t = 7.5 hours). The back of CME2a then starts to accelerate and reaches a speed of about 750~km~s$^{-1}$ at time t = 9 hours. CME2a drives a fast magnetosonic shock, which propagates with an initial speed in excess of 2200~km~s$^{-1}$ and decelerates to 2000~km~s$^{-1}$ after about 25 minutes. The time-height data for CME2a is shown in red in the right panel of Figure~2.

\subsection{Shock-CME1a Interaction}
The shock driven by CME2a reaches the back of CME1a at 8.67 hours at a distance of 18.5~$R_\odot$, which marks the beginning of the interaction phase (noted by the dashed vertical line in the right panel of Figure~2). The shock passes the center of CME1a at 10.5 hours and the front of CME1a at 11.17 hours at a distance of 38~$R_\odot$. The shock driven by CME2a merges with the shock driven by CME1a at a distance of 48~$R_\odot$ (0.22~AU) at 13.5 h. The new merged shock is shown with a dashed purple line in the right panel of Figure~2.

Before the interaction, the shock speed is about 1600~km~s$^{-1}$ and the speed of the back of CME1a is about 550~km~s$^{-1}$. By 9.33 hours, the back of CME1a has a speed of about 850~km~s$^{-1}$, faster than the speed of the front of the CME. This results in a contraction of the CME in the radial direction from 11~$R_\odot$ at 9 hours to 9.5~$R_\odot$ at 12 hours. The center of CME1a accelerates to 900~km~s$^{-1}$ by 11 hours and the front of CME1a to about 750~km~s$^{-1}$ by 11.5 hours. The CME angular width remains relatively constant at 28$^\circ$, which means that the contraction in the radial direction is not associated with an increase in the latitudinal direction. However, because of the compression in the radial direction, the aspect ratio of CME1a increases during the interaction from 1.2 to 1.5 at 10.67 hours and to 2 at 12.33 hours. 

The shock driven by CME2a undergoes large speed variations as it propagates inside CME1, as the upstream magnetic field and density change rapidly. The magnetic field strength is relatively uniform inside CME1a but there is a large density enhancement at the back half of CME1a corresponding to mass added initially, whereas the front half of CME1a is part of the low density cavity (as shown in the left panel of Figure~2). The Alfv{\'e}nic speed in the front half of CME1a is a factor of 5 to 6 higher than in the back half. %Consequently, the  overtaking shock decelerates in the back of CME1a and accelerate as it propagates inside the front of CME1a. 
As the overtaking shock propagates inside the dense sheath of CME1a, it decelerates to a speed of about 750~km~s$^{-1}$. A more complete analysis of the variation in the speed, Mach number and compression ratio of the shock driven by an overtaking CME was performed in \citet{Lugaz:2005b}, with generally similar results.
As noted in previous studies, the shock may be nearly impossible to image with STEREO/HIs as it propagate inside CME1a, since the compression ratio is low and the density inside CME1a is overall low. However, it has a very significant impact on the kinematics of CME1a. Its presence might be inferred from the acceleration of different parts of CME1a.

%%%%%%%%%%%%%%%%%%%%%%%
\begin{figure*}[ht*]
\centering
{\includegraphics*[width=6.5cm]{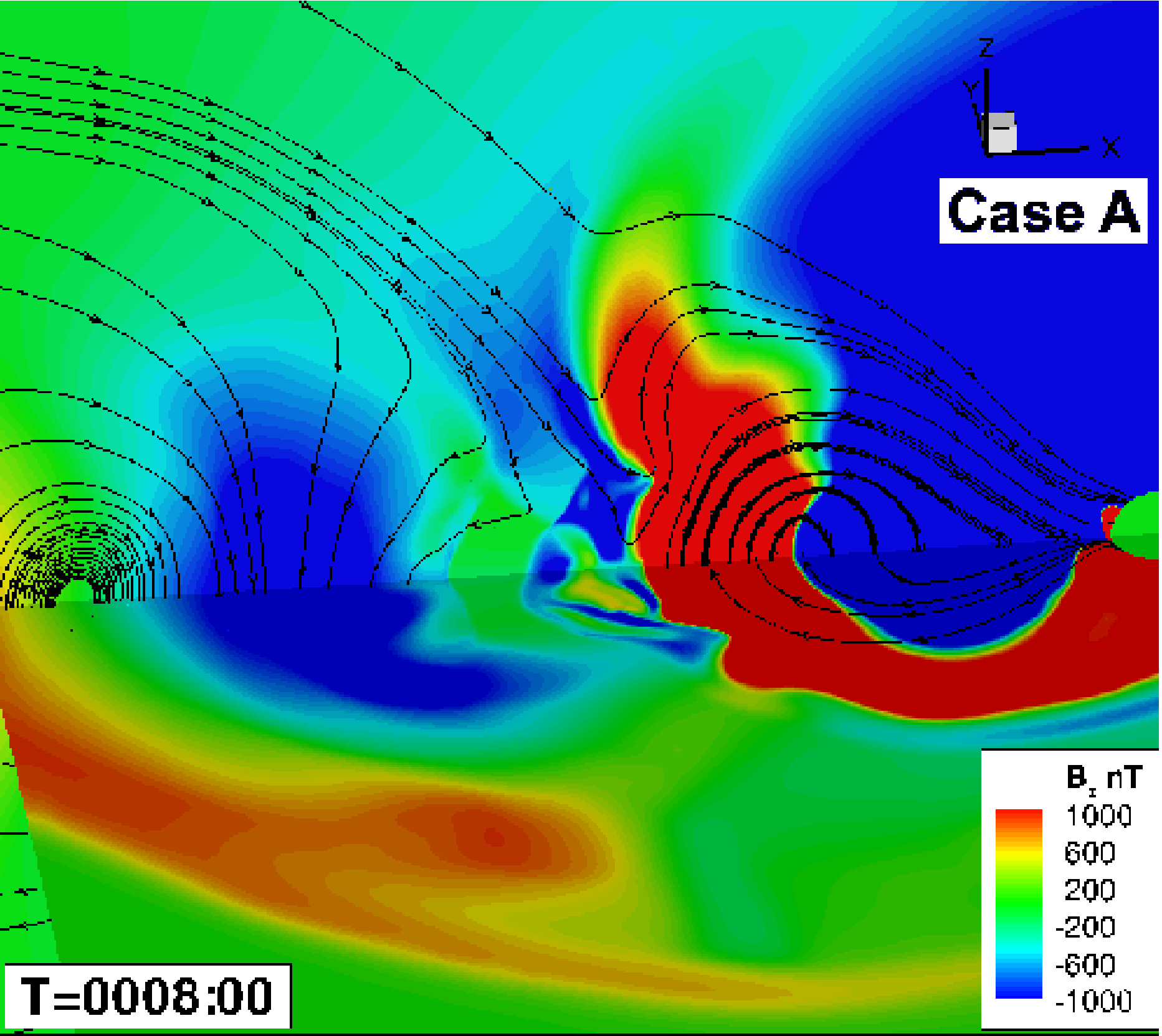}}
{\includegraphics*[width=6.5cm]{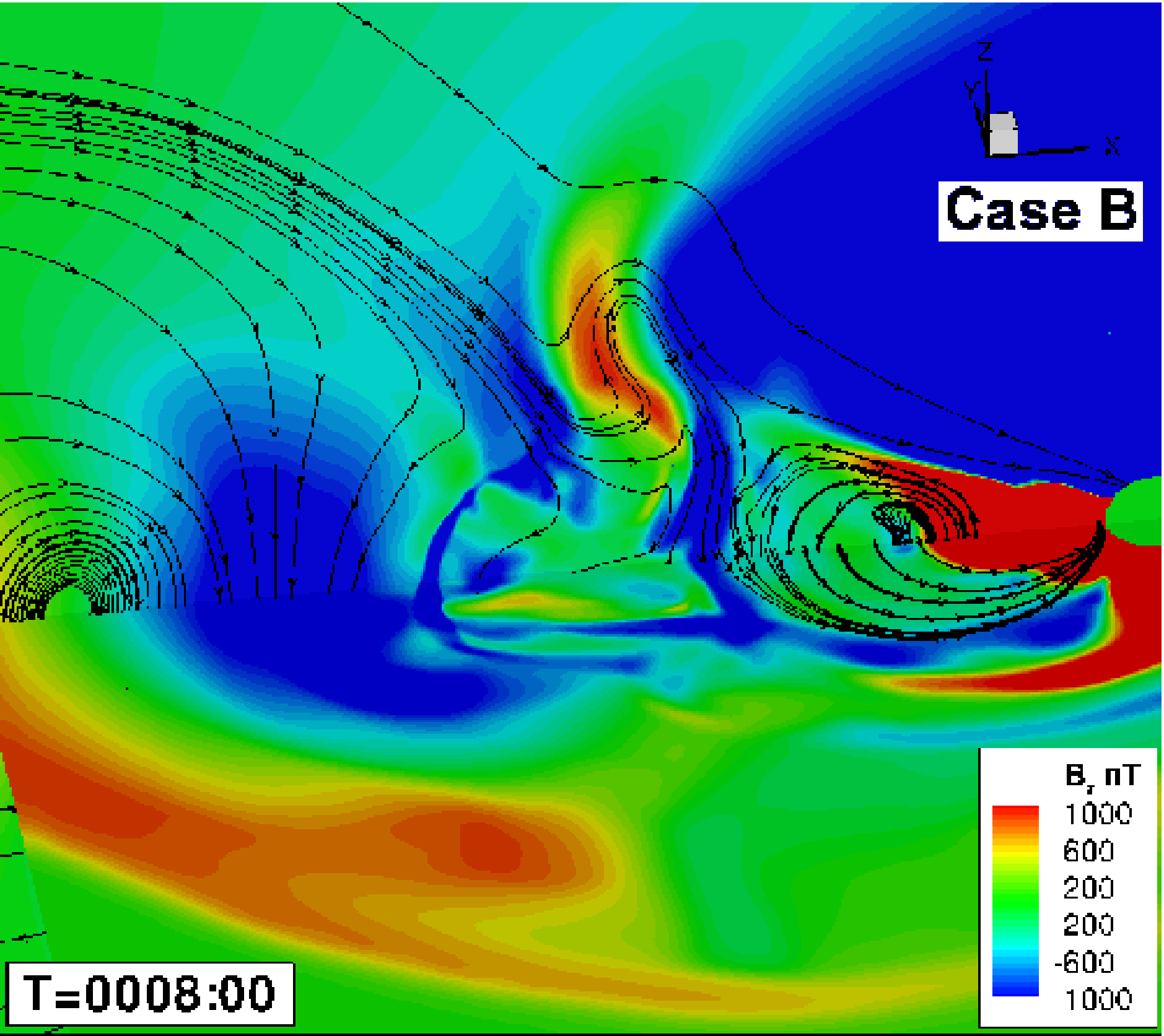}}\\
{\includegraphics*[width=6.5cm]{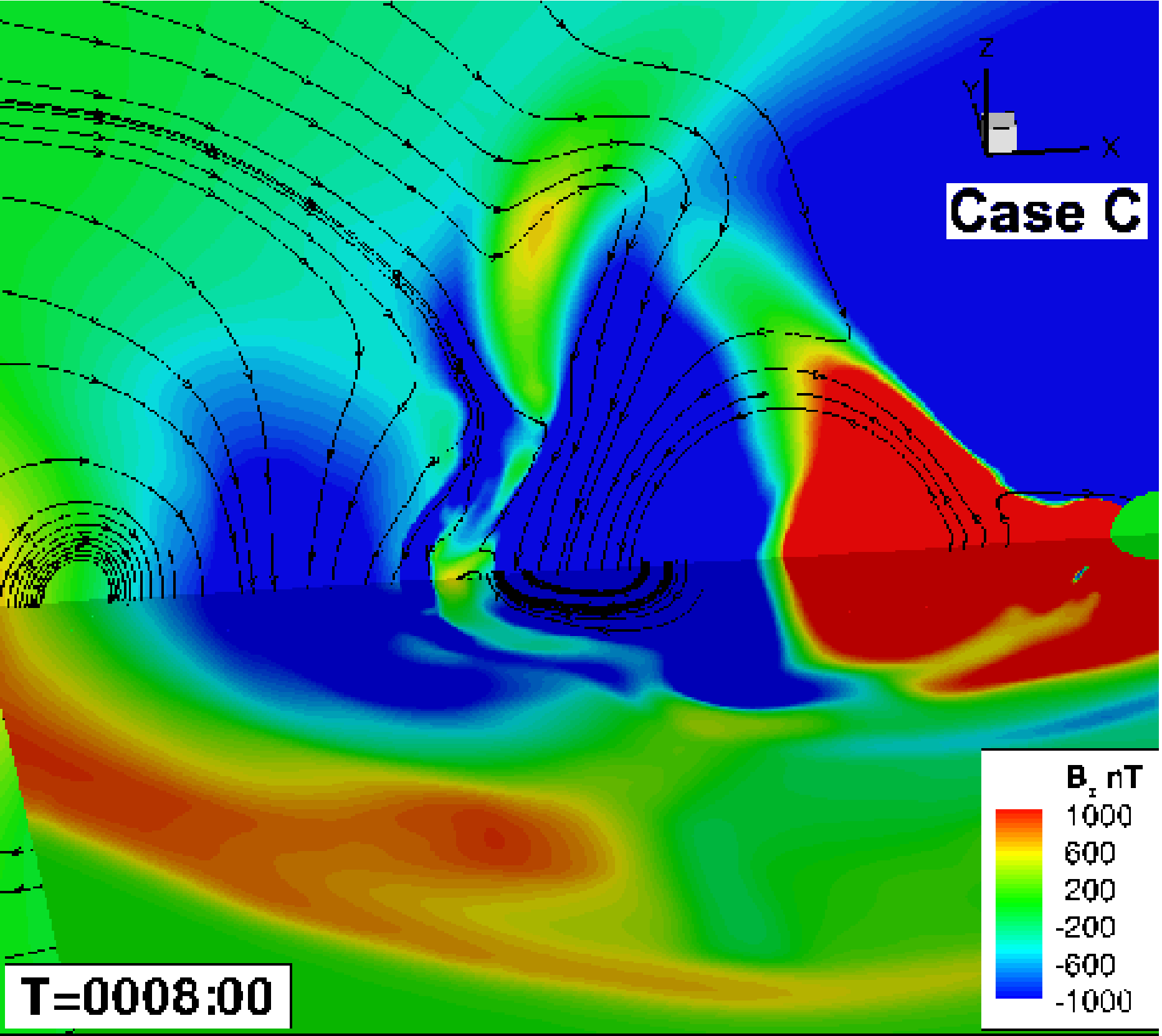}}
{\includegraphics*[width=6.5cm]{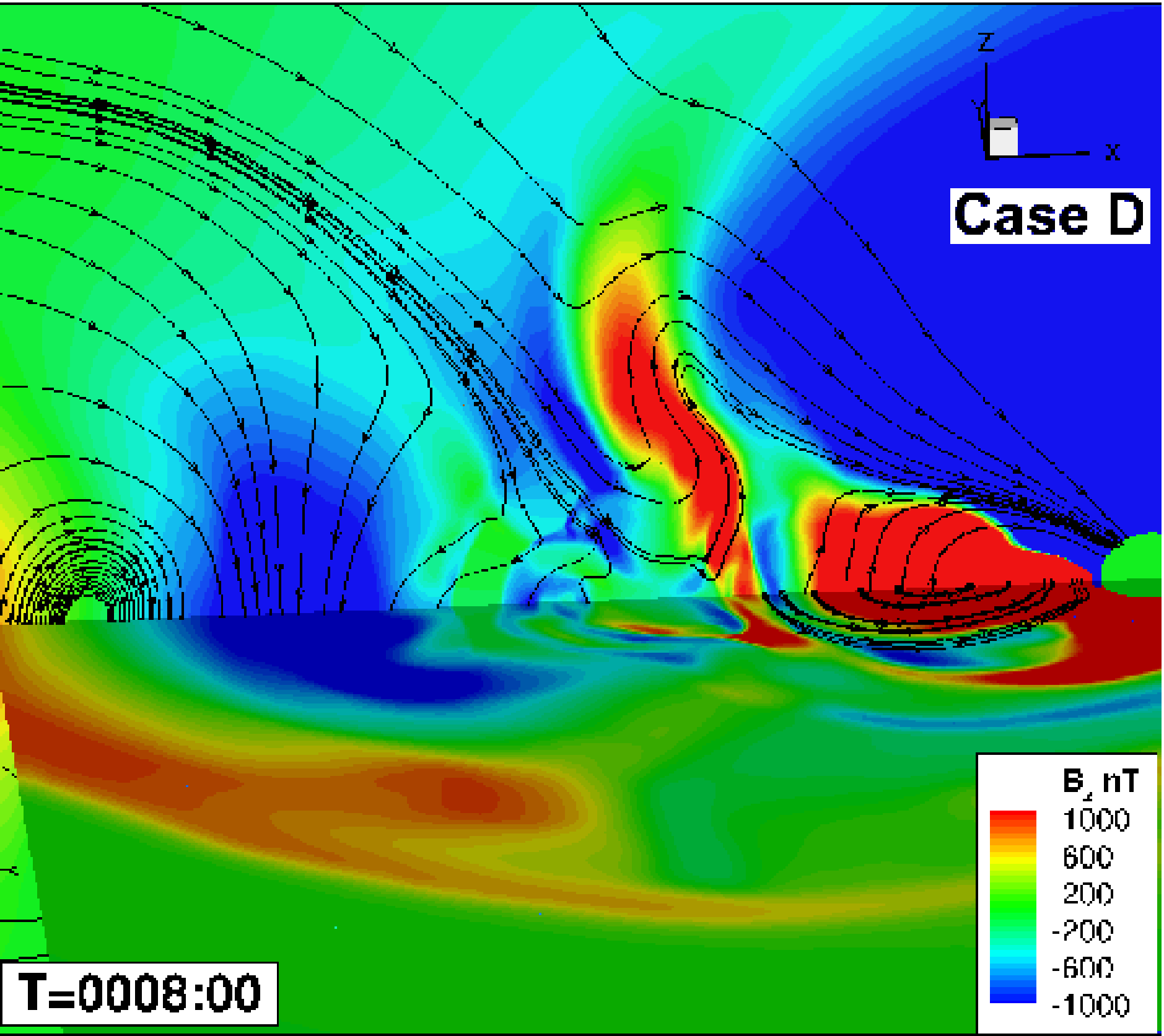}}
\caption{The four cases 1 hour after the launch of the second CME. The panels show the north-south component of the magnetic field, $B_z$ and 2-D projections of the magnetic field lines in the same two planes as Figure~1. Cases A, B, C and D are on the top left, top right, bottom left and bottom right, respectively.}
\end{figure*}

\subsection{CME1a Evolution During the Interaction}
Soon after the shock driven by CME2a enters the back of CME1a, the two CMEs ``collide''. In fact, the minimum distance between the back of CME1a and the front of CME2a is reached at 10.67 hours with a distance of 1.05~$R_\odot$. In between the two CMEs, there is a region of negative $B_y$, which is, initially part of the back of CME1a (see images in Figure~3). At its minimum size, this interaction region is resolved with more than 16 cells, and it is therefore physical to discuss its width. This is where reconnection occurs between the two CMEs. The region increases in size to 4.2~$R_\odot$ by 14 hours and then decreases to 1.5~$R_\odot$ at 24 hours. As the two CMEs have their axial field parallel to each other, reconnection occurs between the poloidal fields of the two CMEs (south at the back of CME1a and north at the front of CME2a). Note that determining the boundaries of CME1a and CME2a is not obvious during the interaction. A view of the two CMEs 14 hours into the simulation is shown in the top panel of Figure~5. This Figure also illustrates the uniformization of the speeds inside the two CMEs. While the speed of the two CMEs differ by a factor of two before the interaction, the resulting complex ejecta after the interaction travels with a uniform speed, which decreases throughout the magnetic ejecta as found for real events by \citet{Burlaga:2003}. This process is already ongoing at time t = 14 hours.

Reconnection erodes CME1a \citep[similarly to what has been discussed for isolated events by][]{Ruffenach:2012}. The evolution of CME1a radial width is plotted with black symbols in the left panel of Figure~6. CME1a continues to contract radially even after the shock has passed the front of CME1a. The CME minimum size in the radial direction is reached at time t = 14.33 hours, shortly after the reconnection region has reached its maximum extent. Thereafter, CME1a radial size increases but at a slower rate than before the interaction, reaching a width of 10.5~$R_\odot$ at time t = 24h. The reason for the slow expansion is simple: the front of CME1a propagates at a faster speed than the center of CME1a (770 vs. 690 km~s$^{-1}$) but the back of CME1a, which is still ``pushed'' by CME2a, propagates at a faster speed than its center (710~km~s$^{-1}$). In some sense, CME1a is over-expanding at its front, at a faster rate than before the interaction, but it is also being compressed at its back, resulting in a positive but limited radial expansion. %The evolution of the radial width of CME1a starting at 14.33 hours can be fitted as $ W = 0.107~r^{0.28}$. 
The angular width of CME1a remains more or less constant between 27$^\circ$ and 30$^\circ$, and the decreasing and then slowly increasing radial width results in a monotonic increase of the aspect ratio of the CME to a value of about 4 at the end of the simulation. The aspect ratio of CME1a is plotted with black diamonds in the right panel of Figure~6.

%%%%%%%%%%%%%%%%%%%%%%%
\begin{figure}[t*]
\centering
{\includegraphics*[width=6.5cm]{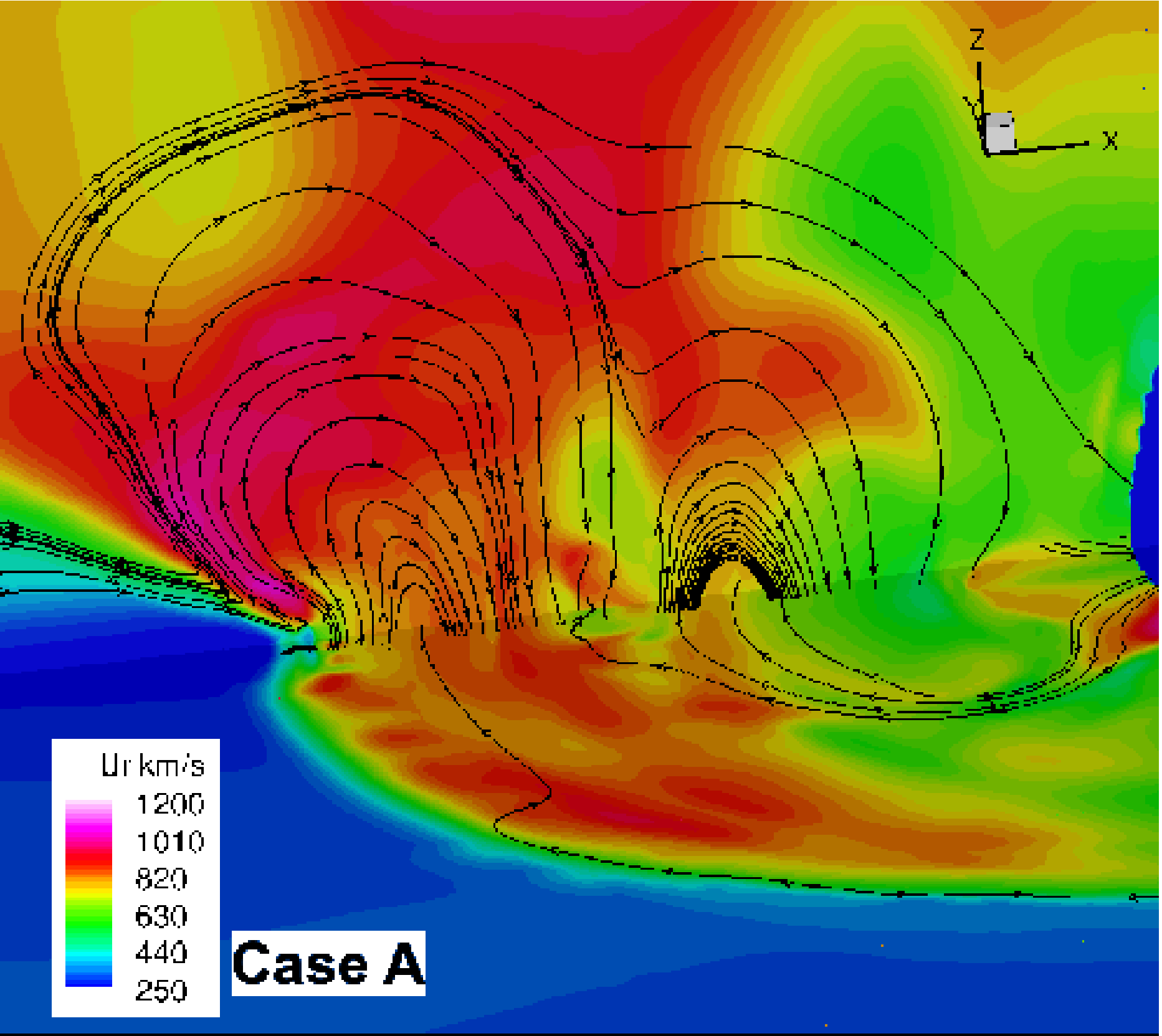}}\\
{\includegraphics*[width=6.5cm]{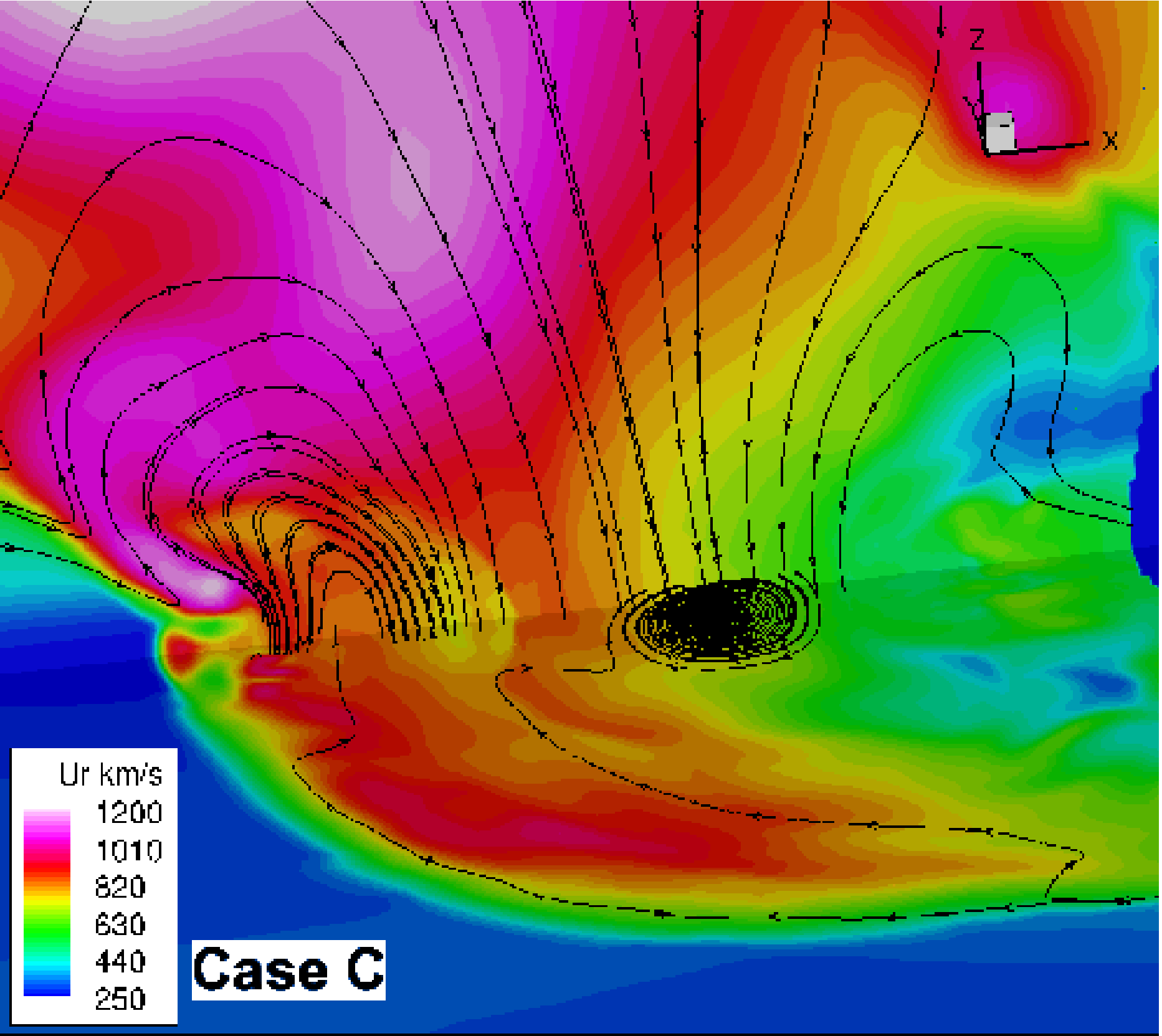}}
\caption{Cases A and C, 14 hours into the simulation. The two panels show the radial velocity with 2-D projections of the magnetic field lines in the ecliptic and a meridional plane containing the Sun-Earth line.}
\end{figure}

\subsection{CME2a Evolution During the Interaction}
The initial evolution of CME2a has some similarities to that of CME1a, with for example a decrease of the aspect ratio from 1.5 to 0.85 at 7.5 h, followed by an increase until the start of the interaction with CME1a (see black stars in the right panel of Figure~6). However, in contrast to CME1a, CME2a angular width increases during the early phase of propagation to reach a value of about 44$^\circ$. Its decreasing aspect ratio comes from the fact that the radial expansion is faster than the latitudinal expansion. As the front of CME2a collides with the back of CME1a, it decelerates to the speed of 750-800~km~s$^{-1}$ and CME2a contracts slightly in the radial direction from a maximum width of 12.1~$R_\odot$ at 9.67 hours to a minimum width of 10.4~$R_\odot$ at 12.33 hours (see black diamonds in the middle panel of Figure~6). Afterwards, CME2a expands again in the radial direction but at a relatively low rate reaching a maximum width of about 16.4~$R_\odot$ at the end of the simulation. During the interaction, the angular width of CME2a increases from 44$^\circ$ to 55$^\circ$. It appears that CME2a, since it cannot expand in the ecliptic plane because of the presence of CME1a, goes around CME1a by over-expanding in the latitudinal direction. This results in a large increase of the aspect ratio from 1.5 at the beginning of the interaction to more than 6 at the end of the simulation. At the end of the simulation, the speed in the center of CME2a is almost identical to that in the front of CME1a or about 770~km~s$^{-1}$. The top right panel of Figure~7 shows the distribution of $\sqrt{B_x^2 + B_z^2}/|B_y|$ in a meridional cut passing through the Sun-Earth line. It clearly shows the larger angular extent of CME2a as compared to CME1a as well as the reconnection occurring at the interface of CME1a and CME2a and clearly visible in the total magnetic field strength (middle panel) and the north-south ($B_z$) component (right panel). The bright diamonds in the left panel are related to the low resolution away from the Sun-Earth line and the fact that a ratio is plotted.

%%%%%%%%%%%%%%%%%%%%%%%
\begin{figure*}[ht*]
{\includegraphics*[width=5.64cm]{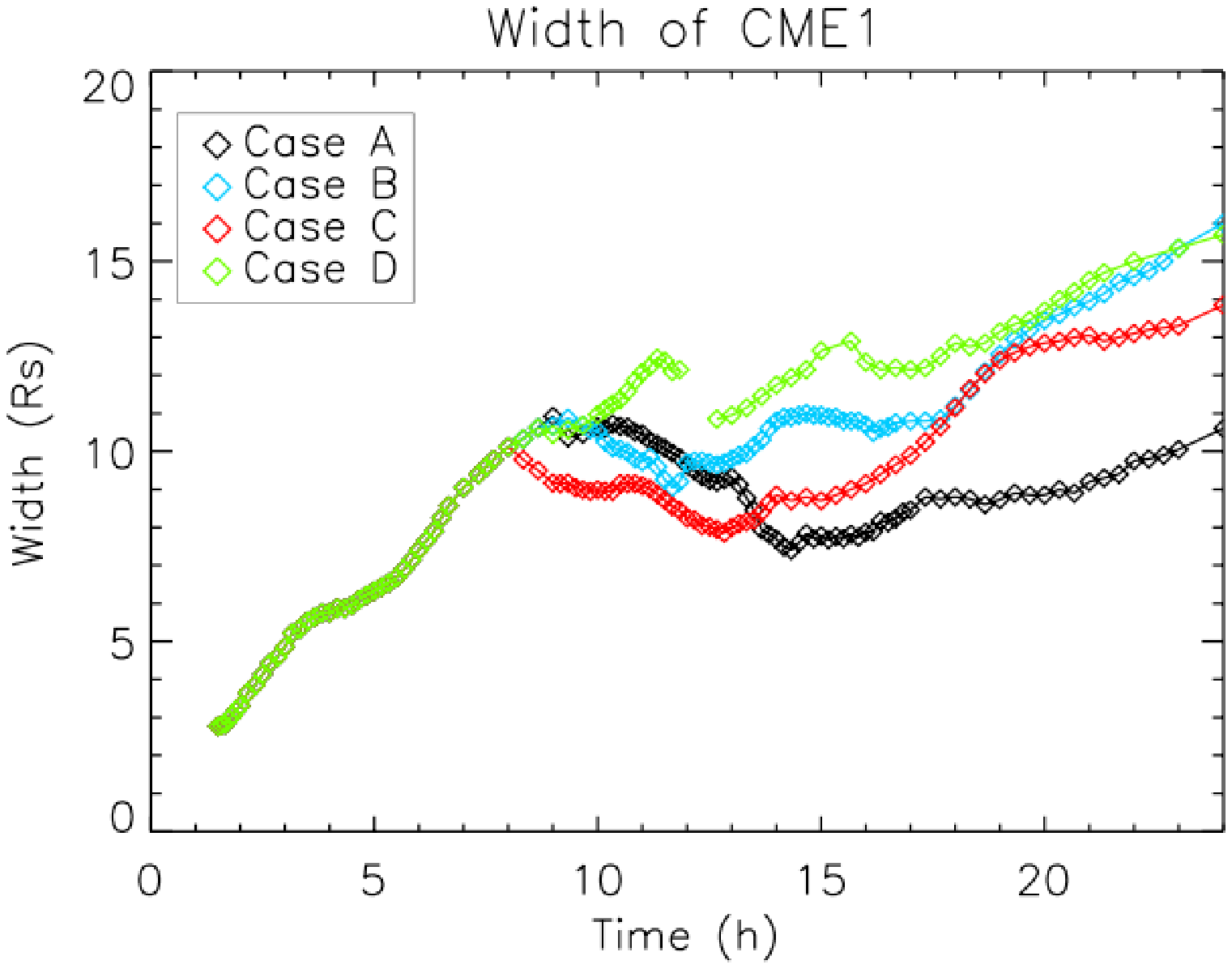}}
{\includegraphics*[width=5.64cm]{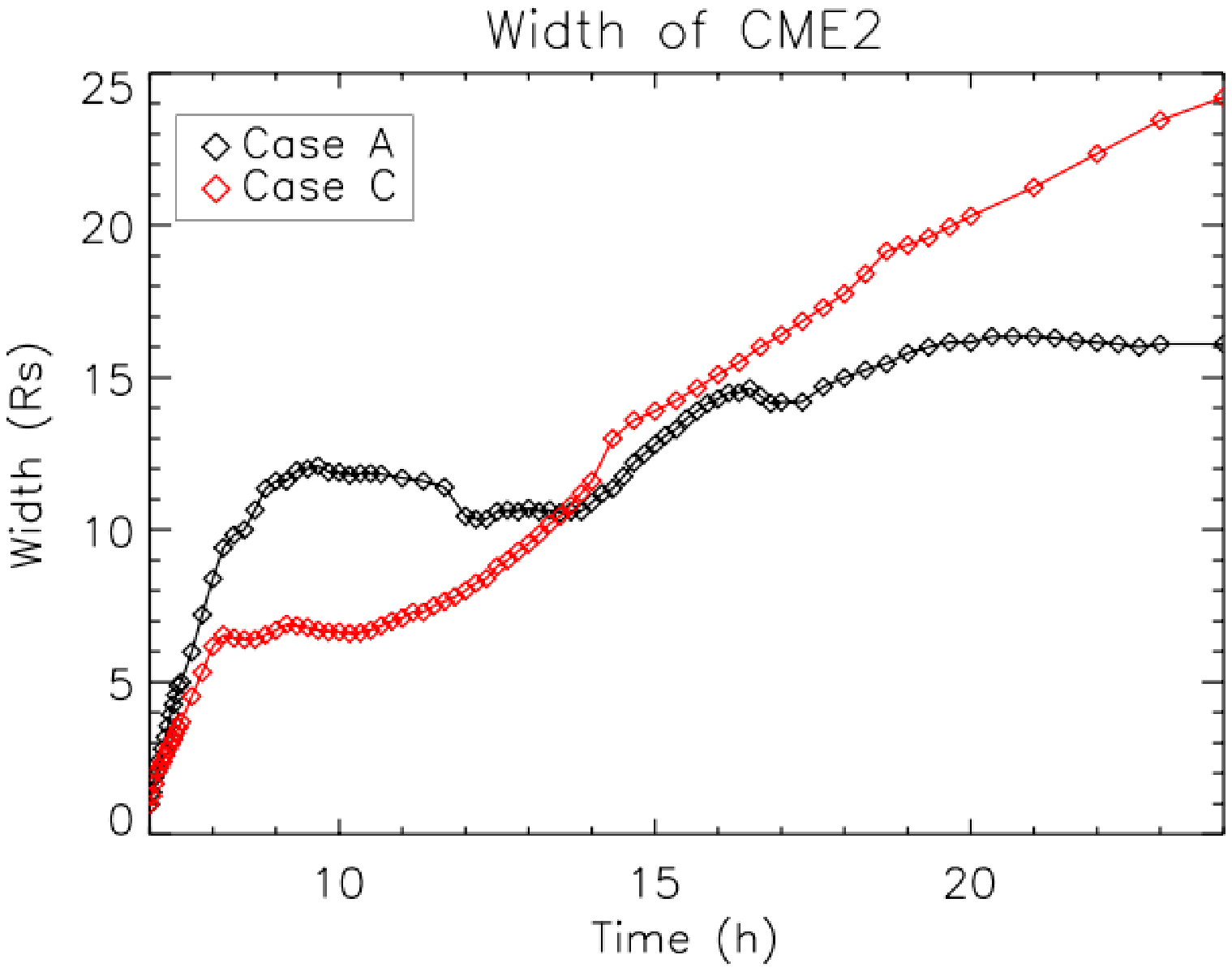}}
{\includegraphics*[width=5.64cm]{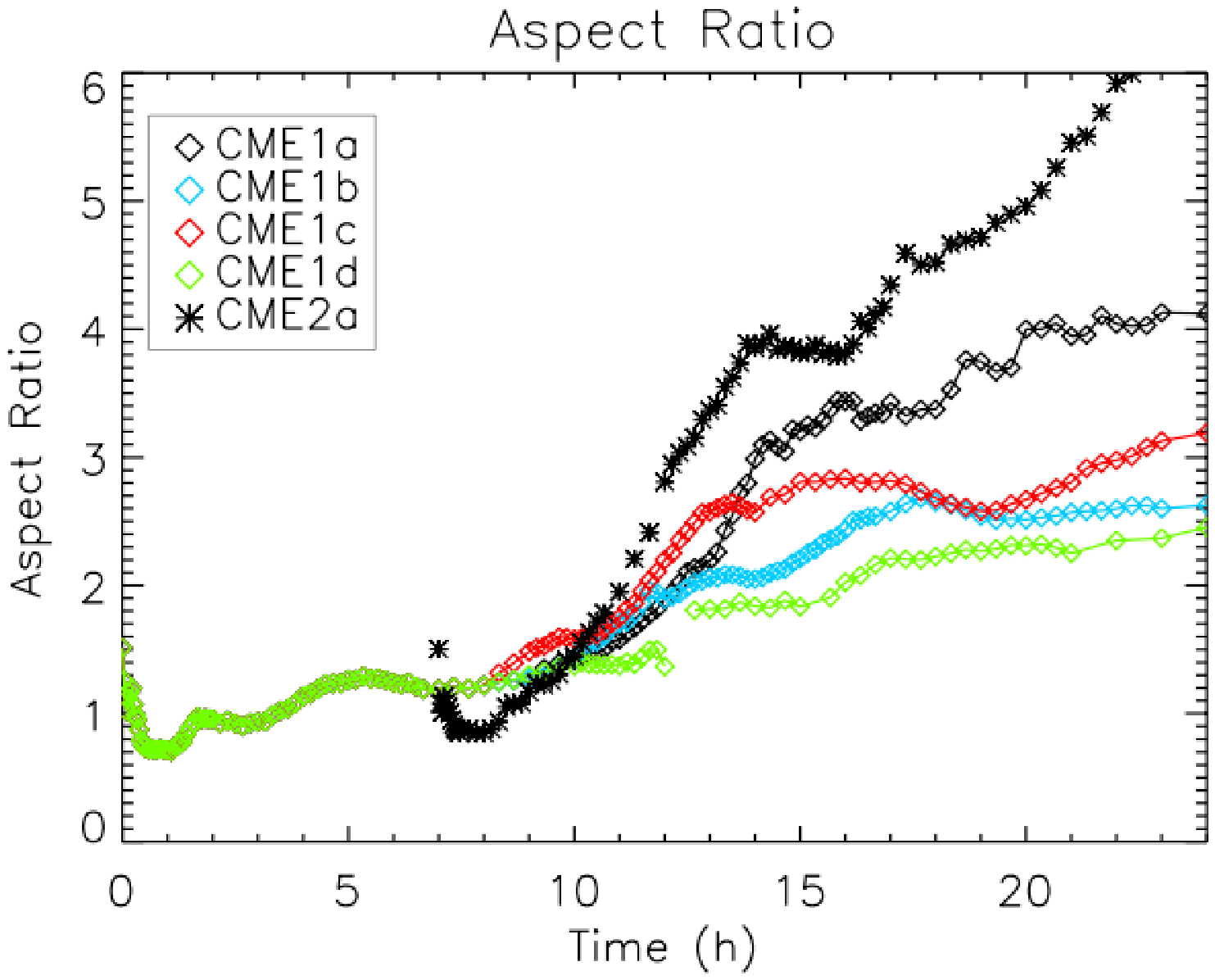}}
%\end{minipage}
%\vspace{-.4cm}
\caption{Left: Time-evolution of the radial width of CME1 in the four cases. %Two widths are plotted for Case B after 10 hours corresponding to the two maxima visible in the left panel. 
The discontinuity in the width of CME1d is due to reconnection between the two CMEs. Center: Width of CME2 for Cases A and C. Right: Aspect ratio for all four cases for CME1 and for Case A for CME2.}
\end{figure*}
%%%%%%%%%%%%%%%%%%%%%%%%

\section{Cases B, C and D: Different Orientations} \label{CME2bd}

Here, we give an overview of the evolution of the two CMEs in the three other cases studied. We do not necessarily get into as much detail as the analysis of CME1a and CME2a but mostly focus on the differences in the kinematics and expansion between the different cases. Plots comparing the four cases are shown in Figures~3 and 4 before the interaction, and in Figures~7 and 8 during the interaction.

\subsection{Case B: Opposite Orientation}

This case corresponds to a second CME with a SWN orientation, a low-inclination CME. In addition to having an axial field anti-parallel to that of CME1b, the poloidal field in the back of CME2b is anti-parallel to the global dipole of the Sun. This orientation results in a large amount of reconnection between the second CME and the Sun on the one hand and between CME1b and CME2b on the other. 

The interaction between CME1b and CME2b has primarily an effect on the back half of CME1b. Starting around 9.5 hours, CME1b has a ``normal'' shape similar to that of CME1a as well as an extended tail of southward magnetic field (see left panel of Figure~6 and second line of Figure~7). In this region, reconnection occurs with the magnetic field from CME2b, which has an axial magnetic field an anti-parallel to that of CME1b but a poloidal field parallel to that of CME1b. This region maintains a (albeit weaker) eastward axial field, characteristic of CME1b. Starting at 12 hours, there are, inside CME1b, two clear local maxima of $B_y$ positive and of the function $\sqrt{B_x^2 + B_z^2}/|B_y|$ which we use to determine the CME boundaries (see left panel in the second line of Figure~7). We consider that the boundary of CME1b is the first maximum, with a large reconnection region between the two CMEs to keep a definition consistent with that used for CME1a.  Note that because the global magnetic field rotation is NESWN with the southward magnetic field corresponding to the back of CME1b and the front of CME2b, it may be reasonably expected that determining the boundaries between the two CMEs is complicated.

Using this definition of the boundaries of CME1b, it has approximately the same radial width as CME1a at t = 12 hours: 9.7~R$_\odot$ (see blue symbols in the right panel of Figure~6). However, from this time onward, whereas the radial width of CME1a continues to decrease until t = 14.33 hours, that of CME1b increases monotonically to reach a value of about 16~$R_\odot$ at t = 24 hours. The angular width of CME1b is the same as that of CME1a (within the relatively large uncertainties due to the low resolution in the $z$ direction). The aspect ratio of CME1b increases from 1.9 at 12 hours (same as CME1a) to about 2.6 at 24 hours (versus 4 for CME1a). At the end of the simulation, CME1b center propagates with a speed of about 600~km~s$^{-1}$ with the front about 100~km~s$^{-1}$ faster and the back at the same speed as the center of CME1b.

Because of the large amount of reconnection, it is hard to follow CME2b. There is a small region of westward $B_y$, which we considered to be the center of CME2b. It is typically about 6 to 8~$R_\odot$ behind the second boundary of CME1b and propagates with a speed of about 550~km~s$^{-1}$, i.e. slower than the center of CME1b and much slower than the center of CME2a.

%%%%%%%%%%%%%%%%%%%%%%%
\begin{figure*}[ht]
\centering
{\includegraphics*[trim = 1.5cm 1.2cm 2cm 2cm, clip, width=5.4cm]{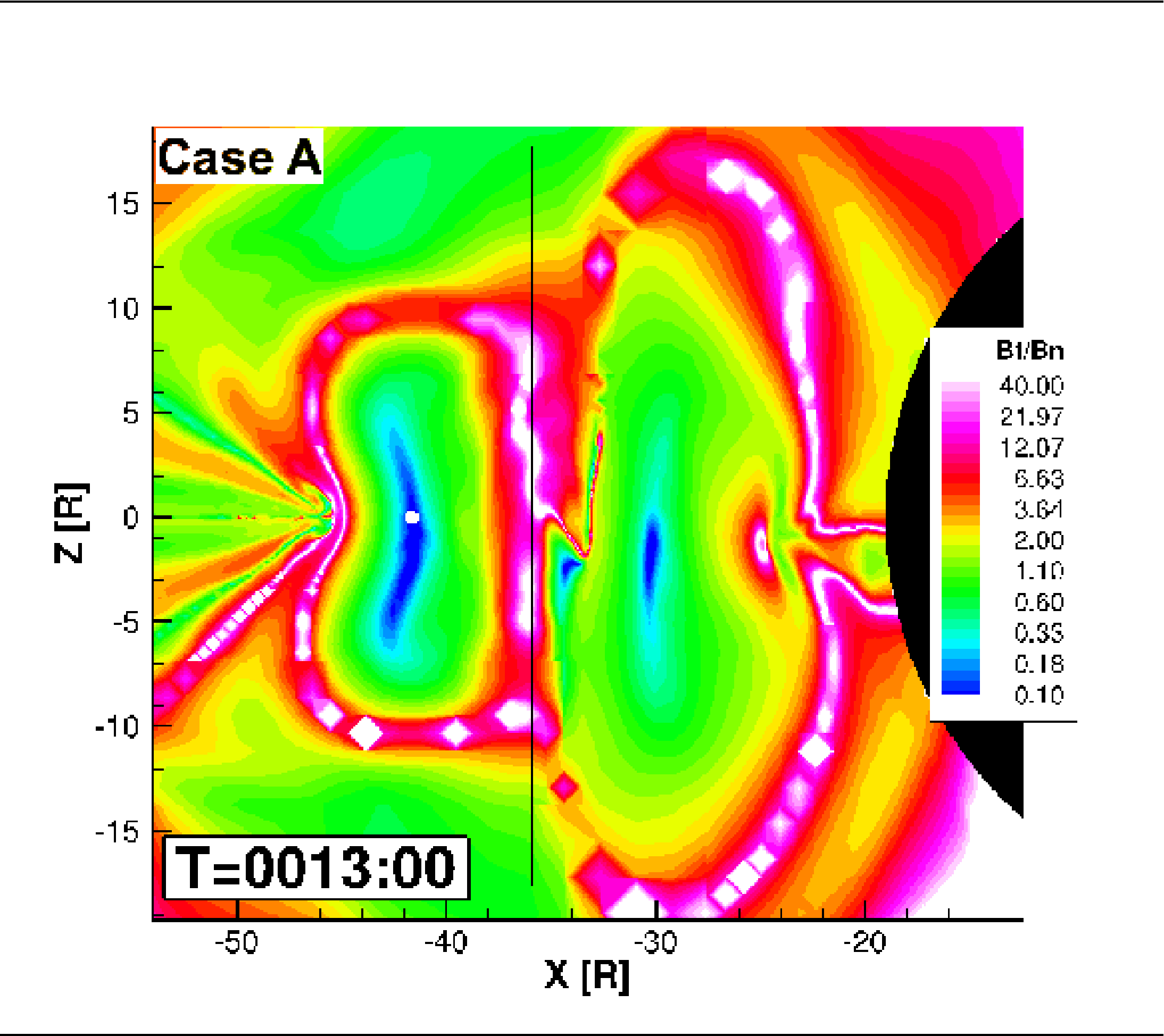}}
{\includegraphics*[trim = 1.5cm 1.2cm 2cm 2cm, clip, width=5.4cm]{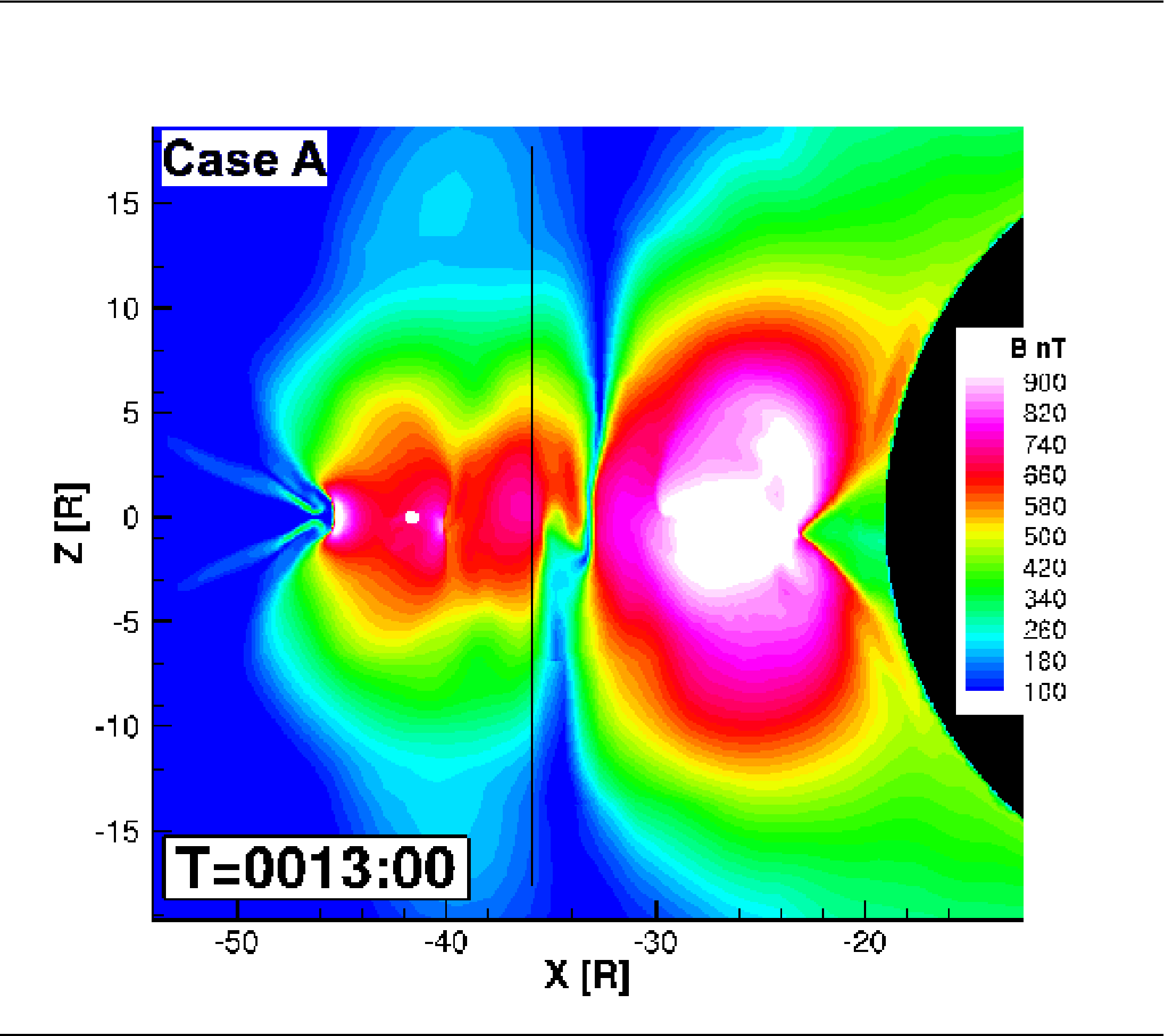}}
{\includegraphics*[trim = 1.5cm 1.2cm 2cm 2cm, clip, width=5.4cm]{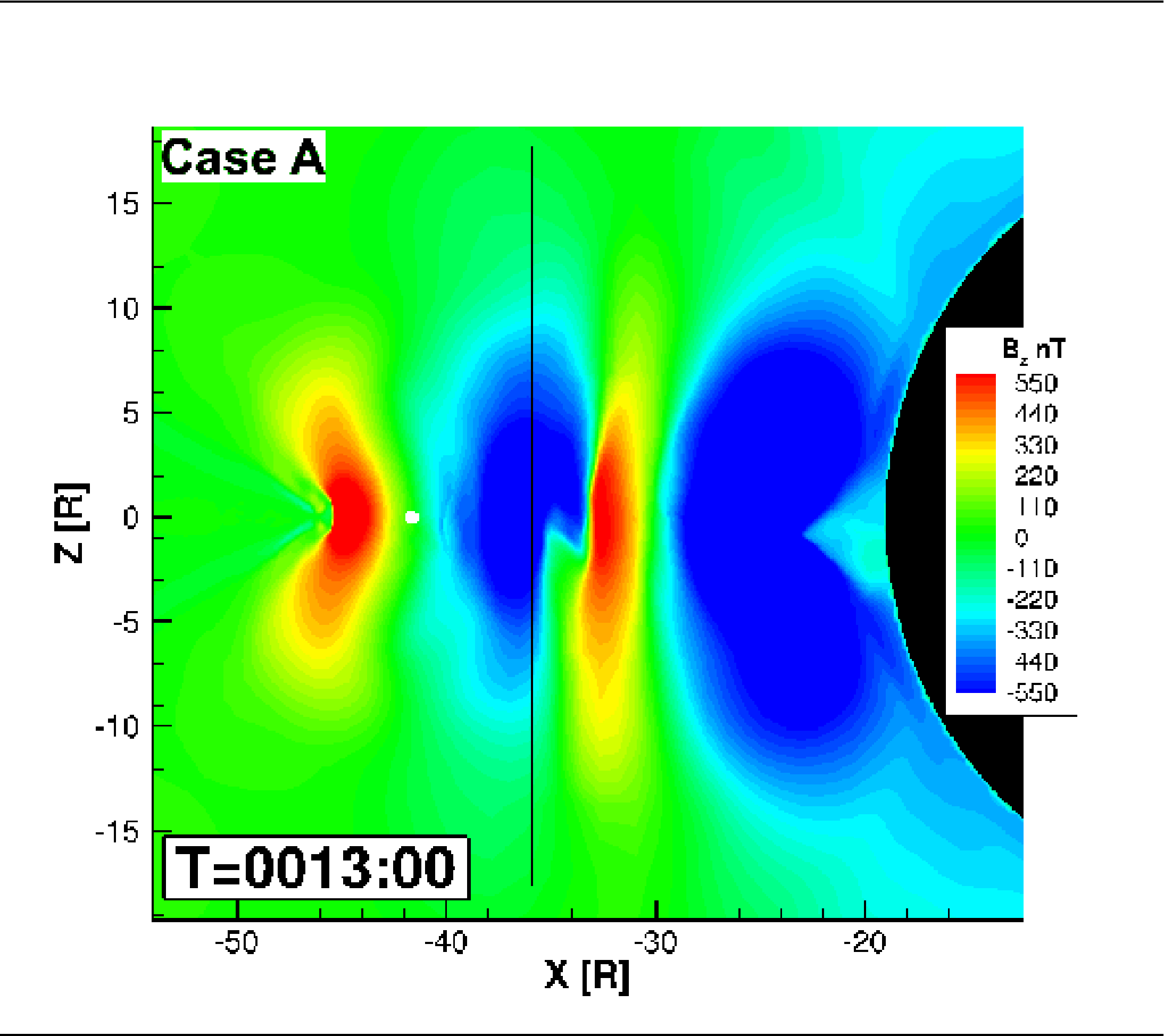}}\\
{\includegraphics*[trim = 1.5cm 1.2cm 2cm 2cm, clip, width=5.4cm]{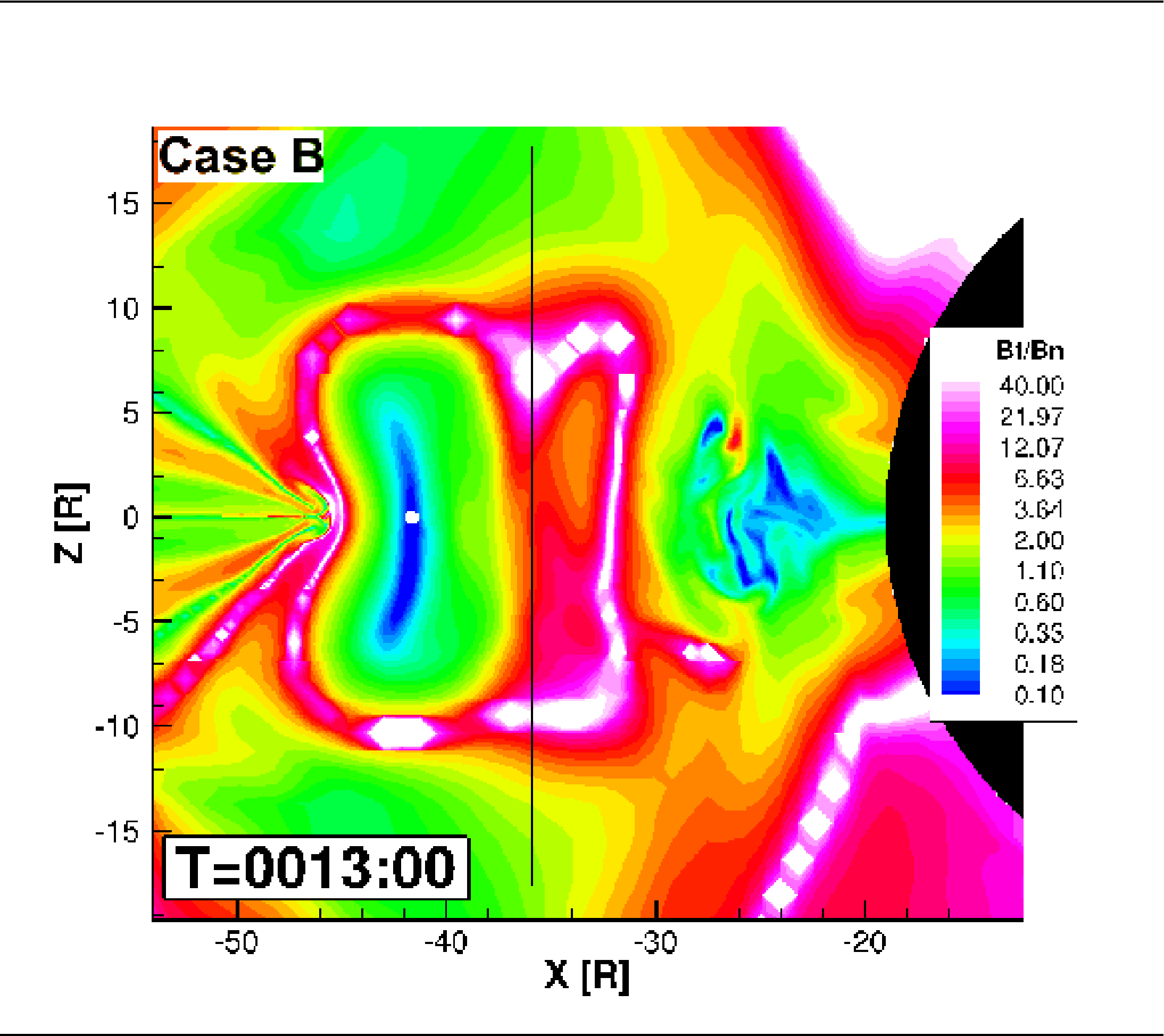}}
{\includegraphics*[trim = 1.5cm 1.2cm 2cm 2cm, clip, width=5.4cm]{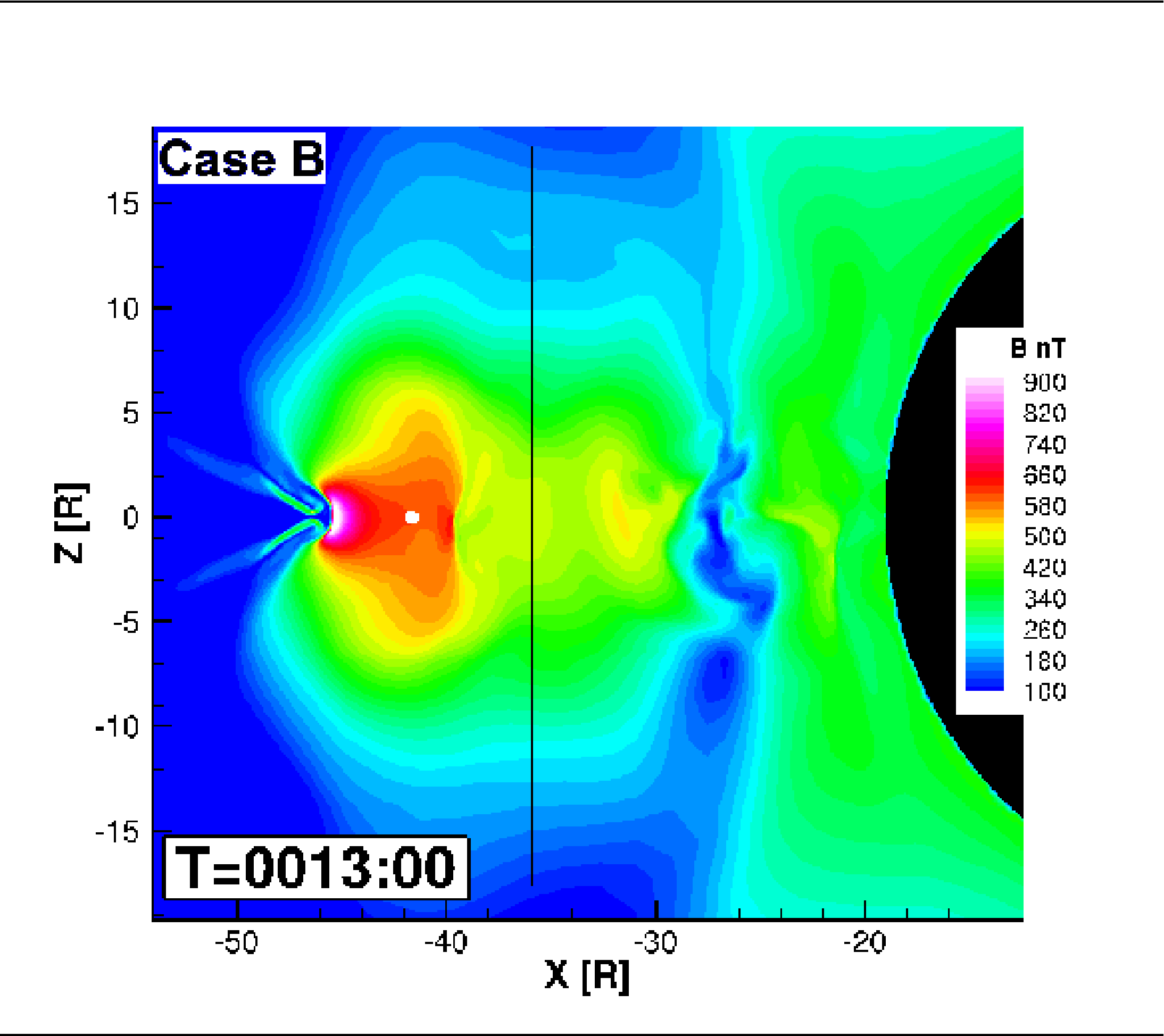}}
{\includegraphics*[trim = 1.5cm 1.2cm 2cm 2cm, clip, width=5.4cm]{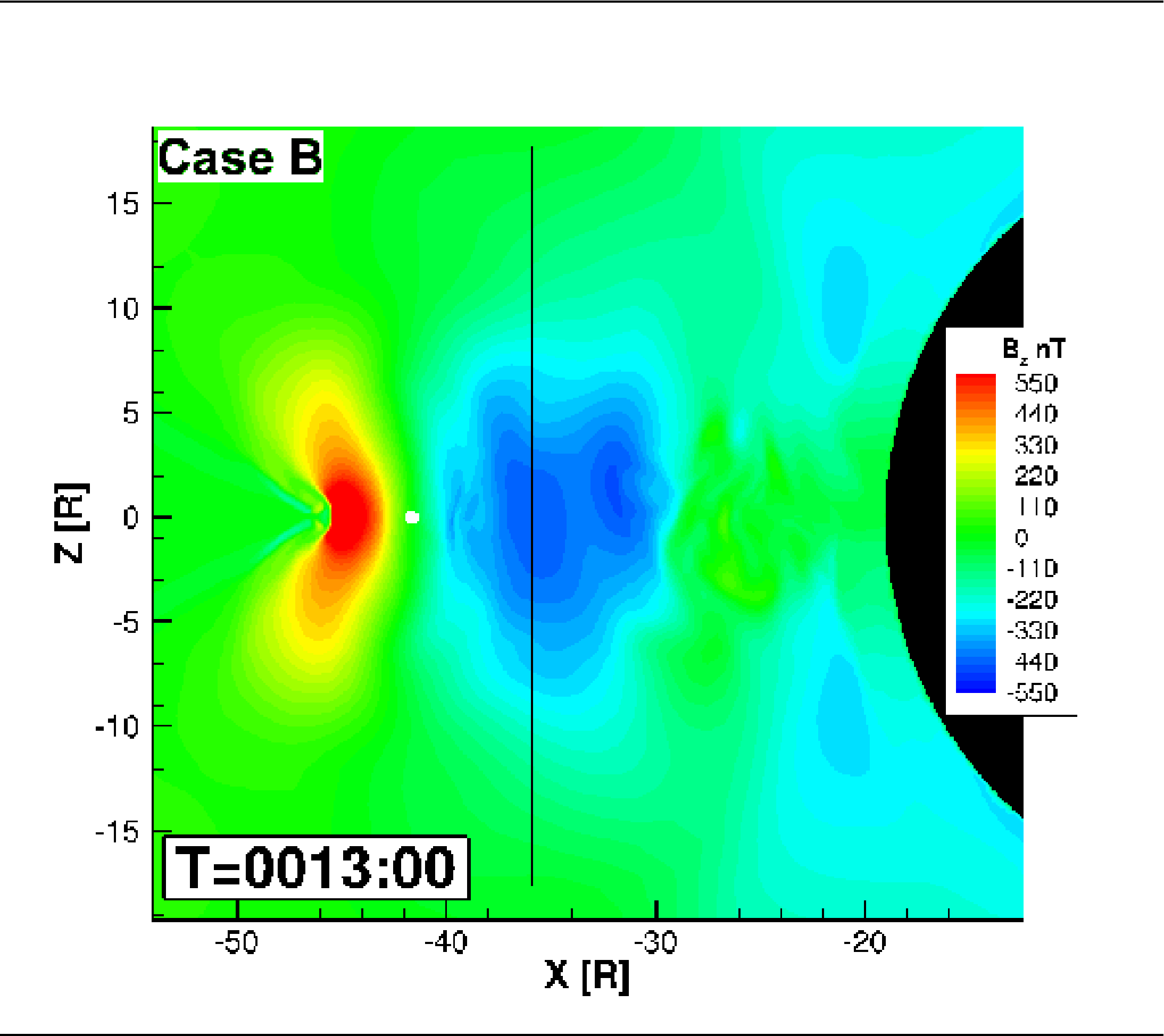}}\\
{\includegraphics*[trim = 1.5cm 1.2cm 2cm 2cm, clip, width=5.4cm]{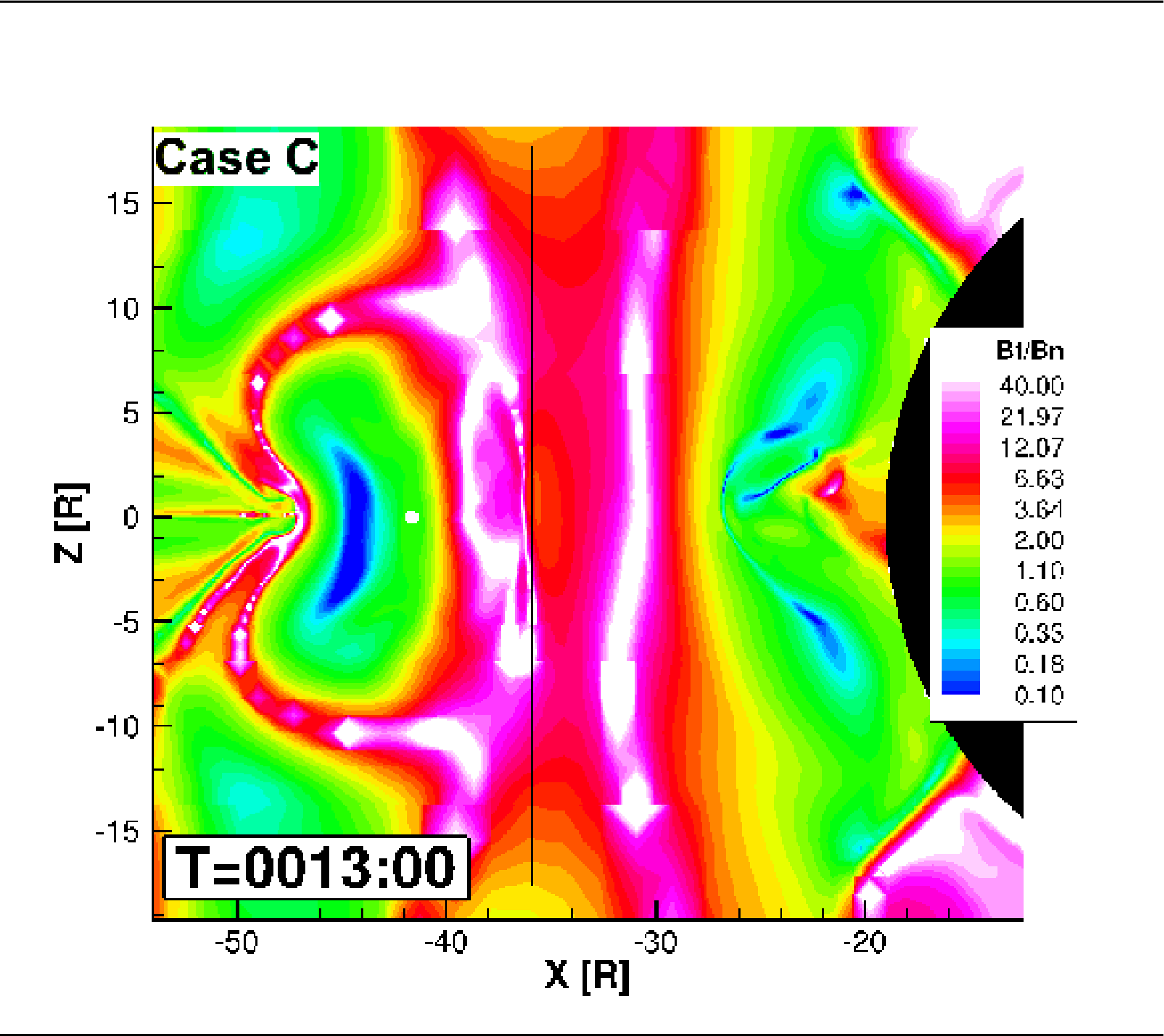}}
{\includegraphics*[trim = 1.5cm 1.2cm 2cm 2cm, clip, width=5.4cm]{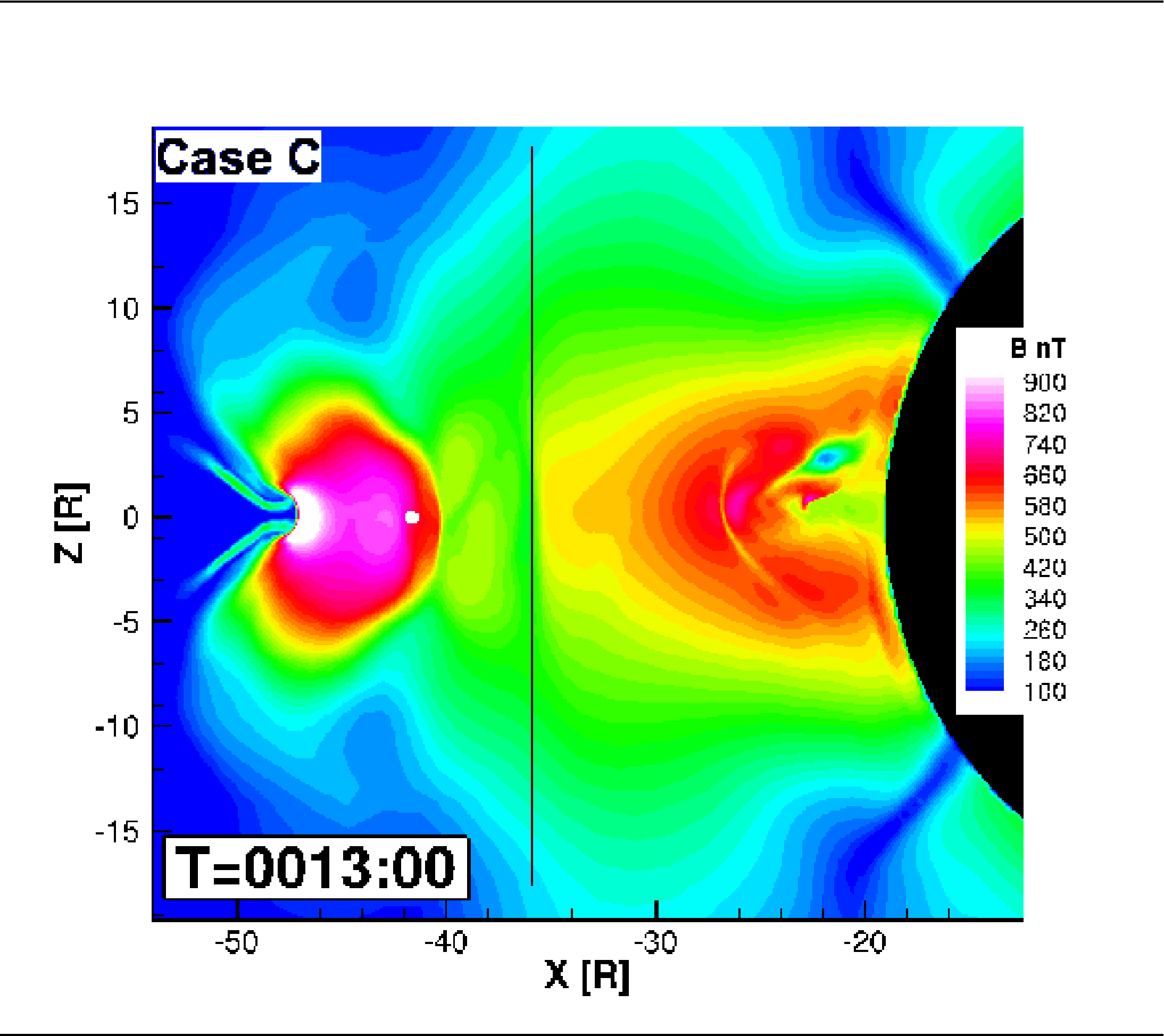}}
{\includegraphics*[trim = 1.5cm 1.2cm 2cm 2cm, clip, width=5.4cm]{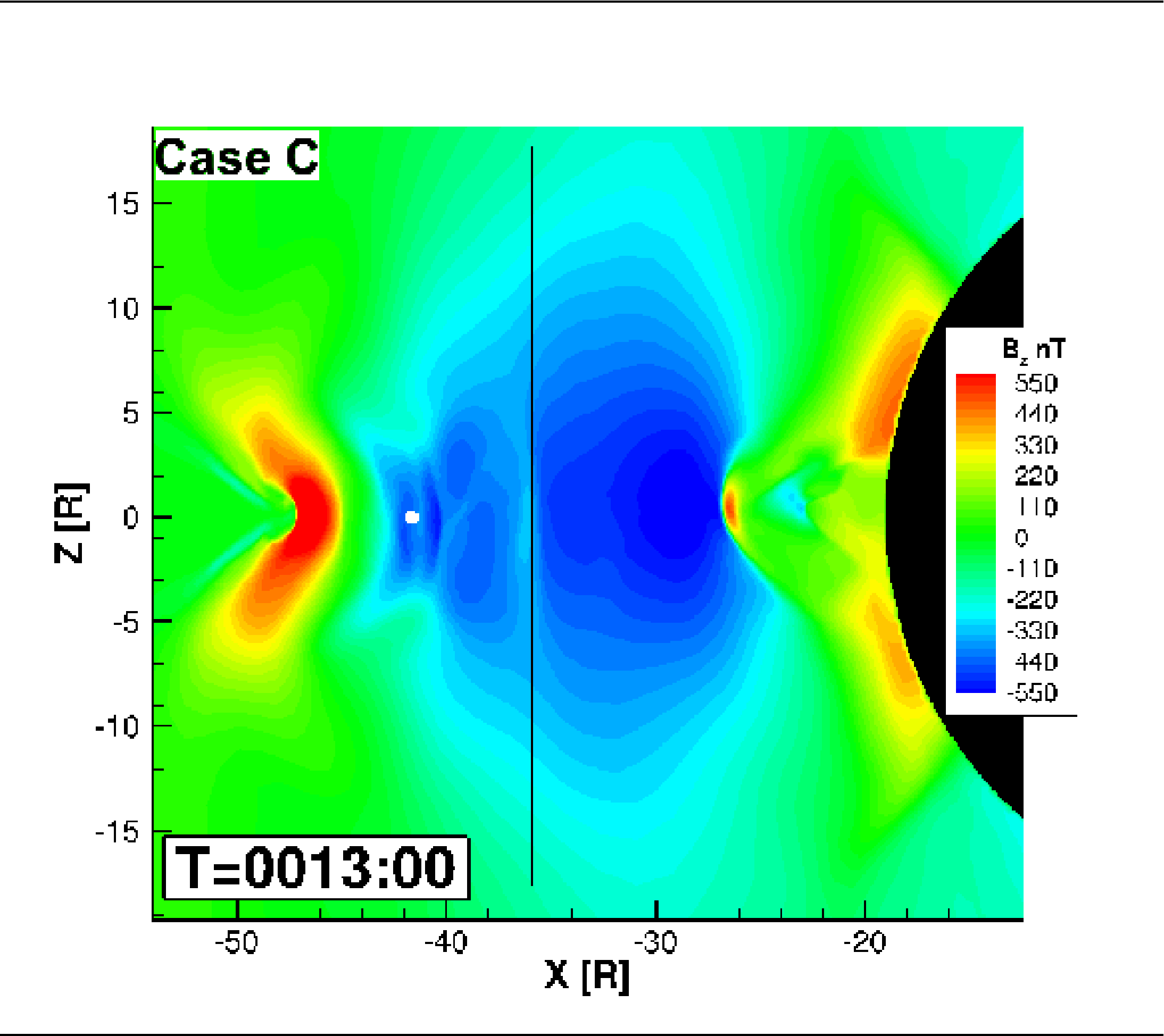}}\\
{\includegraphics*[trim = 1.5cm 1.2cm 2cm 2cm, clip, width=5.4cm]{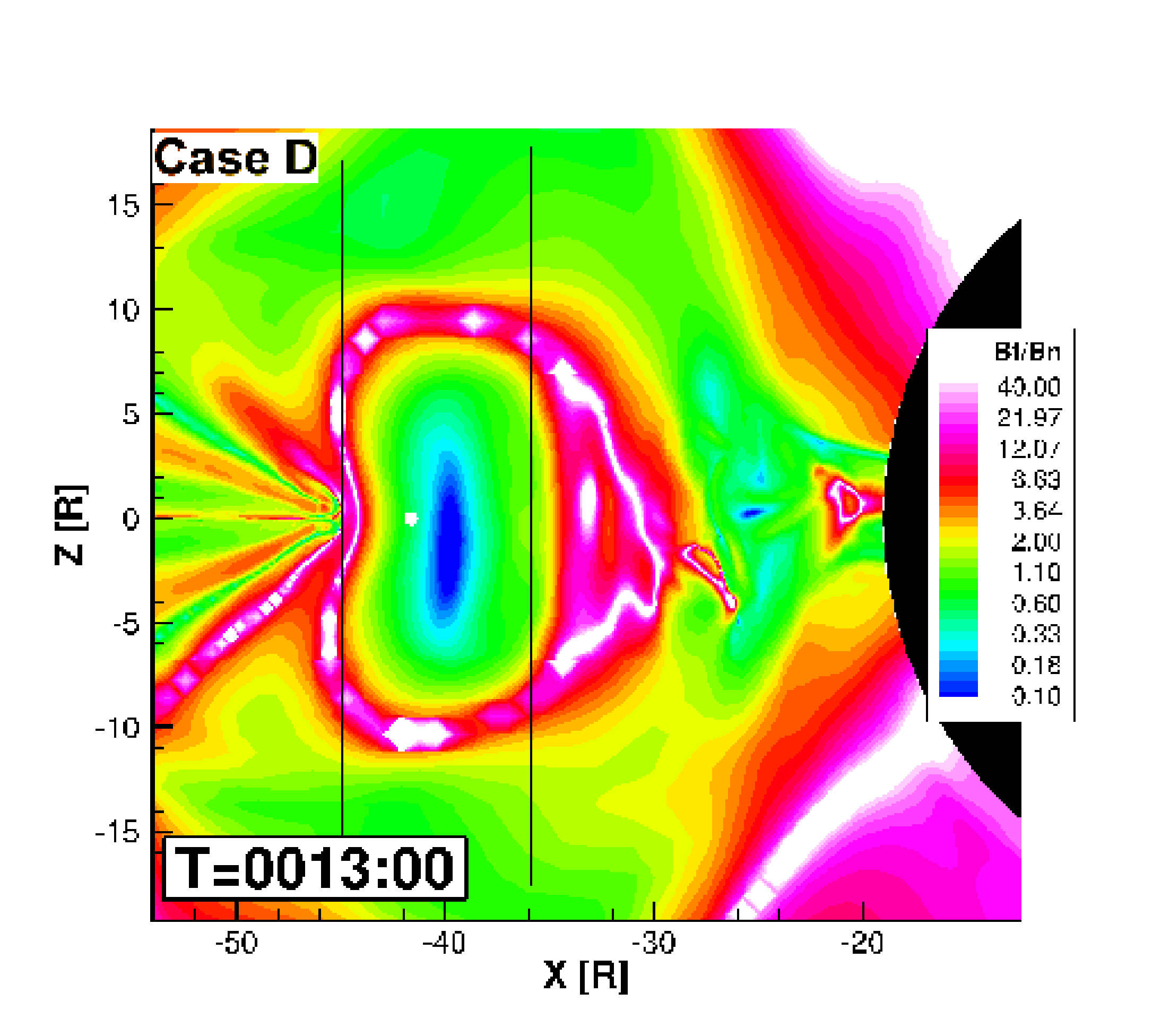}}
{\includegraphics*[trim = 1.5cm 1.2cm 2cm 2cm, clip, width=5.4cm]{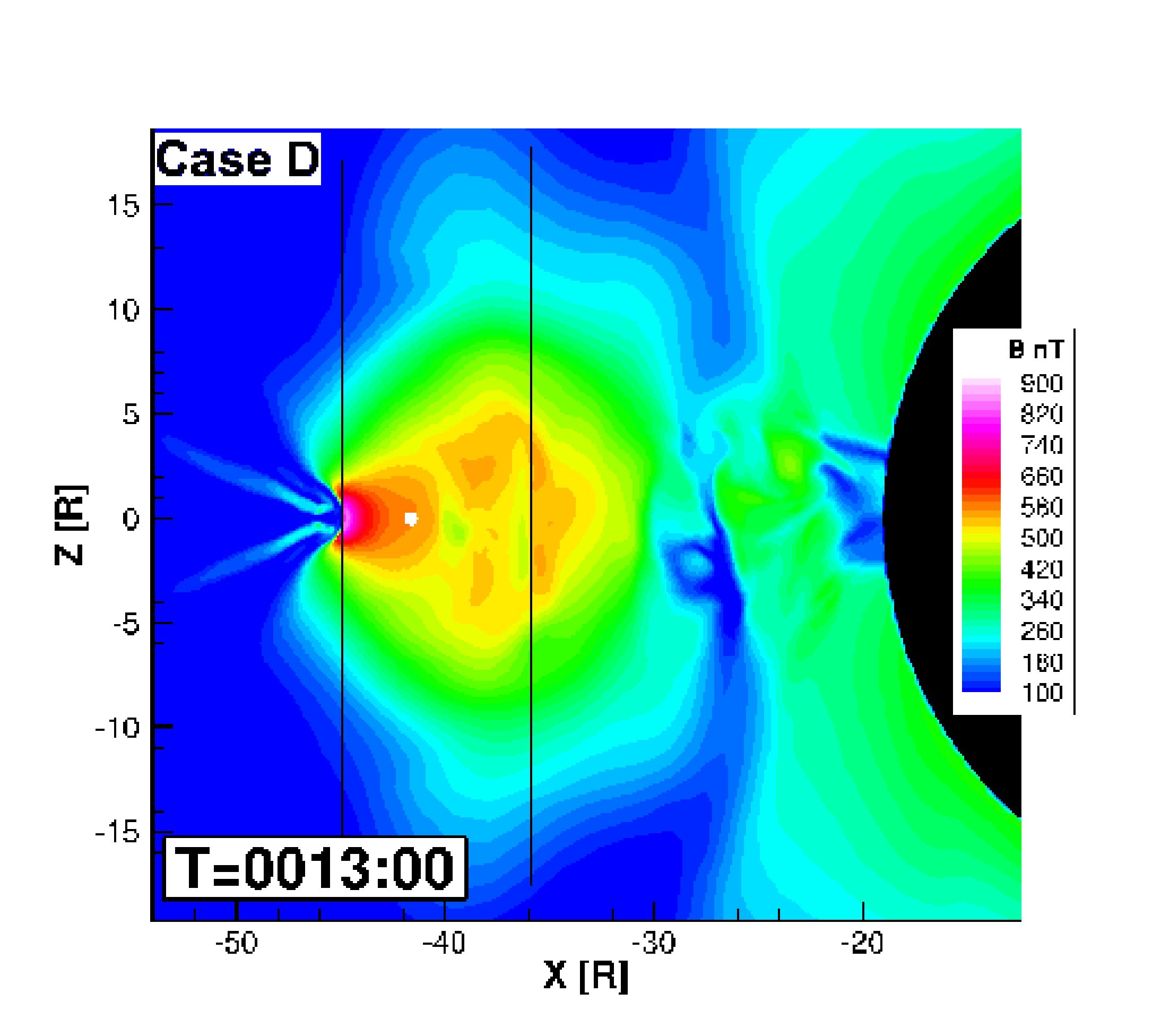}}
{\includegraphics*[trim = 1.5cm 1.2cm 2cm 2cm, clip, width=5.4cm]{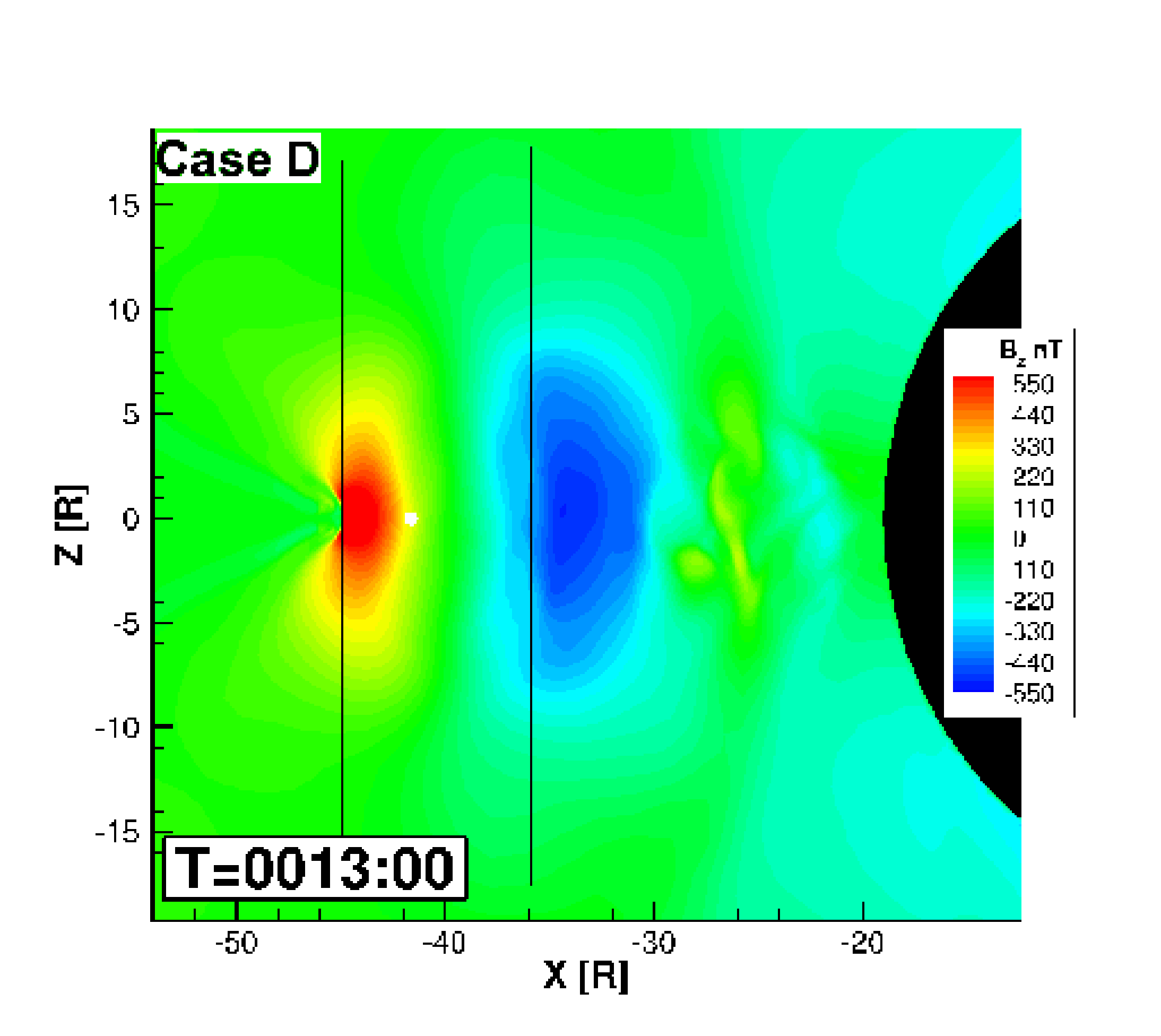}}
\caption{The different columns show, at 13 hours, the ratio of poloidal to axial magnetic field for CME1, the total magnetic field strength  and the north-south component of the poloidal field of CME1, $B_z$, from left to right. Each line corresponds to a different case: A, B, C and D, from top to bottom. The two black lines and white disk mark the beginning, end and center of CME1a. An animated version of some of these panels is available online.}
\end{figure*}
%%%%%%%%%%%%%%%%%%%%%%%%

\subsection{Case C: Large Inclination: Merged CMEs}
This case corresponds to a second CME with a large inclination with respect to the ecliptic and a dominant southward axial field, an ESW type. The interaction of two high-inclination CMEs was recently reported in \citet{Lugaz:2012b} and one WSE CME, the left-handed equivalent to the right-handed CME studied here, was measured {\it in situ} at 1~AU during this event.

CME2c is significantly faster than CME2a. As the only difference with CME2a is its orientation, we believe this is primarily due to the fact that the axis of CME1c is parallel to the overlying coronal field.  Initially, the center of CME1c has a speed of about 2000~km~s$^{-1}$ while the front propagates 400~km~s$^{-1}$ faster and the back about 600~km~s$^{-1}$ slower. The shock driven by CME2c has an initial speed over 3000~km~s$^{-1}$ and decelerates to 2500~km~s$^{-1}$ by 7.5 hours and to 2100~km~s$^{-1}$ by 8 hours. The interaction between CME2c and CME1c starts earlier than for CME1a. At time t = 8.16 hours, the shock driven by CME2c hits the back of CME1c. Then, CME1c starts to contract from a radial width of 10.1~$R_\odot$ at 8 hours to 9.1~$R_\odot$ at 10.5 hours, when the shock passes through the front of CME1c. The contraction continues afterwards, as for CME1a, until the radial width reaches a minimum value of 7.9~$R_\odot$ at 12.83 hours. This minimum value is slightly larger than that for CME1a (barely outside of the uncertainty). After this time, CME1c expands radially and its final width, 13.8~$R_\odot$ at t = 24 hours is significantly larger than that of CME1a. The width of CME1c is plotted with red diamonds in the left panel of Figure~6.

Reconnection between CME1c and CME2c starts soon after the shock collides with CME1c. The poloidal field at the back of CME1c makes an angle of about 90$^\circ$ with the poloidal field in the front of CME2c but it is aligned with the axial field of CME1c. At around t = 18 hours, it is not possible to distinguish between the two CMEs and there is a smooth NESW rotation. The angular width of CME1c is similar to that of CME1a and the aspect ratio increases monotonically but more slowly to a value of about 3.1 at t = 24 hours (shown with red diamonds in the right panel of Figure~6). The speed of CME1c after 24 hours is similar to that of CME1a, except that the back is slower than the center, with its front, center and back propagating at about 750~km~s$^{-1}$, 700~km~s$^{-1}$ and 670~km~s$^{-1}$, respectively.

Because CME2c has a different inclination, we track its boundaries by plotting the function $\sqrt{B_x^2 + B_y^2}/|B_z|$ and we determine its aspect ratio in the ecliptic plane since the axis is perpendicular to it. The width of CME2c is plotted with red diamonds in the middle panel of Figure~6. 
Starting at 8.33 hours, the radial width of CME2c stops to increase as the two CMEs collide. It plateaus to a value of about 6.5~$R_\odot$ for 2 hours before starting to increase again and reaches a value of about 24~$R_\odot$ at the end of the simulation. The time variation of the width of CME2c is highly reminiscent of theoretical and statistical works by \citet{Gulisano:2010} with a phase of over-expansion following a phase of contraction and the final width of the CME being close to what would have been expected in the absence of interaction. This is also what we deduced in \citet{Lugaz:2012b} from the combination of  remote-sensing and {\it in situ} observations. The aspect ratio of CME2c reaches an approximately constant value of about 2.5 after 12.5 hours. The front of CME2c decelerates to 1100~km~s$^{-1}$ by 8.5 hours and to continue to decelerate to reach its final speed of about 720~km~s$^{-1}$. Its center decelerates to 900~km~s$^{-1}$ at 8.5 hours, and to about 650~km~s$^{-1}$ after 10 hours, and the back of CME2c to 800~km~s$^{-1}$ at 9.5 hours, 550~km~s$^{-1}$ at 12 hours and 480~km~ss$^{-1}$ from 14 hours onward.

\subsection{Case D: Large Inclination: Reconnection of CME2}
This case corresponds to a second CME with a large inclination with respect to the ecliptic and a dominant northward axial field, an WNE type. The axial field is anti-parallel with the global magnetic field of the Sun and this can, somewhat similarly to Case B yields extensive reconnection with the background solar wind and coronal magnetic field. As can be seen from Figure~4, 1 hour after its launch, there is only little axial magnetic field left in CME2d. Also, the axial field of CME2d is anti-parallel to the poloidal field at the back of CME1, resulting in additional reconnection. 

CME2d is slower than CME2c but faster than CME2a, with a initial center speed of about 1500~km~s$^{-1}$ and a shock speed of about 2000~km~s$^{-1}$. We believe this is due to the fact that the axis of the CME is parallel to the overlying dipolar coronal field as for CME2c, but, contrary to CME2c, there is extensive reconnection. Interaction between the shock and CME1d starts at 8.66 hours and the radial width of CME1d plateaus at a value of about 10.5~$R_\odot$ from this time to 10 hours. After this time, the behavior of the back boundary of CME1d is somewhat similar to that of CME1b: there is two local maxima of the ratio of poloidal to axial magnetic fields (see bottom left panel of Figure~7). However, for Case D, it is clear that where the boundary is and there is a reconnection occurring around 11.5 hours, around which time, we consider that the boundary goes from the second maximum to the first one (see discontinuity in the green symbols in the right panel of Figure~6. The position of the boundary is very similar to the inner one for Case B and the radial width evolves in a similar way to reach a maximum value of 15.7~$R_\odot$ at time t = 24 hours corresponding to a aspect ratio of 2.45.

%%%%%%%%%%%%%%%%%%%%%%%
\begin{figure*}[ht*]
\centering
{\includegraphics*[width=6.5cm]{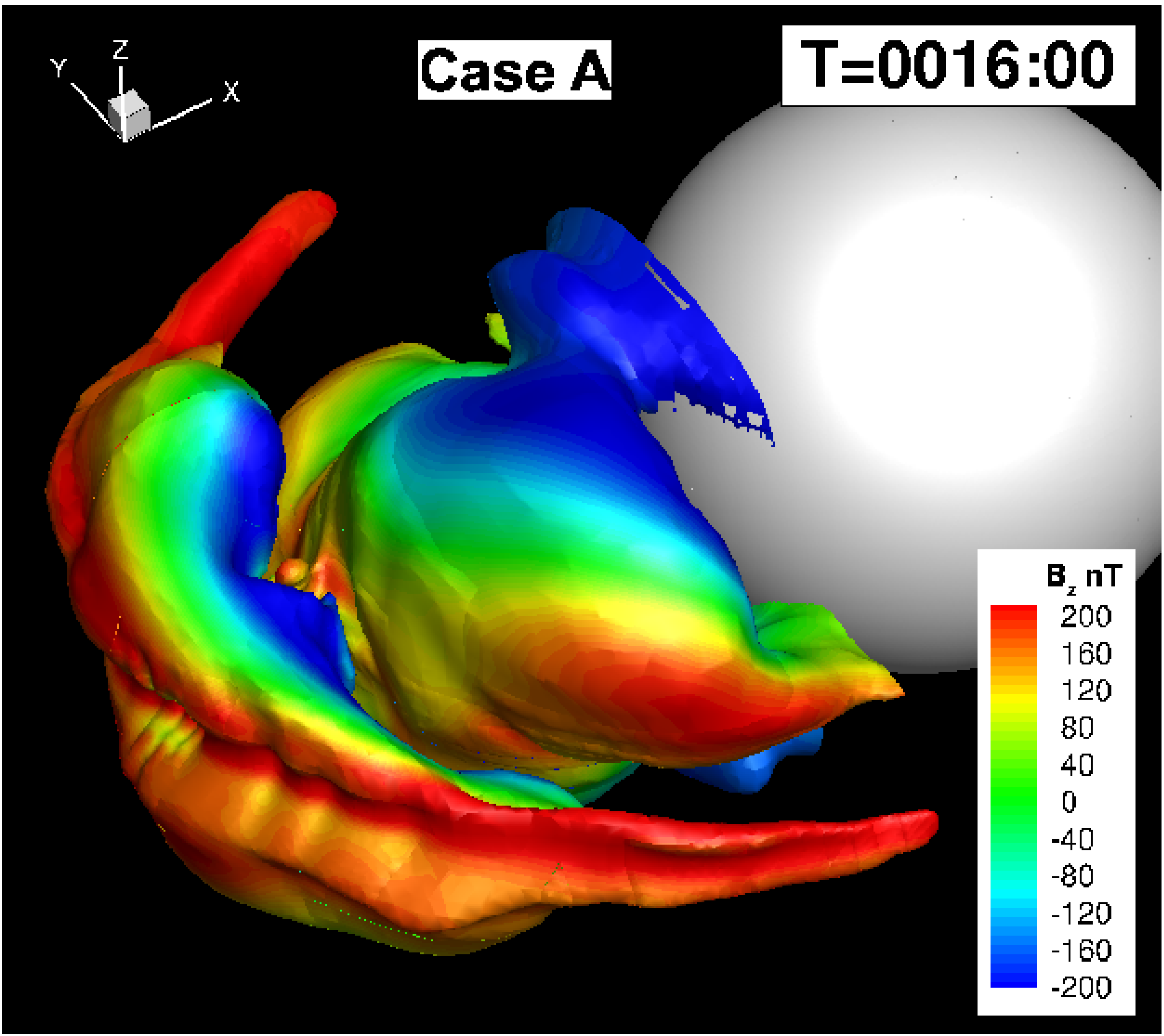}}
{\includegraphics*[width=6.5cm]{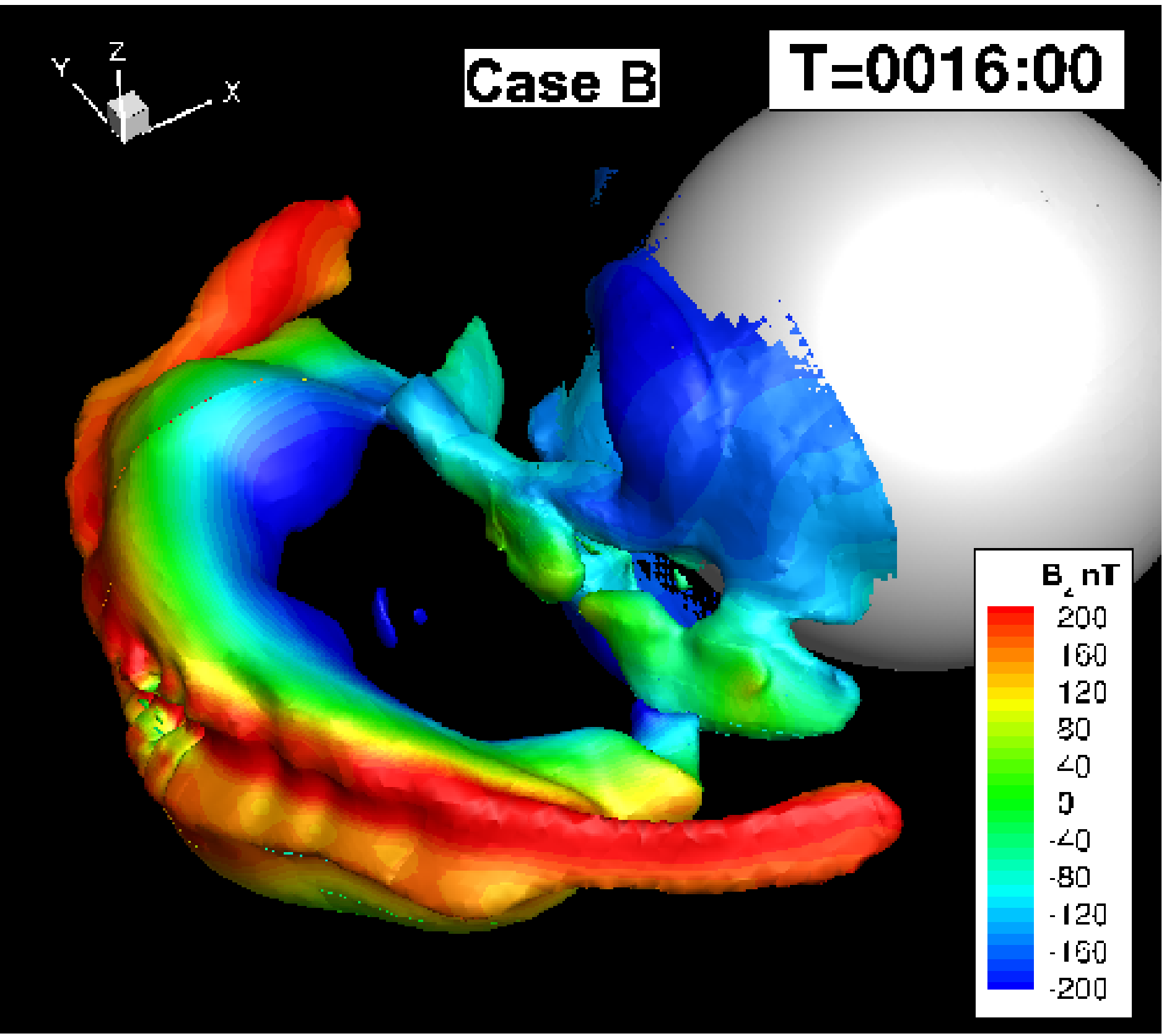}}\\
{\includegraphics*[width=6.5cm]{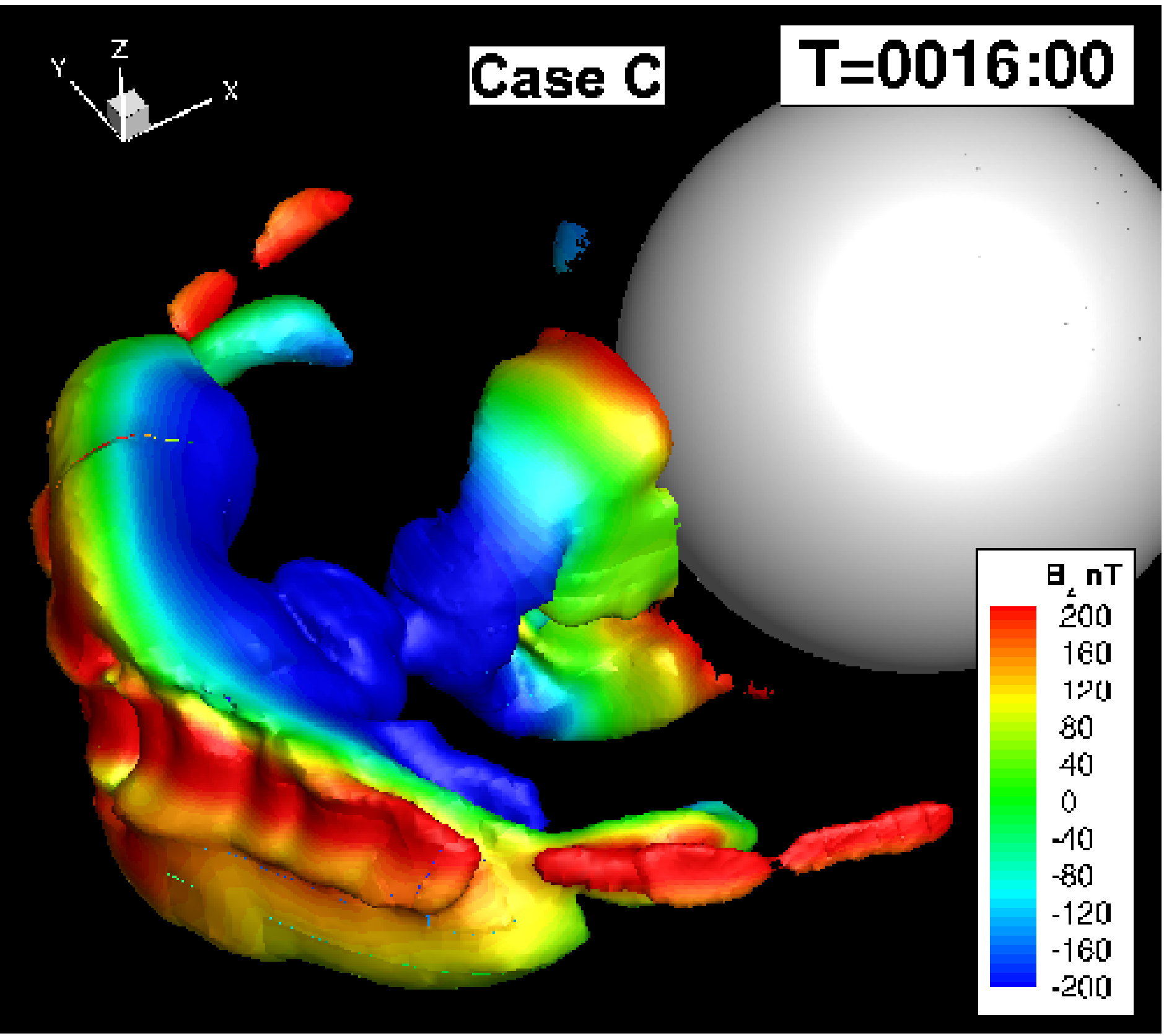}}
{\includegraphics*[width=6.5cm]{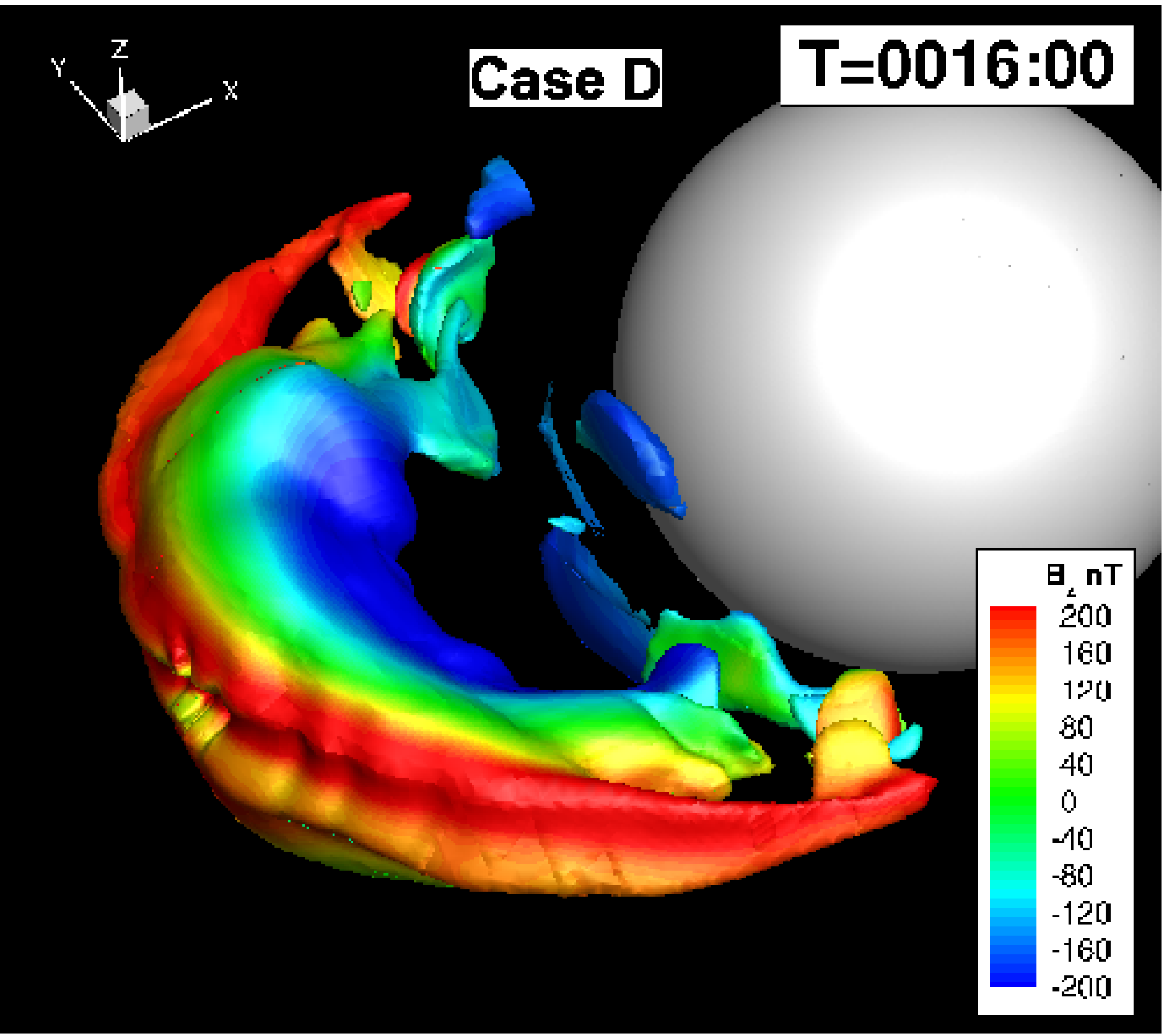}}
\caption{The four cases after 16 hours. The panels show a 3-D isosurface of total magnetic field equal to 230~nT color-coded with the north-south component, $B_z$. Cases A, B, C and D are on the top left, top right, bottom left and bottom right, respectively.}
\end{figure*}
%%%%%%%%%%

\subsection{Comparing the Four Cases}
Figure~8 shows a 3-D isosurface of the magnetic field (equals to 230 nT) at 16 hours, color-coded with the north-south component of the magnetic field, $B_z$ for the four cases. It shows many of the results discussed in the previous sections. CME1 appears the most compressed in Case A and least compressed in Case D. In Cases B and D, CME2 has strongly reconnected with CME1 and the IMF and is barely visible. The large inclination of CME2 is clearly visible for Case C and the resulting long period of southward $B_z$ corresponding to the back of CME1 and the front of CME2. In Case A, CME2 appears to have a larger angular width as compared to CME1. In the next section, we discuss synthetic satellite measurements during and after the interaction. 

\section{Synthetic Satellite Measurements}\label{satellite}

\subsection{During the Interaction at 34~$R_\odot$}
We first extract synthetic satellite measurements at 0.16 AU (34~$R_\odot$), which correspond to a distance where the shock driven by CME2 has progressed to the front half of the magnetic ejecta of CME1.  Results are shown in Figure~9 comparing Cases A and B (low inclination) in the left panel and Cases C and D (high inclination) in the right panel.
At 0.16~AU, focusing on CME1, the synthetic measurements are typical of a shock propagating inside a magnetic cloud (MC). The front of CME1 is at a lower speed than its back, except for Case D, consistent with the compression found for the 3 other cases between 10 and 12 hours.

%%%%%%%%%%%%%%%%%%%%%%%%
\begin{figure*}[ht*]
\centering
{\includegraphics*[width=8.1cm]{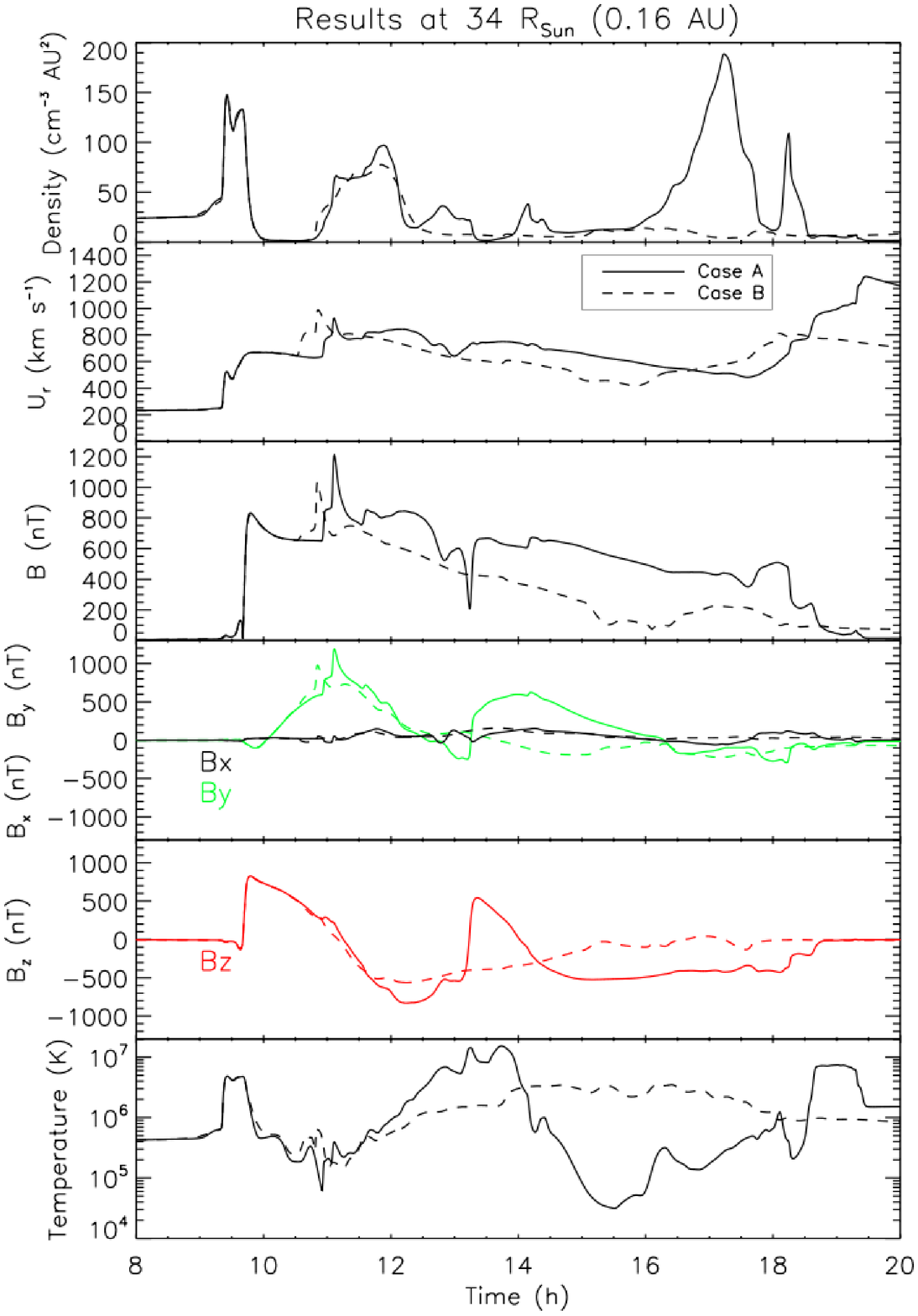}}
{\includegraphics*[width=8.1cm]{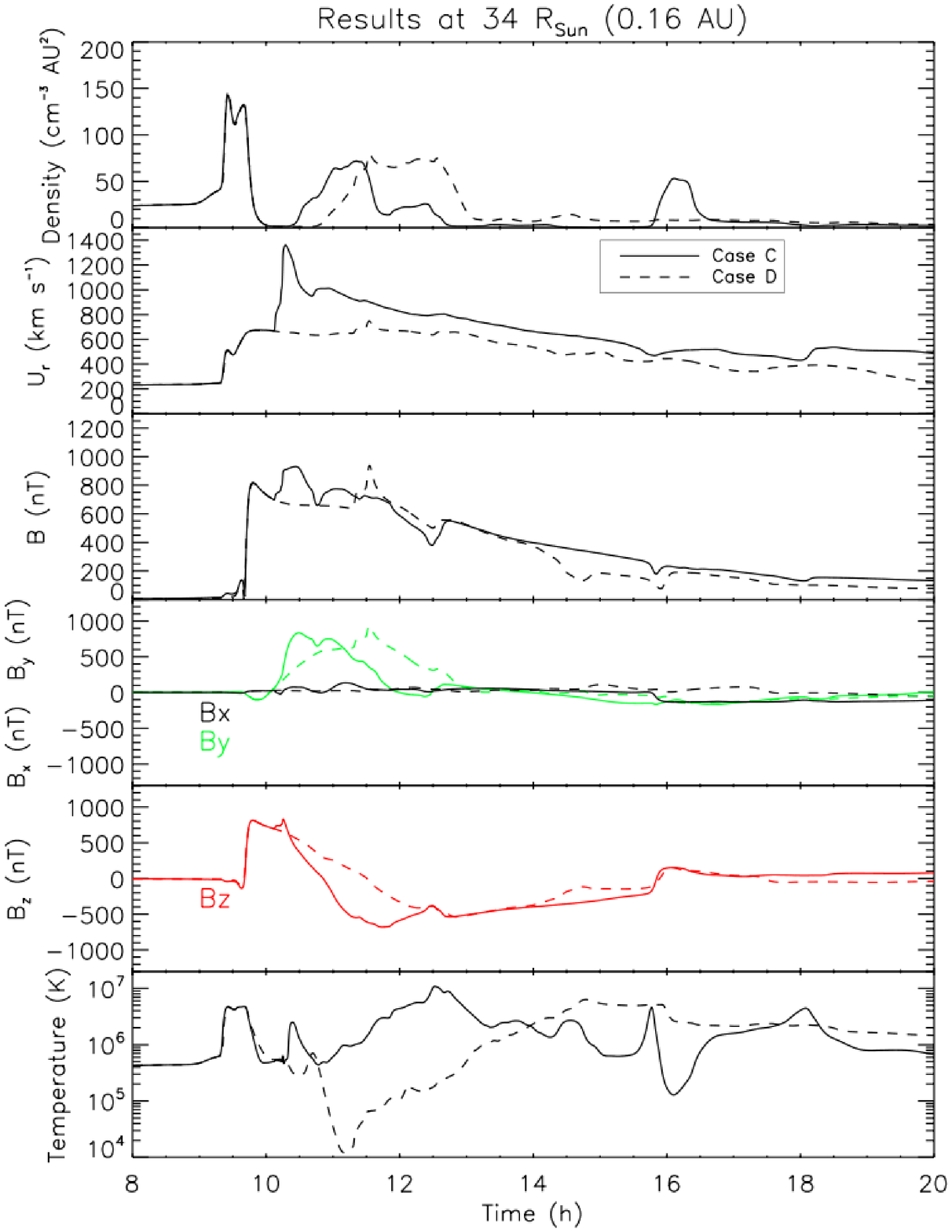}}
\caption{Synthetic satellite data showing the results of the four cases at 34~$R_\odot$ with Cases A and B in the left panel and Cases C and D in the right panel. Each plot shows the density scaled by 1/$r^2$, the radial velocity, the magnetic field strength and $x$,$y$ and $z$ components and the temperature, from top to bottom.}
\end{figure*}
%%%%%%%%%%%%%%%%%%%%%%%%

For Case A, the magnetic field measurements are typical of a multiple-MC event, with two smooth variations of the magnetic field vector and a region of enhanced temperature and lower magnetic field strength at the interface of the two clouds at 13.33 hours, similar to what has been described for real events by \citet{Wang:2003} or in our previous simulation \citep[]{Lugaz:2005b}. This region corresponds to the interface between the two magnetic ejecta and is associated with reconnection.%As described above, this region is characterized by a negative $B_y$. 

For Case B, the magnetic field measurements show a smooth NESW rotation with an extended period of southward $B_z$ corresponding to the back of CME1 and the front of CME2 and during which the axial field rotates from east to west. There is no extended period of northward magnetic field at the back of the complex event, and it is consistent with the reconnection found with the IMF and coronal magnetic field. As is shown in Figure~4, already at time t = 8 hours, the northward magnetic field of CME2 is almost fully reconnected. The amount of westward magnetic field is also significantly reduced as compared to the amount of eastward magnetic field in CME1.

For Case C, the magnetic field measurements also show a smooth NESW rotation with an extended period of southward $B_z$. However, in this case, a significant portion of the southward $B_z$ period occurs while $B_y$ is close to 0. There is also a clear remnant of an interface around time t = 12.5 hours, characterized once again by a weaker magnetic field strength and a hotter temperature, and which could be associated with reconnection. For Case A and Case C, the southward magnetic field is stronger in CME1 and it is probably due to its compression by the shock driven by CME2 as discussed in more details in \citet{Lugaz:2005b}. 

For Case D, there are few, if any, indications of the presence of the second CME, except for the shock inside the first CME. There is no extended period of northward magnetic field except at the very back of the CME. It is very similar to Case B except for the presence of some northward field around time t = 16 hours and a weaker westward field at the back of the CME.

\begin{figure*}[ht*]
\centering
{\includegraphics*[width=8.1cm]{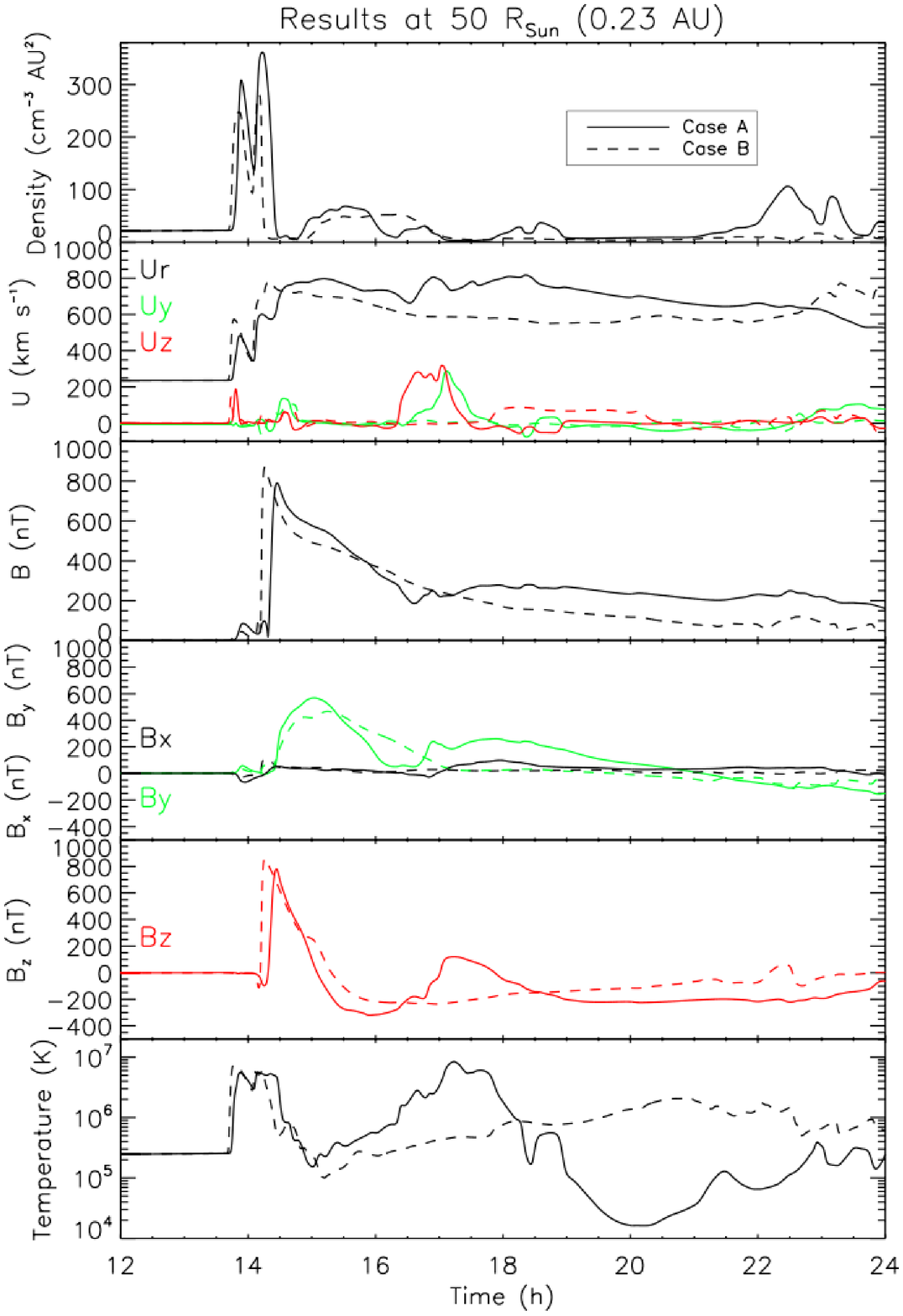}}
{\includegraphics*[width=8.1cm]{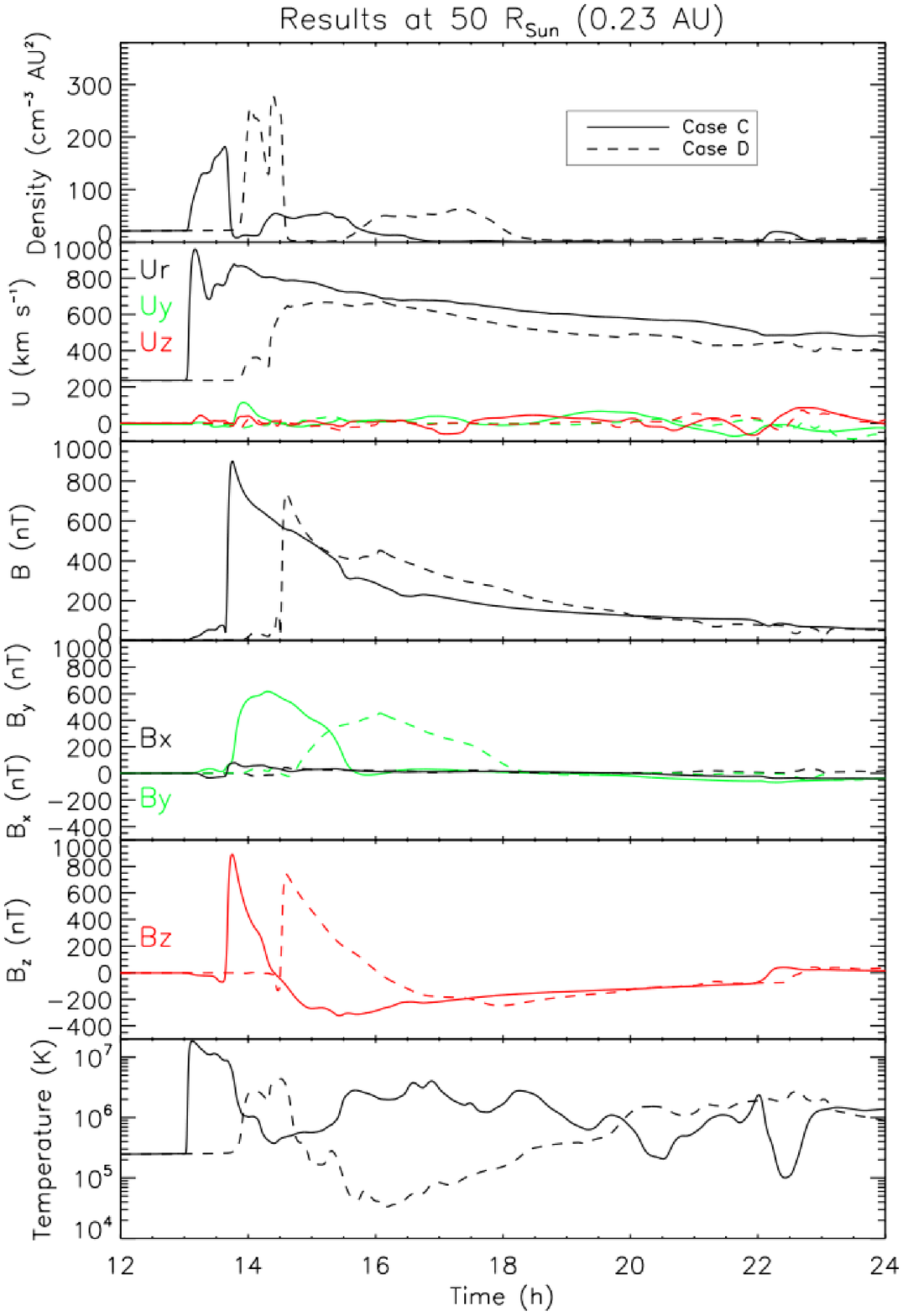}}
\caption{Same as Figure~10 but at 50~$R_\odot$. Note that the magnetic field panel is not symmetric with respect to 0. In the velocity panel, we also plot the east-west ($U_y$) and north-south ($U_z$) components of the velocity.}
\end{figure*}
%%%%%%%%%%%%%%%%%%%%%%%%

\subsection{After the Interaction at 50~$R_\odot$}
We then extract synthetic satellite measurements at 0.23 AU (50~$R_\odot$), which correspond to a time shortly after the end of the main part of the interaction.  Results are shown in Figure~10 comparing Cases A and B  in the left panel and Cases C and D in the right panel.
The two shocks have merged and the new shock reaches 50~$R_\odot$ at the same time in Cases A and B and almost at the same time for Case D. In Case C, it arrives almost 1 hour earlier, consistent with the fact that CME2c was the fastest event and also with the fast shock measured at 0.16~AU. %However, in all these cases, the sheath region behind this shock has a different width (shorter in Case B, wider in Cases A and D).

Case A shows a typical example of multiple-MC event with two low-temperature, high-magnetic field, ejecta with a nearly uniformly decreasing speed profile. As is clear from the presence of a region of higher temperature and decreased magnetic field strength, the front of the second CME is reconnecting. We also plot the north-south and east-west components of the velocity. This shows further evidence of the reconnection between the two magnetic ejecta with strong non-radial flows from 16.33 hours to 17.5 hours, corresponding to the time of higher temperature and reduced magnetic field strength. Note that these non-radial flows, which are highly reminiscent of reconnection jets are only present in Case A. %however, we have not performed a tangential stress balance test or Wallen test \citep[]{Sonnerup:1981}.

Case B and C resemble an isolated CME with an extended tail as the westward magnetic field (axial field of CME2b, poloidal field of CME2c) has almost fully disappeared. The period of $B_z$ negative is longer for Case C and there is a stronger remnant of the westward (negative) $B_y$ in Case B. These events could be characterized as MC-like and may be mistaken for a glancing encounter with a MC \citep[]{Marubashi:2007}. For Case D, there are little evidence left of any type of interaction as the period of eastward magnetic field (axial field of CME1) has a much wider radial extent and is characterized by a low temperature. 

Lastly, we determine the dimensionless expansion rate $\xi$ following \citet{Demoulin:2009} and \citet{Gulisano:2010} and defined as:
$$ \xi = \frac{\Delta V_r}{\Delta t} \frac{D}{V_c^2}$$
where $-\Delta V_r/ \Delta t$ is the slope of the velocity profile obtained from {\it in situ} measurements taken at a distance $D$ from the Sun, and $V_c$ is the speed at the center of the magnetic ejecta. This measure of the expansion has been shown to depend only weakly on the distance from the Sun and the ejecta size, especially for non-perturbed ejecta \citep[]{Gulisano:2010}. This is obviously not the case here, since the expansion rate and velocity decrease through the CMEs are strongly influenced by the collision, interaction and reconnection between the two CMEs. 

Typical values of $\xi$ for non-perturbed ejecta in the inner heliosphere have been reported to be $0.91 \pm 0.23$, whereas perturbed ejecta typically exhibit much slower radial expansion and a much greater standard deviation with behaviors ranging from contraction ($\xi < 0$) to over-expansion ($\xi > 1$).  The average value of $\xi$ for perturbed ejecta (based on 16 events) has been reported to be 0.48 $\pm 0.79$. If $\xi$ is independent of the distance, the width $W$ of a magnetic ejecta evolves as $W \propto r^\xi$. 

We derive $\xi$ for CME1 in the four cases at 0.23~AU. Boundaries are identified visually with the help from the 3-D simulation. Results are summarized in columns 2--4 of Table~1. We also fit the width measurements described in the previous sections with respect to the distance of the center of the CME. The results are shown in the fifth column. We list two fits for Cases B and C, depending on the time period fitted, the shortest period is in parentheses. Based on the $\xi$ values, CME1a has a slow increase characteristics of perturbed CMEs, CME1b and CME1c are over-expanding and CME1d has an expansion rate within the limit of unperturbed clouds.

\begin{table}[ht*]
\centering
\begin{tabular}{|c|ccc|c|}
\hline

CME & $\Delta V_r/ \Delta t $ & $V_c$  & $\xi$ & Fit of width\\
\hline
CME1a & 17 &  770 & 0.28 & 0.08 $r^{0.54}$ \\
CME1b & 60 & 690 & 1.2 & 0.16 $r^{0.78}$ (0.24 $r^{1.16}$)\\
CME1c & 72 & 800 & 1.09 & 0.16 $r^{0.93}$ (0.19 $r^{1.09}$) \\
CME1d & 31 & 675 & 0.67 & 0.16 $r^{0.74}$\\
\hline
\end{tabular}
\caption{Dimensionless expansion rate at 50~$R_\odot$.  \newline The second column is the slope of the velocity in km~s$^{-1}~h^{-1}$, the third is the speed at the center in km~s$^{-1}$.}
\end{table}

In fact, looking at the radial width of CME1a as plotted in Figure~6, it is clear that the CME has almost no expansion between 14 and 16 hours which is when it passes the synthetic spacecraft. During this time, the width increases from 7.6 to 7.9~$R_\odot$. However, we do not have enough datapoints to perform an adequate fitting during such a short time period. If we perform a fit on all the datapoints following the contraction (starting at 14.33 hours), we obtain a radial dependence of the width as r$^{0.54}$, i.e. characteristics of a perturbed CME but larger than what is obtained from the snapshot provided by a time series. In any case, both diagnostics indicate that the expansion of CME1a has been perturbed due to its interaction, even though we are more than 5 hours after the start of the interaction. For CME1d, the two diagnostics appear to again agree, pointing towards a CME close to unperturbed, which is consistent with what we have discussed above for Case D.

For CME1b, there is initially disagreement between the two diagnostics: the fitting points to an expansion close to unperturbed, while the dimensionless parameter points to a over-expanding width. The difference might be due to the difficulties in determining the boundaries of CME1b, as discussed in section 4.1. The value given here is for all points after 14 hours (the front, center and back of CME1b passes 50~$R_\odot$ at 14.33, 15.33 and 17.33 hours, respectively). If we instead fit all points after 16 hours, we find a relation consistent with over-expansion as $ W \propto r^{1.16}$. This indicates that CME1b does indeed over-expand, as found from the dimensionless parameter.

For CME1c, the value of $\xi$ equal to 1.09 corresponds to over-expansion. This is consistent with the width measurements as described in section 4.2. Fitting the width-distance measurements during the time of the passage over the synthetic spacecraft (from 14 to 19 hours) leads a width variation as $r^{1.09}$, in agreement with the value of $\xi$. As expected after this period of over-expansion, the radial expansion slows down and our best fit to the data after 12.83 hours is a radial dependency of $r^{0.93}$, which is close to that of a normal CME.

\section{Discussions and Conclusion} \label{conclusion}

\subsection{Radial Expansion of Perturbed CMEs}

We have determined the width and expansion of the CMEs from synthetic {\it in situ} measurements following the methods of \citet{Gulisano:2010} and from the 3-D simulations. The dimensionless parameter, $\xi$, derived from {\it in situ} measurements at a single point should be related to the global behavior of the width as $W \propto r^\xi$ as long as $\xi$ is invariant with distance. This is obviously not the case here due to the collision, interaction and reconnection between the two CMEs, which create a strong time and distance dependency in the width and its expansion. We find nonetheless that $\xi$ gives a good approximation to the behavior of the first CME around the time of the spacecraft crossing: slowly expanding CMEs were recognized as such and so were over-expanding CMEs. A value of $\xi$ between 0.6 and 0.9 appears to be consistent with a undisturbed CMEs as found by \citet{Gulisano:2010} for heliocentric distances less than 1~AU. 

We have also found that the width of perturbed CMEs and their expansion rate strongly depend on the possibility of reconnection between the two magnetic ejecta. For two CMEs with the same orientation (Case A), the fact that the axial fields are parallel limits the reconnection and hinders the over-expansion of the first CME after the interaction. This confirms our findings from \citet{Lugaz:2005b} and the CME width at 1~AU may be as small as 0.08~AU, which would correspond to a crossing time of 4--8 hours. In the other cases, where the axial fields make a non-zero angle, the reconnection between the two CMEs allows the first CME to expand after the interaction. We also find a short  phase of over-expansion for CME1 in Case B and C after the interaction, which is consistent with the scenario proposed by \citet{Gulisano:2010} for the expansion of perturbed CMEs.

\subsection{Change in the Speed of the CMEs During the Interaction}

During the interaction, in all cases, the speed of the two CMEs becomes more uniform and the speed profile of the complex ejecta at 50~$R_\odot$ is already similar to what would be expected from a single CME event (see Figure~10). Overall, the center of CME1 gets accelerated from about 600~km~s$^{-1}$ before the interaction to [750, 650, 800, 620]~km~s$^{-1}$ for Cases [A,B,C,D] at time t = 24~hours. Although Case D does not appear to get accelerated, the simple fact that the center of CME1d does not decelerate in 16 hours is akin to an acceleration. Meanwhile, the center of CME2 has a pre-interaction speed of [1500, 1000, 2000, 1400]~km~s$^{-1}$ and a post-eruption speed of [700,650,725,625]~km~s$^{-1}$. 

Determining the mass of the two CMEs is not straight-forward, as simulations and recent observations \citep[]{Lugaz:2005a,deForest:2013} show that it can increase by a factor of two or more from the upper corona to 1~AU due to the snowplow effect. Since the CMEs are initiated with the same parameters, we assume that their mass is similar at the time of the interaction. Even if CME1 were three times as massive as CME2, the result below would hold.

As a first approximation, it appears that the collision reduces the speed of CME2 below that from the inelastic collision limit, $V_{\mathrm{inelast.}} = \frac{1}{2} (V_{\mathrm{CME1}} + V_{\mathrm{CME2}})$, which is [1050, 800, 1300, 1000]~km~s$^{-1}$. This result is similar to that of \citet{Temmer:2012} for the ejection of 2010 August 1. Here, the kinetic energy does not appear to increase during the interaction as found recently by \citet{CShen:2012} and \citet{FShen:2013}. In fact, there are a number of problems in determining the type of collision for CME-CME interaction: (i) there are more than 16 hours between the start of the interaction and the end of the simulation and it is unclear when exactly we should consider that the interaction is finished;  (ii) the deceleration and mass accretion of the CMEs due to the interaction with the solar wind cannot be neglected during the interaction; (iii) the expansion speeds change drastically during the interaction and the expansion is still perturbed more than 10 hours after the collision, (iv) the effect of the reconnection jets on the mass of the event should also be taken into consideration. An in-depth analysis of the variation of the different energies taking into account all these effects in the four cases studied here is beyond the scope of this paper and it is left for follow-up analyses.

\subsection{Conclusions}

In this article, we have reported and analyzed numerical simulations of the interaction of two CMEs in four different cases, where the relative orientation of the two  CMEs was varied. The CMEs are initiated with GL flux ropes and the second CME is launched seven hours after the first one from the same latitude and longitude at the solar surface. To the best of our knowledge, this is the first parametric study of interacting CMEs performed in 3-D and it is performed with higher resolution (but much fewer cases) than the previous 2.5-D numerical parametric study of \citet{Xiong:2006}. \citet{Schmidt:2004} also performed two simulations in 2.5-D corresponding to different orientations (180$^\circ$ apart). Recent works have found that all orientations of magnetic clouds are equally common \citep[e.g., see][]{Janvier:2013} and the four cases presented here should therefore, occur with the same frequency (except if CMEs from the same active region are the main cause of CME-CME interaction, then Case A should occur more frequently). 

We have studied in detail the radial width and expansion of the CMEs during the interaction. We have found that there is a radial compression of the first CME irrespectively of their orientation, as the two CMEs collide and the shock driven by CME2 propagates through CME1. Afterwards, the rate of expansion is determined, in part, by the rate of reconnection between the two CMEs. In the case where the axial fields of the two CMEs is parallel, there is limited expansion following the interaction; in other cases, there is often a short period of over-expansion and the expansion rate eight or more hours after the interaction is similar to that before the interaction (but with a smaller width than expected). The compression in radial direction is not associated with an increase in the angular width; the aspect ratio of CME1 increases but it is purely due to the compression in the radial direction. The notable exception is for CME2 in Case A (two CMEs with the same orientation), where the angular width of CME2 increases during the interaction. This might be due to an inefficient rate of reconnection between two CMEs with the same orientation, as compared to the time needed to transfer magnetic flux.

We have found that multiple-MC events are just one of the possible results of the interaction of two CMEs. Other resulting configurations include a complex event with a seemingly smooth rotation over more than 180$^\circ$ of the magnetic vector. In the configuration chosen here, it corresponds to a long tail of southward magnetic field which should have an impact on the geo-effectiveness of these complex ejecta. \citet{Dasso:2009} reported the passage of a long (40 hours) and complex event in May 2005, which they analyzed as two ``non-merging'' left-handed magnetic clouds with different orientations, a low-inclination first cloud followed by a larger high-inclination cloud, similar to our Case C.  The variations of the magnetic field vector was found to be smooth  (see their Figure~1). The event resulted in a minimum Dst of $-263$ nT due to large southward magnetic field in the compressed first magnetic cloud. \citet{Marubashi:2007} identified 17 MCs  with duration longer than 30 hours; five of these exhibited rotation of the magnetic field vector over 180$^\circ$. The authors associated these five events with the crossing through the leg of a flux rope, but some of them may also be associated with unrecognized CME-CME interactions.  %Also see \citet{Steed:2011}.\\

Our simulation work is based on solving the ideal MHD equations. As such, it does not treat reconnection in a physically correct manner;  reconnection should be studied with kinetic codes. However, current computing capabilities do not make it possible to simulate the Sun-to-Earth propagation of CMEs in 3-D with such codes, and, while the coupling between kinetic and MHD codes is a promising area of research, it is still in its early steps. We believe that, in our simulations, the reconnection occurs at the correct location and time but the reconnection rate, being controlled by the numerical resistivity, is not correct. The main effect will be on the exact expansion of the CMEs after the interaction. We note however, that we found behavior in agreement with previous theoretical works \citep[]{Gulisano:2010}. Follow-up studies should also propagate these complex events to 1~AU to study their geo-effectiveness and compare the simulation results to actual spacecraft measurements. More cases will also be needed in order to study the effect of different handedness for the two interacting CMEs, study the interaction of slower CMEs, not preceded by shocks, and consider different directions and speeds of the interacting CMEs.

\begin{acknowledgments}
The authors would like to thank P.~D{\'e}moulin and the reviewer for their useful comments on the manuscript.
The research for this manuscript was supported by the following grants: NSF AGS-1239699 and NASA NNX13AH94G.
N.~L. was also partially supported by the following NSF grants: AGS-1239704 and AGS-1135432. \\ %and W.~M. was supported by the following grants: \\
The simulations were performed on the {\it Pleiades} system provided by NASAÕs High-End Computing Program under awards SMD-12-3360 and SMD-13-3919. 
\end{acknowledgments}

\bibliographystyle{apj}

\begin{thebibliography}{68}
\expandafter\ifx\csname natexlab\endcsname\relax\def\natexlab#1{#1}\fi

\bibitem[{{Bothmer} \& {Schwenn}(1998)}]{Bothmer:1998}
{Bothmer}, V., \& {Schwenn}, R. 1998, Annales Geophysicae, 16, 1

\bibitem[{{Burlaga} {et~al.}(2003){Burlaga}, {Berdichevsky}, {Gopalswamy},
  {Lepping}, \& {Zurbuchen}}]{Burlaga:2003}
{Burlaga}, L., {Berdichevsky}, D., {Gopalswamy}, N., {Lepping}, R., \&
  {Zurbuchen}, T. 2003, J. Geophys. Res., 108, 2

\bibitem[{{Burlaga} {et~al.}(1987){Burlaga}, {Behannon}, \&
  {Klein}}]{Burlaga:1987}
{Burlaga}, L.~F., {Behannon}, K.~W., \& {Klein}, L.~W. 1987, J. Geophys. Res.,
  92, 5725

\bibitem[{{Burlaga} {et~al.}(2002){Burlaga}, {Plunkett}, \&
  {St.~Cyr}}]{Burlaga:2002}
{Burlaga}, L.~F., {Plunkett}, S.~P., \& {St.~Cyr}, O.~C. 2002, J. Geophys.
  Res., 107, 1

\bibitem[{{Cohen} {et~al.}(2008){Cohen}, {Sokolov}, {Roussev}, {Lugaz},
  {Manchester}, {Gombosi}, \& {Arge}}]{Cohen:2008a}
{Cohen}, O., {Sokolov}, I.~V., {Roussev}, I.~I., {et~al.} 2008, J. Atmos.
  Solar-Terr. Phys., 70, 583

\bibitem[{{Dasso} {et~al.}(2009){Dasso}, {Mandrini}, {Schmieder}, {Cremades},
  {Cid}, {Cerrato}, {Saiz}, {D{\'e}moulin}, {Zhukov}, {Rodriguez}, {Aran},
  {Menvielle}, \& {Poedts}}]{Dasso:2009}
{Dasso}, S., {Mandrini}, C.~H., {Schmieder}, B., {et~al.} 2009, J. Geophys.
  Res., 114, 2109

\bibitem[{{DeForest} {et~al.}(2013){DeForest}, {Howard}, \&
  {McComas}}]{deForest:2013}
{DeForest}, C.~E., {Howard}, T.~A., \& {McComas}, D.~J. 2013, Astrophys. J.,
  769, 43

\bibitem[{{D{\'e}moulin} \& {Dasso}(2009)}]{Demoulin:2009}
{D{\'e}moulin}, P., \& {Dasso}, S. 2009, Astron. Astrophys., 507, 969

\bibitem[{{Evans} {et~al.}(2012){Evans}, {Opher}, {Oran}, {van der Holst},
  {Sokolov}, {Frazin}, {Gombosi}, \& {V{\'a}squez}}]{Evans:2012}
{Evans}, R.~M., {Opher}, M., {Oran}, R., {et~al.} 2012, Astrophys. J., 756, 155

\bibitem[{{Eyles} {et~al.}(2009){Eyles}, {Harrison}, {Davis}, {Waltham},
  {Shaughnessy}, {Mapson-Menard}, {Bewsher}, {Crothers}, {Davies}, {Simnett},
  {Howard}, {Moses}, {Newmark}, {Socker}, {Halain}, {Defise}, {Mazy}, \&
  {Rochus}}]{Eyles:2009}
{Eyles}, C.~J., {Harrison}, R.~A., {Davis}, C.~J., {et~al.} 2009, Solar Phys.,
  254, 387

\bibitem[{{Farrugia} \& {Berdichevsky}(2004)}]{Farrugia:2004}
{Farrugia}, C., \& {Berdichevsky}, D. 2004, Annales Geophysicae, 22, 3679

\bibitem[{{Farrugia} {et~al.}(2006{\natexlab{a}}){Farrugia}, {Jordanova},
  T{Thomsen}, {Lu}, {Cowley}, \& {Ogilvie}}]{Farrugia:2006}
{Farrugia}, C.~J., {Jordanova}, V.~K., T{Thomsen}, M.~F., {et~al.}
  2006{\natexlab{a}}, J. Geophys. Res., 111

\bibitem[{{Farrugia} {et~al.}(2006{\natexlab{b}}){Farrugia}, {Matsui},
  {Kucharek}, {Jordanova}, {Torbert}, {Ogilvie}, {Berdichevsky}, {Smith}, \&
  {Skoug}}]{Farrugia:2006b}
{Farrugia}, C.~J., {Matsui}, H., {Kucharek}, H., {et~al.} 2006{\natexlab{b}},
  Advances in Space Research, 38, 498

\bibitem[{{Gibson} \& {Low}(1998)}]{Gibson:1998}
{Gibson}, S.~E., \& {Low}, B.~C. 1998, Astrophys. J., 493, 460

\bibitem[{{Gombosi} {et~al.}(2001){Gombosi}, {De Zeeuw}, Groth, {Powell},
  {Clauer}, \& Song}]{Gombosi:2001}
{Gombosi}, T.~I., {De Zeeuw}, D.~L., Groth, C.~P.~T., {et~al.} 2001, Geophys.
  Monograph, 125, 169

\bibitem[{{Gopalswamy} {et~al.}(2001){Gopalswamy}, {Yashiro}, {Kaiser},
  {Howard}, \& {Bougeret}}]{Gopalswamy:2001}
{Gopalswamy}, N., {Yashiro}, S., {Kaiser}, M.~L., {Howard}, R.~A., \&
  {Bougeret}, J.-L. 2001, Astrophys. Journ. Lett., 548, L91

\bibitem[{{Gulisano} {et~al.}(2010){Gulisano}, {D{\'e}moulin}, {Dasso}, {Ruiz},
  \& {Marsch}}]{Gulisano:2010}
{Gulisano}, A.~M., {D{\'e}moulin}, P., {Dasso}, S., {Ruiz}, M.~E., \& {Marsch},
  E. 2010, Astron. Astrophys., 509, A39

\bibitem[{{Harrison} {et~al.}(2012){Harrison}, {Davies}, {M{\"o}stl}, {Liu},
  {Temmer}, {Bisi}, {Eastwood}, {de Koning}, {Nitta}, {Rollett}, {Farrugia},
  {Forsyth}, {Jackson}, {Jensen}, {Kilpua}, {Odstrcil}, \&
  {Webb}}]{Harrison:2012}
{Harrison}, R.~A., {Davies}, J.~A., {M{\"o}stl}, C., {et~al.} 2012, Astrophys.
  J., 750, 45

\bibitem[{{Howard} {et~al.}(2012){Howard}, {DeForest}, \&
  {Reinard}}]{THoward:2012b}
{Howard}, T.~A., {DeForest}, C.~E., \& {Reinard}, A.~A. 2012, Astrophys. J.,
  754, 102

\bibitem[{{Intriligator}(1976)}]{Intriligator:1976}
{Intriligator}, D.~S. 1976, Space Sci. Rev., 19, 629

\bibitem[{{Ivanov}(1982)}]{Ivanov:1982}
{Ivanov}, K.~G. 1982, Space Sci. Rev., 32, 49

\bibitem[{{Janvier} {et~al.}(2013){Janvier}, {Demoulin}, \&
  {Dasso}}]{Janvier:2013}
{Janvier}, M., {Demoulin}, P., \& {Dasso}, S. 2013, Astron. Astrophys., 556, A50

%\bibitem[{{Jin} {et~al.}(2012){Jin}, {Manchester}, {van der Holst},
%  {Gruesbeck}, {Frazin}, {Landi}, {Vasquez}, {Lamy}, {Llebaria}, {Fedorov},
 % {Toth}, \& {Gombosi}}]{Jin:2012}
%{Jin}, M., {Manchester}, W.~B., {van der Holst}, B., {et~al.} 2012, Astrophys.
 % J., 745, 6

\bibitem[{{Jin} {et~al.}(2013){Jin}, {Manchester}, {van der Holst},
  {Oran}, {Sokolov}, {Toth}, {Liu}, {Sun}, \& {Gombosi}}]{Jin:2013}
{Jin}, M., {Manchester}, W.~B., {van der Holst}, B., {et~al.} 2013, Astrophys.
  J., 773, 50


\bibitem[{{Kaiser} {et~al.}(2008){Kaiser}, {Kucera}, {Davila}, {St.~Cyr},
  {Guhathakurta}, \& {Christian}}]{Kaiser:2008}
{Kaiser}, M.~L., {Kucera}, T.~A., {Davila}, J.~M., {et~al.} 2008, Space Sci.
  Rev., 136, 5

\bibitem[{{Liu} {et~al.}(2005){Liu}, {Richardson}, \& {Belcher}}]{Liu:2005}
{Liu}, Y., {Richardson}, J.~D., \& {Belcher}, J.~W. 2005, Planet. Space Sci.,
  53, 3

\bibitem[{{Liu} {et~al.}(2012){Liu}, {Luhmann}, {M{\"o}stl},
  {Martinez-Oliveros}, {Bale}, {Lin}, {Harrison}, {Temmer}, {Webb}, \&
  {Odstrcil}}]{Liu:2012}
{Liu}, Y.~D., {Luhmann}, J.~G., {M{\"o}stl}, C., {et~al.} 2012, Astrophys.
  Journ. Lett., 746, L15

\bibitem[{{Loesch} {et~al.}(2011){Loesch}, {Opher}, {Alves}, {Evans}, \&
  {Manchester}}]{Loesch:2011}
{Loesch}, C., {Opher}, M., {Alves}, M.~V., {Evans}, R.~M., \& {Manchester}, IV,
  W.~B. 2011, J. Geophys. Res., 116, 4106

\bibitem[{{Lugaz} {et~al.}(2012){Lugaz}, {Farrugia}, {Davies}, {M{\"o}stl},
  {Davis}, {Roussev}, \& {Temmer}}]{Lugaz:2012b}
{Lugaz}, N., {Farrugia}, C.~J., {Davies}, J.~A., {et~al.} 2012, Astrophys. J.,
  759, 68

\bibitem[{{Lugaz} {et~al.}(2005{\natexlab{a}}){Lugaz}, {Manchester}, \&
  {Gombosi}}]{Lugaz:2005b}
{Lugaz}, N., {Manchester}, W.~B., \& {Gombosi}, T.~I. 2005{\natexlab{a}},
  Astrophys. J., 634, 651

\bibitem[{{Lugaz} {et~al.}(2005{\natexlab{b}}){Lugaz}, {Manchester}, \&
  {Gombosi}}]{Lugaz:2005a}
---. 2005{\natexlab{b}}, Astrophys. J., 627, 1019

\bibitem[{{Lugaz} {et~al.}(2007){Lugaz}, {Manchester}, {Roussev}, {T{\'o}th},
  \& {Gombosi}}]{Lugaz:2007}
{Lugaz}, N., {Manchester}, W.~B., {Roussev}, I.~I., {T{\'o}th}, G., \&
  {Gombosi}, T.~I. 2007, Astrophys. J., 659, 788

\bibitem[{{Lugaz} {et~al.}(2009){Lugaz}, {Vourlidas}, {Roussev}, \&
  {Morgan}}]{Lugaz:2009b}
{Lugaz}, N., {Vourlidas}, A., {Roussev}, I.~I., \& {Morgan}, H. 2009, Solar
  Phys., 256, 269

\bibitem[{{Lynch} \& {Edmondson}(2013)}]{Lynch:2013}Lynch, B.J.~and 
Edmondson, J.K. 2013, Astrophys. J., 764, 87

\bibitem[{{Manchester} {et~al.}(2004{\natexlab{a}}){Manchester}, {Gombosi},
  {Roussev}, {De Zeeuw}, {Sokolov}, {Powell}, {T{\' o}th}, \&
  {Opher}}]{Manchester:2004a}
{Manchester}, W.~B., {Gombosi}, T.~I., {Roussev}, I., {et~al.}
  2004{\natexlab{a}}, J. Geophys. Res., 109, 1102

\bibitem[{{Manchester} {et~al.}(2004{\natexlab{b}}){Manchester}, {Gombosi},
  {Roussev}, {Ridley}, {De Zeeuw}, {Sokolov}, {Powell}, \& {T{\'
  o}th}}]{Manchester:2004b}
---. 2004{\natexlab{b}}, J. Geophys. Res., 109, 2107

\bibitem[{{Manchester} {et~al.}(2012){Manchester}, {van der Holst}, {T{\'o}th},
  \& {Gombosi}}]{Manchester:2012}
{Manchester}, IV, W.~B., {van der Holst}, B., {T{\'o}th}, G., \& {Gombosi},
  T.~I. 2012, Astrophys. J., 756, 81

\bibitem[{{Marubashi} \& {Lepping}(2007)}]{Marubashi:2007}
{Marubashi}, K., \& {Lepping}, R.~P. 2007, Annales Geophysicae, 25, 2453

\bibitem[{{Nieves-Chinchilla} {et~al.}(2012){Nieves-Chinchilla}, {Colaninno},
  {Vourlidas}, {Szabo}, {Lepping}, {Boardsen}, {Anderson}, \&
  {Korth}}]{Nieves:2012}
{Nieves-Chinchilla}, T., {Colaninno}, R., {Vourlidas}, A., {et~al.} 2012, J.
  Geophys. Res., 117, 6106

\bibitem[{{Odstrcil} {et~al.}(2003){Odstrcil}, {Vandas}, {Pizzo}, \&
  {MacNeice}}]{Odstrcil:2003}
{Odstrcil}, D., {Vandas}, M., {Pizzo}, V.~J., \& {MacNeice}, P. 2003, in
  American Institute of Physics Conference Series, Vol. 679, Solar Wind Ten,
  ed. {M.~Velli, R.~Bruno, F.~Malara, \& B.~Bucci}, 699--702

\bibitem[{{Patsourakos} {et~al.}(2010){Patsourakos}, {Vourlidas}, \&
  {Stenborg}}]{Spiro:2010b}
{Patsourakos}, S., {Vourlidas}, A., \& {Stenborg}, G. 2010, Astrophys. Journ.
  Lett., 724, L188

\bibitem[{{Pomoell} \& {Vainio}(2012)}]{Pomoell:2012}Pomoell, J.~and 
Vainio, R. 2012, Astrophys. J., 745, 151

\bibitem[{{Roussev} {et~al.}(2003){Roussev}, {Forbes}, {Gombosi}, {Sokolov},
  {DeZeeuw}, \& {Birn}}]{Roussev:2003a}
{Roussev}, I.~I., {Forbes}, T.~G., {Gombosi}, T.~I., {et~al.} 2003, Astrophys.
  Journ. Lett., 588, L45

\bibitem[{{Ruffenach} {et~al.}(2012){Ruffenach}, {Lavraud}, {Owens}, {Sauvaud},
  {Savani}, {Rouillard}, {D{\'e}moulin}, {Foullon}, {Opitz}, {Fedorov},
  {Jacquey}, {G{\'e}not}, {Louarn}, {Luhmann}, {Russell}, {Farrugia}, \&
  {Galvin}}]{Ruffenach:2012}
{Ruffenach}, A., {Lavraud}, B., {Owens}, M.~J., {et~al.} 2012, J. Geophys.
  Res., 117, 9101

\bibitem[{{Savani} {et~al.}(2011){Savani}, {Owens}, {Rouillard}, {Forsyth},
  {Kusano}, {Shiota}, \& {Kataoka}}]{Savani:2011a}
{Savani}, N.~P., {Owens}, M.~J., {Rouillard}, A.~P., {et~al.} 2011, Astrophys.
  J., 731, 109

\bibitem[{{Savani} {et~al.}(2009){Savani}, {Rouillard}, {Davies}, {Owens},
  {Forsyth}, {Davis}, \& {Harrison}}]{Savani:2009}
{Savani}, N.~P., {Rouillard}, A.~P., {Davies}, J.~A., {et~al.} 2009, Annales
  Geophysicae, 27, 4349

\bibitem[{{Savani} {et~al.}(2013){Savani}, {Vourlidas}, {Shiota}, {Linton},
  {Kusano}, {Lugaz}, \& {Rouillard}}]{Savani:2013}
{Savani}, N.~P., {Vourlidas}, A., {Shiota}, D., {et~al.} 2013, Astrophys. J.,
  {\it in revision}

\bibitem[{{Schmidt} \& {Cargill}(2004)}]{Schmidt:2004}
{Schmidt}, J., \& {Cargill}, P. 2004, Annales Geophysicae, 22, 2245

\bibitem[{{Shen} {et~al.}(2012){Shen}, {Wang}, {Wang}, {Liu}, {Liu},
  {Vourlidas}, {Miao}, {Ye}, {Liu}, \& {Zhou}}]{CShen:2012}
{Shen}, C., {Wang}, Y., {Wang}, S., {et~al.} 2012, Nature Physics, 8, 923

\bibitem[{{Shen} {et~al.}(2011){Shen}, {Feng}, {Wang}, {Wu}, {Song}, {Guo}, \&
  {Zhou}}]{FShen:2011}
{Shen}, F., {Feng}, X.~S., {Wang}, Y., {et~al.} 2011, J. Geophys. Res., 116,
  9103

\bibitem[{{Shen} {et~al.}(2013){Shen}, {Shen}, {Wang}, {Feng}, \&
  {Xiang}}]{FShen:2013}
{Shen}, F., {Shen}, C., {Wang}, Y., {Feng}, X., \& {Xiang}, C. 2013, Geophys.
  Res. Lett., 40, 1457

\bibitem[{{Temmer} {et~al.}(2012){Temmer}, {Vr{\v s}nak}, {Rollett}, {Bein},
  {de Koning}, {Liu}, {Bosman}, {Davies}, {M{\"o}stl}, {{\v Z}ic}, {Veronig},
  {Bothmer}, {Harrison}, {Nitta}, {Bisi}, {Flor}, {Eastwood}, {Odstrcil}, \&
  {Forsyth}}]{Temmer:2012}
{Temmer}, M., {Vr{\v s}nak}, B., {Rollett}, T., {et~al.} 2012, Astrophys. J.,
  749, 57

\bibitem[{{Titov} \& {D{\'e}moulin}(1999)}]{Titov:1999}
{Titov}, V.~S., \& {D{\'e}moulin}, P. 1999, Astron. Astrophys., 351, 707

\bibitem[T{\"o}r{\"o}k {et 
al.}(2011)]{Torok:2011}T{\"o}r{\"o}k, T., Panasenco, O., Titov, 
V.S., Miki{\'c}, Z., Reeves, K.K., Velli, M., Linker, J.A., and De Toma, 
G. 2011, Astrophys. Journ. Lett., 739, L63 

\bibitem[{{T{\'o}th} {et~al.}(2007){T{\'o}th}, {De Zeeuw}, {Gombosi},
  {Manchester}, {Ridley}, {Roussev}, \& {Sokolov}}]{Toth:2007}
{T{\'o}th}, G., {De Zeeuw}, D.~L., {Gombosi}, T.~I., {et~al.} 2007, Space
  Weather, 5, {S06003}

\bibitem[{{T{\'o}th} {et~al.}(2012){T{\'o}th}, {van der Holst}, {Sokolov}, {de
  Zeeuw}, {Gombosi}, {Fang}, {Manchester}, {Meng}, {Najib}, {Powell}, {Stout},
  {Glocer}, {Ma}, \& {Opher}}]{Toth:2012}
{T{\'o}th}, G., {van der Holst}, B., {Sokolov}, I.~V., {et~al.} 2012, Journal
  of Computational Physics, 231, 870

\bibitem[{{van der Holst} {et~al.}(2010){van der Holst}, {Manchester},
  {Frazin}, {V{\'a}squez}, {T{\'o}th}, \& {Gombosi}}]{Holst:2010}
{van der Holst}, B., {Manchester}, W.~B., {Frazin}, R.~A., {et~al.} 2010,
  Astrophys. J., 725, 1373

\bibitem[{{Vandas} {et~al.}(1997){Vandas}, {Fischer}, {Dryer}, {Smith},
  {Detman}, \& {Geranios}}]{Vandas:1997}
{Vandas}, M., {Fischer}, S., {Dryer}, M., {et~al.} 1997, J. Geophys. Res., 102,
  22295

\bibitem[{{Wang} {et~al.}(2002){Wang}, {Wang}, \& {Ye}}]{Wang:2002}
{Wang}, Y.~M., {Wang}, S., \& {Ye}, P.~Z. 2002, Solar Phys., 211, 333

\bibitem[{{Wang} {et~al.}(2003{\natexlab{a}}){Wang}, {Ye}, \&
  {Wang}}]{Wang:2003}
{Wang}, Y.~M., {Ye}, P.~Z., \& {Wang}, S. 2003{\natexlab{a}}, J. Geophys. Res.,
  108, 6

\bibitem[{{Wang} {et~al.}(2003{\natexlab{b}}){Wang}, {Ye}, {Wang}, \&
  {Xue}}]{Wang:2003a}
{Wang}, Y.~M., {Ye}, P.~Z., {Wang}, S., \& {Xue}, X.~H. 2003{\natexlab{b}},
  Geophys. Res. Lett., 30, 33

\bibitem[{{Webb} {et~al.}(2009){Webb}, {Howard}, {Fry}, {Kuchar}, {Odstrcil},
  {Jackson}, {Bisi}, {Harrison}, {Morrill}, {Howard}, \&
  {Johnston}}]{Webb:2009}
{Webb}, D.~F., {Howard}, T.~A., {Fry}, C.~D., {et~al.} 2009, Solar Phys., 256,
  239

\bibitem[{{Webb} {et~al.}(2013){Webb}, {M{\"o}stl}, {Jackson}, {Bisi},
  {Howard}, {Mulligan}, {Jensen}, {Jian}, {Davies}, {de Koning}, {Liu},
  {Temmer}, {Clover}, {Farrugia}, {Harrison}, {Nitta}, {Odstrcil}, {Tappin}, \&
  {Yu}}]{Webb:2013}
{Webb}, D.~F., {M{\"o}stl}, C., {Jackson}, B.~V., {et~al.} 2013, Solar Phys.,
  285, 317

\bibitem[{{Wu} {et~al.}(2007){Wu}, {Fry}, {Wu}, {Dryer}, \& {Liou}}]{WuCC:2007}
{Wu}, C.-C., {Fry}, C.~D., {Wu}, S.~T., {Dryer}, M., \& {Liou}, K. 2007, J.
  Geophys. Res., 112, {A09104}

\bibitem[{{Wu} {et~al.}(2002){Wu}, {Wang}, \& {Gopalswamy}}]{Wu:2002}
{Wu}, S.~T., {Wang}, A.~H., \& {Gopalswamy}, N. 2002, in ESA SP-505: SOLMAG
  2002. Proceedings of the Magnetic Coupling of the Solar Atmosphere
  Euroconference, 227--230

\bibitem[{{Xie} {et~al.}(2006){Xie}, {Gopalswamy}, {Manoharan}, {Lara},
  {Yashiro}, \& {Lepri}}]{Xie:2006}
{Xie}, H., {Gopalswamy}, N., {Manoharan}, P.~K., {et~al.} 2006, J. Geophys.
  Res., 111, 1103

\bibitem[{{Xiong} {et~al.}(2009){Xiong}, {Zheng}, \& {Wang}}]{Xiong:2009}
{Xiong}, M., {Zheng}, H., \& {Wang}, S. 2009, J. Geophys. Res., 114, A11101

\bibitem[{{Xiong} {et~al.}(2006{\natexlab{a}}){Xiong}, {Zheng}, {Wang}, \&
  {Wang}}]{Xiong:2006}
{Xiong}, M., {Zheng}, H., {Wang}, Y., \& {Wang}, S. 2006{\natexlab{a}}, J.
  Geophys. Res., 111, {A08105}

\bibitem[{{Xiong} {et~al.}(2006{\natexlab{b}}){Xiong}, {Zheng}, {Wang}, \&
  {Wang}}]{Xiong:2006b}
---. 2006{\natexlab{b}}, J. Geophys. Res., 111, {A11102}

\bibitem[{{Xiong} {et~al.}(2007){Xiong}, {Zheng}, {Wu}, {Wang}, \&
  {Wang}}]{Xiong:2007}
{Xiong}, M., {Zheng}, H., {Wu}, S.~T., {Wang}, Y., \& {Wang}, S. 2007, J.
  Geophys. Res., 112, 11103

\end{thebibliography}

\end{document}